%% file: main.tex
\newif\ifabstract
\abstracttrue
 \abstractfalse %Uncommenting this line gives the full version
\newif\iffull
\ifabstract \fullfalse \else \fulltrue \fi
\documentclass[11pt]{article}
\usepackage{amsfonts}
\usepackage{amssymb}
\usepackage{amstext}
\usepackage{amsmath}
\usepackage{xspace}
\usepackage{theorem}
\usepackage{graphicx}
\usepackage{url}
\usepackage{graphics}
\usepackage{colordvi}
\usepackage{colordvi}
\usepackage{subfigure}

\usepackage{natbib}
\usepackage{wasysym}

\usepackage{subeqnarray}
\usepackage{xcolor}
\usepackage{moreverb} 
\usepackage{textcomp} 
\usepackage{epsfig}
\usepackage{color} 
\usepackage{upgreek}
\usepackage{bbm}

\usepackage{dcolumn}% Align table columns on decimal point
\usepackage{bm}% bold math
\newcommand{\overbar}[1]{\mkern 1.5mu\overline{\mkern-1.5mu#1\mkern-1.5mu}\mkern 1.5mu}

\advance \oddsidemargin by -0.8in
\newcommand{\myparskip}{3pt}
\parskip \myparskip

\newsavebox{\astrutbox}
\sbox{\astrutbox}{\rule[-5pt]{0pt}{20pt}}

\begin{document}

\title{Limiting regimes of turbulent horizontal convection. Part I: Intermediate and low Prandtl numbers}

\author{Pierre-Yves Passaggia\thanks{University of Orl\'eans, INSA-CVL, PRISME, EA 4229, 45072, Orl\'eans, France, Email: pierre-yves.passaggia@univ-orleans.fr {\tt pierre-yves.passaggia@univ-orleans.fr}. Also affiliated: Carolina Center for Interdisciplinary Applied Mathematics, Dept. of Mathematics, University of North Carolina, Chapel Hill, NC 27599, USA}
\and Alberto Scotti\thanks{Department of Marine Sciences, University of North Carolina, Chapel Hill, NC 27599, USA. Email: {\tt ascotti@unc.edu}.} 
\and Brian L. White\thanks{Department of Marine Sciences, University of North Carolina, Chapel Hill, NC 27599, USA. Email: {\tt bwhite@unc.edu}.}}

% \begin{titlepage}
\maketitle

\thispagestyle{empty}

\begin{abstract}
We report the existence of two new limiting turbulent regimes in horizontal convection (HC) using direct numerical simulations at intermediate to low Prandtl numbers. The flow driven by a horizontal gradient along a horizontal surface, perpendicular to the acceleration of gravity is shown to transition to turbulence in the plume and the core, modifying the rate of heat and momentum transport. These transitions set a sequence of scaling laws blending the theoretical arguments from both the Shishkina, Grossmann \& Lohse (SGL) theory \cite{ShishkinaGL16} with the Hughes, Griffiths \& Mullarney (HGM) regime \cite{Hughes07}. These results embed the HGM model in the SGL theory, agree, and extend the known regime diagram of horizontal convection \cite{hughes2008horizontal} at high Rayleigh numbers. In particular, we show that HC and Rayleigh-B\'enard share similar turbulent characteristics at low-Prandtl numbers, where HC is shown for the first time to be ruled by its core dynamics and turbulent boundary layers. This new scenario confirms that fully turbulent HC enhances the transport of heat and momentum with respect to previously reported regimes at high Rayleigh numbers. This work provides new insights on the applicability of horizontal convection for geophysical flows such as overturning circulations found in the atmosphere, the oceans, and flows near the earth's inner core.
\end{abstract}

%\pacs{}
%\keywords{}

%\maketitle

%\section{}
%\subsection{}
%\subsubsection{}
%

%\maketitle

%\tableofcontents

\section{Introduction}\label{sec:Intro}

Since the early work of Sandstr{\"o}m \cite{Sandstrom08} on marine glacial discharges in Norwegian Fjords and his pioneering work on ocean circulation through fresh/cold and salty/warm water differential inputs, attempts to link differential heating \cite{Rossby65,Rossby98} and/or salt and fresh water input to a deep meridional circulation capable of driving the world's ocean circulation led to series of results predicting that differential buoyancy forcing on a horizontal surface alone could not explain the deep water cycle that, over a a millennial time scale, conveys the world's ocean waters around the globe \cite{defant1961physical}. The amount of circulation, which can be sought in terms of dissipation of kinetic energy in natural convection, is bounded by the amount of heat uptake absorbed at the surface \cite{PaparellaY02}. Natural convection driven by a buoyancy gradient along a geopotential iso-surface is a particular flow archetype. 
Paparella \& Young\cite{PaparellaY02} derived a bound on the amount of dissipation and argued that when viscosity vanishes, the turbulent dissipation in HC also vanishes, unlike, Rayleigh-B\'enard convection where it is expected to reach a finite value. Although that "anti-turbulence" theorem has been established, horizontal convection was shown to undergo turbulence\cite{ScottiW11,Gayen14} but the transport of heat and momentum between the buoyancy the sources and sinks is expected to follow scaling exponents which are essentially lower than Rayleigh-B\'enard convection\cite{ShishkinaGL16} in the Prandtl-Rayleigh landscape (cf. except in the particular case of laminar steady HC flows at high Prandtl numbers\cite{ShishkinaW16}).
This is in part due to the rate at which the energy of the flow is dissipated which, so far, proved to consistently follow laminar-type scaling laws with respect to the magnitude of the forcing \cite{Rossby98,Hughes07,ShishkinaW16}, as long as the buoyancy gradient was unidirectional \cite{griffiths2015turbulent}, and this, despite the flow being unstable with respect to two- and three-dimensional perturbations \cite{Gayen14,PassaggiaSW17}. An analogue of Rayleigh-B\'enard theory was recently proposed and applied to horizontal convection with the aim of characterising the regime diagram of laminar and turbulent regimes \cite{ShishkinaGL16}. However the validity of a such map is currently under investigation \cite{ShishkinaW16,PassaggiaSW18}. In this work, we investigate numerically the low-Prandtl region of this regime diagram. Our aim is to investigate which limiting regimes are effectively observed and where turbulent horizontal convection starts to appear. According to the recent work of Shishkina {\it et al.} \cite{ShishkinaGL16}, the transition to the limiting turbulent regime, appearing at sufficiently high Rayleigh numbers, should be observed first at low Prandtl numbers and we show here that it is the case.\\

Griffiths \& Gayen \cite{griffiths2015turbulent} considered the problem of horizontal convection forced by spatially periodic forcing. Their result showed that  
horizontal convection would become turbulent in the core. Their forcing, localised on a length scale smaller than the depth of the domain, and with variation in both horizontal
directions show turbulence throughout the domain, a regime transition to a dominant domain-scale circulation, and a region of logarithmic velocity in the boundary layer. The same geometry was further analysed by Rosevear {\it et al.} \cite{rosevear2017turbulent} where they observed that the non-dimensional heat flux, denoted by the Nusselt number had a steeper scaling with respect to the Rayleigh number than the (laminar) Rossby scaling\cite{Rossby65}. Their scaling analysis suggest that for deep enough domains, the flow is fully driven by the core (i.e. the interior) of the flow, located between the boundary layer and the opposite side of the domain.
One interesting fact is that despite the existence of a log-layer in their direct numerical simulation, they did not observe log-type corrections in the scaling for the heat transfer. This is relatively surprising since it is now well established in Rayleigh-B\'enard convection that heat transfers are buffered through the log-layer\cite{GrossmannL11,ahlers2012logarithmic,ahlers2014logarithmic}. 

Recent work by Shishkina \& Wagner\cite{ShishkinaW16} report a similar exponent in the case of large Prandlt number and low Rayleigh numbers. Their study shows that when the boundary layer extends all the way to the bottom of the domain, horizontal convection was highly effective at transporting heat. In addition, they report the dependence on Prandtl numbers and in their study which follows either their new regime, denoted by $I^*_l$ in their study or the Rossby regime denoted by $I_l$. In our analysis we follow the same nomenclature in an attempt to unify the results of both aforementioned groups. Note that these results are also observed experimentally in the companion paper \cite{Passaggia2019LimitigB}.

Very recently, Reiter \& Shishkina \cite{reiter2020classical} analysed classical and symmetrical horizontal convection in Rayleigh numbers up to $10^{12}$ and three Prandtl numbers. They found that for large Rayleigh numbers at $10^{11}$ and large-aspect ration domains, the plume detaches and exhibits low-frequency oscillations while the Nusselt exhibits locally a steeper scaling. In this paper, we confirm these results in a different setup and extend the Rayleigh number range by three orders of magnitude, up to $10^{15}$.

While the ratio of viscosity to heat diffusion, taken here as the Prandtl number ($\rm Pr$) is $O(1)$ or larger in atmospheric and oceanic applications, 
Horizontal Convection (HC) at low Prandtl numbers has interesting geophysical applications, such as in the highly thermally conductive part of the mantle. Although a lot of attention has been devoted to Rayleigh-B\'enard Convection (RBC) for the outer core's dynamics, it is only very recently that HC has attracted the attention of planetary scientists \cite{alboussiere:12}. For example, Takehiro\cite{takehiro:11} suggest that it could be a potential mechanism to drive zonal heat and momentum near the inner core through the Joule effect due to Earth's magnetic field. 
%Lateral motions in the outer core are responsible for driving large scale horizontal flows in the mantle and have long been thought to be at the origin of striping and faulting of tectonic plates and strike-slip earthquakes \cite{Knopoff63}. 
At the edge of Earth's inner core, horizontal regions of thermally stable (crystallising) and unstable (melting) stratified layers explain the East-West asymmetry of the inner core\cite{alboussiere:12}. However, only very little is known about the properties of the turbulent horizontal flows generated in these regions and HC appears as an interesting candidate to analyse such flows.\\

In this study, we report Direct Numerical Simulation (DNS) results on how the Reynolds number ($\rm{Re}$) and the Nusselt number ($\rm{Nu}$) depend on the Rayleigh number ($\rm{Ra}$) and the Prandtl number ($\rm{Pr}$) in turbulent HC at low to intermediate $\rm{Pr}$ for values characteristic of convection in gases where $0.1<\rm{Pr}<1$ \cite{Roche02,Taylor13},
and liquid metals where $\rm{Pr}=\mathcal{O}(10^{-2})$ (see ref.\cite{takehiro:11}). The results are in agreement with the scaling power laws recently derived by Shishkina {\it et al.} \cite{ShishkinaGL16} based on the original work of Grossmann \& Lohse \cite{GL00} (GL) and numerical simulations of Takehiro \cite{takehiro:11}. Furthermore, we provide evidence that the regime observed by Rosevear {\it et al.} \cite{rosevear2017turbulent} generalises to horizontal convection over a monotonic temperature profile with a turbulent log-layer which indeed acts as a buffer to heat transfer and slightly decreases the exponent previously reported. It also provides for the first time, a connection between the GL theory and the plume driven dynamics derived by Hughes {\it et al.} \cite{Hughes07}.
%THIS MAYBE A PET PEEVE OF MINE, BUT I FOUND IT JARRING THAT WE REPORT THE SCALING OF A DIMENSIONAL QUANTITY (EPSILON)... IF WE RESCALE WITH THE SAME VELOCITY AND LENGTH SCALE THAT WE USE FOR THE REYNOLDS NUMBER... 
% Indeed, there is no need of introducing epsilon here. Furthermore, it is only used dissipation to check the quality of the DNS. 

Our simulations cover the laminar Rossby regime $I_l$ (see ref.\cite{Rossby65}), the high-$\rm{Pr}$ laminar regime $I^*_l$ recently reported by Shishkina  \& Wagner \cite{ShishkinaW16} and a new low-$\rm{Pr}$ turbulent regimes named $II_l$ (see ref.\cite{shishkina2017scaling,GL00} for theoretical predictions of HC and RBC), 
which is a new turbulent limiting regime reported in HC. We also observe the plume dominated flow regime of Hughes {\it et al.} (see ref.\cite{Hughes07}), that we name $II_u$ according to the SGL theory. We also report the existence of the turbulent interior-dominated regime $IV_u$ at high Rayleigh number amended with the appropriate log-type corrections. An important contribution of our work is that these results agree and extend the regime diagram of horizontal convection proposed in Hughes \& Griffiths (see ref.\cite{hughes2008horizontal}) to fit within the theoretical prediction Shishkina {\it et al.} \cite{ShishkinaGL16}, blending all known regimes of horizontal convection (See the companion paper \cite{Passaggia2019LimitigB}).

In the final section, we further explore the relation between the Reynolds number characterising the magnitude of the overturning flow and turbulent dissipation. This analysis allows for condensing this complex regime transitions diagram into a more traditional laminar, transitional, soft turbulence, and hard turbulence diagram, solely dependent on the Reynolds number. In this section, we further confirm that a hard turbulent regime cannot be observed for the Prandtl numbers considered in this work. This assumption is justified theoretically with a bound on the minimum Richardson in the stably stratified layer which cannot even approach the threshold for instabilities.

Similarly to Shishkina \& Wagner \cite{ShishkinaW16}, we exploit the idea that in turbulent thermal convection, the time- and volume-averaged thermal and viscous dissipation rates are determined to leading order by their bulk or Boundary Layer (BL) contributions. For the ease of comparison, we follow the same presentation as Shishkina \& Wagner \cite{ShishkinaW16}.

\section{Problem description}

We consider here the problem of convection in the Boussinesq limit, where the density difference $\Delta\rho=\rho_{max}-\rho_{min}$ across the horizontal surface is a small deviation from the reference density $\rho_{min}$. In this limit, the equations of fluid motion are
\begin{subeqnarray}
\frac{D \mathbf{u}}{D t} &=& -\nabla p + b\mathbf{e_z}  + \left(\frac{\rm{Pr}}{\rm{Ra}}\right)^{1/2}\nabla^2 \mathbf{u},\\
\nabla\cdot\mathbf{u}&=&0, \\
\frac{D b}{D t} &=& \left(\rm{Pr}\,\rm{Ra}\right)^{-1/2}\nabla^2 b,
\label{NS}
\end{subeqnarray}
where $D/Dt$ denotes the material derivative, $\mathbf{u}=(u,v,w)^T$ is the velocity vector, $b=-g(\rho-\rho_{min})/\rho_{min}$ is the buoyancy, $g$ is the acceleration of gravity along the vertical unit vector $\mathbf{e}_z$ and $p$ is the hydrodynamic pressure.

The Prandtl number is given by $\rm{Pr}=\nu/\kappa$ where $\nu$ and $\kappa$ are the viscosity and stratifying agent's diffusivity respectively.
The Rayleigh number is defined such that ${\rm Ra}=\Delta L^3/(\nu\kappa)$ where $L$ is the horizontal length scale of the domain and $\Delta=-g(\rho_{max}-\rho_{min})/\rho_{min}$. The computational domain is a parallelepiped of aspect ratio $\Gamma=L/H=4$ with dimensions $[L,W,H]=[1,1/8,1/4]$ where $W$ is the width of the computational domain \cite{Scotti08}. A buoyancy profile is imposed at the surface $z=H$ where $H$ is the height of the domain using a buoyancy profile such that $b(x)|_{z=H}=(1+\tanh(9.5x))/2$, which is a smoothed version of the sharp profile used in our previous calculations \cite{ScottiW11,PassaggiaSW17}. This proved to be necessary in order to keep the numerical strategy stable in the forthcoming numerical simulations.
The equations are non-dimensionalized using the length of the box $L$ as reference length and the buoyancy difference imposed along the non-isolating horizontal boundary such that
\begin{equation}
\mathbf{x}=\mathbf{x}^*/L, \;\; b=b^*/\Delta , \;\; \mathbf{u}=\mathbf{u}^*/\sqrt{L\Delta}.
\end{equation}

In what follows, we define integral values to be linked with the control parameters $\rm Ra$ and $\rm Pr$: the magnitude of the large-scale flow is defined $\rm{Re}={(\overbar{\mathbf{u}\cdot\mathbf{u}})}^{1/2}L/\nu$ where the overbar denotes the spatio-temporal average over the computational domain, similarly the P\'eclet number $\rm{Pe}={(\overbar{\mathbf{u}\cdot\mathbf{u}})}^{1/2}L/\kappa$. For the Nusselt, we use ${\rm Nu}=\overbar{\partial b/\partial z}|_{z=H,b=1}/\Phi_c$, where $\Phi_c=\overbar{\partial b_c/\partial z}|_{z=H,b_c=1}$ is the average gradient in the purely conducting case (i.e. when ${\rm Ra}<10^3$)\cite{siggers2004bounds}, though other definitions have been considered \cite{rocha2019heat}. Its value depends only on the geometry of the domain and of the boundary conditions. For the geometry considered here, its value was found numerically to be $\Phi_c=0.53\, \Delta L$. 

 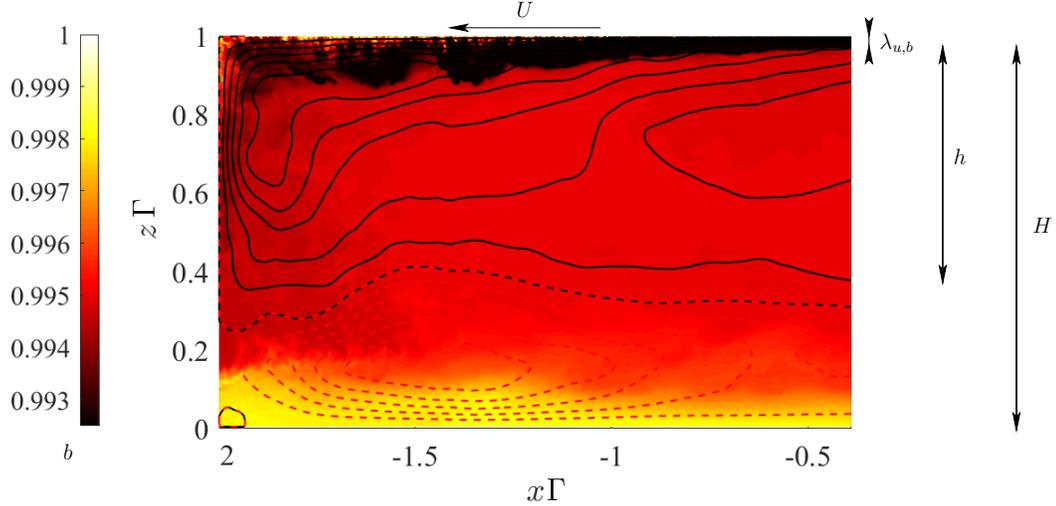
\begin{figure*}[t]
 \centering
 \scalebox{0.7}{\input{schematics.tex}}\put(-187,4.5){\large$\Gamma$}\put(-343.5,111.5){\large\rotatebox{90}{$\Gamma$}}%
 \caption{Schematic of the present setup using a snapshot of a simulation performed at $\rm Ra=6.4\, 10^{14}$ and $\rm Pr=0.1$ where the colour is the value of the buoyancy $b$, the solid line is counter-clockwise rotating streamfunction whereas the red dashed line represents the clockwise rotating part. 
 To the right, the different length scales of the thermal BL $\lambda_b$ and the kinetic BL $\lambda_u$ are shown together with the full depth $H$ and the large overturning scale $h$ used for the theoretical prediction.}
\label{schematic}
\end{figure*}

\section{Previously known regimes of laminar and turbulent HC}\label{sec:preregimes}

In this section, we review the existing scaling laws for heat and momentum exchanges in horizontal convection. 
%The known regimes are shown in Fig. \ref{Ra_Pr}. 
Parts of the landscape previously explored using direct numerical simulations and experiments are reported together with the parts of the Rayleigh-Prandtl map investigated in this paper and its companion Part II\cite{Passaggia2019LimitigB}. %The points shown in the top right part occupy a previously unexplored region of the parameter space. 
The following subsections introduce these known scaling exponents, tied with what is known as the Paparella \& Young \cite{PaparellaY02} inequality which relates the mean mechanical dissipation of the system with the input of heat through the horizontal boundary and opens the door for choices in modelling which lengthscales drive convection in different regimes.

\subsection{Rossby's (1965) original idea}

H. Rossby\cite{Rossby65} explored horizontal convection induced by differential heating in a parallelepipedic container with an aspect ratio $L/H=2.5$ and derived a scaling law relating the Nusselt number as a function of the Rayleigh number. In his original work, Rossby analysed temperature measurements from his experiments to derive a scaling relationship between the thickness of the boundary layer and the streamfunction (cf. pp. 13 in\cite{Rossby65}). Taking the curl of eq. (\ref{NS}a), neglecting the nonlinear terms, and defining the streamfunction $\psi$ such that $\psi_x=-w$ and $\psi_z=u$, the two dimensional Navier-Stokes equations reduce to
\begin{subeqnarray}
\left(\frac{\rm{Pr}}{\rm{Ra}}\right)^{1/2}\nabla^4 \psi &=& \partial_x  b\\
-\partial_x\psi\partial_zb &=& \left(\rm{Pr}\,\rm{Ra}\right)^{-1/2}\nabla^2 b.
\label{stream_func}
\end{subeqnarray}
Near the conducting wall, the flow is governed by the laminar boundary-layer whose thickness is defined by $\lambda$ and eqs. (\ref{stream_func}a),(\ref{stream_func}b) reduce at leading order to
\begin{equation}
\left(\frac{\rm{Pr}}{\rm{Ra}}\right)^{1/2} \frac{\psi}{\lambda^4} \sim \frac{\Delta}{L}, \quad \mbox{and} \quad
\frac{\psi \Delta}{\lambda L} \sim \left(\rm{Pr}\,\rm{Ra}\right)^{-1/2}  \frac{\Delta}{\lambda^2}.
\label{Rossby_scale}
\end{equation}
Combining these equations, Rossby obtained the relation
\begin{equation}
\lambda \sim L Ra^{-1/5}.
\end{equation}
What Rossby had not recognised in his original work was that the thickness of the boundary layer $\lambda$ could be defined using either the thermal boundary layer thickness, denoted by the subscript $_b$ and the kinetic boundary layer denoted by the subscript $_u$, which is one of the important aspects of this study. While this has no implication for the $\rm Ra$-dependence as shown in the next subsection, the Prandtl-number dependence may not be predicted accurately for different values of the Rayleigh number $\rm Ra$ and varying $\rm Pr$.

\subsection{Paparella \& Young (2002) inequality}

Horizontal and Rayleigh-B\'enard convection are both closed systems driven by the heat/buoyancy flux imposed through their boundaries. Paparella and Young (PY) \cite{PaparellaY02} first performed a spatio-temporal average of the kinetic-energy equation (i.e. $\overbar{\mathbf{u}\cdot(\mbox{\ref{NS}a})}$) leading to the equality
\begin{equation}
\overbar{\epsilon_u} \;=\; \overbar{wb}, %\;\leq\; B(\Gamma/2)\nu^{3}L^{-4}\rm{Ra}\rm{Pr}^{-2},
\label{epsu1}
\end{equation}
where $\overbar{\epsilon_u}$ is the mean kinetic-energy-dissipation rate $\overbar{\epsilon_u}\equiv\nu\sum_{i,j}(\partial u_j/\partial x_i)^2$. Another condition can be written using the spatio-temporal average of eq. (\ref{NS}c), and integrating over $z$ leads to
\begin{equation}
\overbar{wb}=\kappa\langle \partial b/\partial z\rangle_H,
\end{equation}
where $\langle\rangle_{H}$ denotes the surface and time average at $z=H$. This equality can be recast into an inequality for the buoyancy between the top an the bottom of the domain which writes
\begin{equation}
\overbar{wb}\leq\kappa(\langle b\rangle_{z=H}-\langle b\rangle_{z=0})/H = B(\Gamma/2)\kappa\Delta/L,
\label{PYbound}
\end{equation}
where $1<B<0$ is an arbitrary constant which depends on the domain geometry and boundary conditions\cite{ShishkinaW16}. The PY inequality thus writes
\begin{equation}
\overbar{\epsilon_u} \;=\; B(\Gamma/2)\nu^{3}L^{-4}\rm{Ra}\rm{Pr}^{-2},
\label{epsu2}
\end{equation}
which, once combined with the original idea of Rossby, opens possibilities for relating the dissipation in the boundary layer or the core with the heat transfer coefficient near the horizontal boundary.\\

One interesting fact is that PY's inequality suggests that as $\rm Ra$ increases while keeping $\rm Pr$ and $\Gamma$ constant, the flow becomes progressively confined under the conducting boundary. This effect is also known as the anti-turbulence theorem and implies that beyond a certain point, the overturning depth scale becomes 
$$h<H,$$
and a zone of stratified fluid nearly at rest will form on the insulating boundary adjacent to the conducting horizontal boundary. 
%In other words, one may recast the PY inequality eq. (\ref{PYbound})
%
%\begin{equation}
%\overbar{\epsilon_u} = B/2(L/h)\nu^{3}L^{-4}\rm{Ra}\rm{Pr}^{-2},
%\label{epsunew}
%\end{equation}
%where dissipation is only bounded to the turbulent core of depth $h$ rather than the entire depth $H$. Note that Shishkina {\it et al.}\citep{ShishkinaGL16} refer to this as the large scale overturning flow in their theory.

This follows Sandstr\"om\cite{Sandstrom16} inference where at large $\rm Ra$ or for high $\rm Pr$, the flow becomes confined to a progressively thinner surface layer and the core becomes a stagnant pool of stratified water\cite{defant1961physical}. Although such regimes were only observed in direct numerical simulations of laminar HC \cite{ilicak2012simulations} at high Pr and theoretically by \cite{chiu2008very} for the same regimes, experiments by Wang \& Huang\cite{wang2005experimental} show the onset of such behaviour at intermediate $\rm Pr$ and relatively low $\rm Ra$.\\

\subsection{Rossby's laminar regime $I_l$}

Rossby's laminar regime can be recast to obtain a more accurate prediction for the Prandtl number dependence. The idea is to start with the steady thermal boundary layer equation, which is obtained from eq. (\ref{NS}c) and write an advection-diffusion balance in the boundary layer
\begin{equation}
u b_x + v b_z = \kappa b_{zz}.
\end{equation}
The dominant terms in this expression reduce to $U\Delta/L = \kappa \Delta/\lambda_b^2$ where $\lambda_b$ is the thickness of the thermal BL, which scales as $\lambda_b \sim \rm{Nu}^{-1}$. Combining the above reduces to the well known thermal-laminar BL scaling
\begin{equation}
\rm Nu=Re^{1/2}Pr^{1/2},
\label{Nu}
\end{equation}
and provides a relation tying $\rm{Nu}$, $\rm{Re}$ and $\rm{Pr}$.
In laminar regimes, the buoyancy variance is essentially concentrated in the boundary layer and writes
\begin{equation}
\overbar{\epsilon_{b, \mathrm{BL}}} \sim \kappa \frac{\Delta^{2}}{\lambda_{b}^{2}} \frac{\lambda_{b}}{h}=\kappa \frac{\Delta^{2}}{h^{2}} \frac{\lambda_{u}}{\lambda_{b}} \operatorname{Re}^{1 / 2},
\label{epsbbl_lam}
\end{equation}
where the dependence on the aspect ratio $\Gamma$ was omitted.  
Noting that the thickness of the laminar boundary layer scales as $\lambda_u/H \sim Re^{-1/2}$, the scaling for the mean dissipation in the particular case of laminar BL\cite{Landau87} is
\begin{equation}
\overbar{\epsilon_{u,BL}}\sim\nu\frac{U^2}{\lambda^2_u}\frac{\lambda_u}{h}=\nu^3h^{-4}{\rm Re}^{5/2}.
\label{epsu_lam}
\end{equation}
Combining (\ref{Nu}), (\ref{epsu2}) and (\ref{epsu_lam}), and assuming that $h=H$, one recovers the laminar scaling \cite{Rossby65,Rossby98,Gayen14,ShishkinaGL16}
\begin{subeqnarray}
\rm{Re} &\sim& \rm{Ra}^{2/5}\rm{Pr}^{-4/5}, \\
\rm{Nu} &\sim& \rm{Ra}^{1/5}\rm{Pr}^{1/10}.
\label{lam_scal}
\end{subeqnarray}
By analogy to the notation in the GL theory for RBC \cite{GL00,ShishkinaGL16}, this scaling regime is denoted as $I_l$, where the subscript $l$ stands for low-$\rm{Pr}$ fluids.

\subsection{Hughes {\bf{\it et al.}}'s (2007) laminar boundary-layer/turbulent plume regime $II_u$}

Increasing $\rm{Ra}$ and for intermediate $\rm{Pr}$, the kinetic boundary layer becomes progressively thinner while the boundary remains relatively thick in comparison. In this case, it is the thermal boundary layer that drives the dynamics and leads to a turbulent plume, detached from the bottom [see Fig. \ref{3D}(b)]. This particular case was theorised by Hughes {\it et al.}\cite{Hughes07} with a plume model inside a filling box.
Here we recast their model according to the SGL theory (i.e. see the plume model definition eq. (2.15)-(2.20) in  ref.\cite{Hughes07}) and the dissipation in the boundary layer is balanced by the ratio between the thermal and the kinetic boundary layer $\lambda_b/\lambda_u$
which writes
\begin{equation}
\overbar{\epsilon_{u}}_{HGM} \sim\nu\frac{U^2}{\lambda^2_u}\frac{\lambda_u}{h}\frac{\lambda_b}{\lambda_u}= \nu^{3}h^{-4}\rm{Re}^{5/2}\rm{Pr}^{-1/2},
\label{epsuh}
\end{equation}
where the dissipation now scales with the thickness of the thermal layer, not the kinetic BL, where $h=H$ is enforced, and is given by 
\begin{equation}
\overbar{\epsilon_{u}}_{HGM}\sim\nu U^2/(\lambda_b L). 
\end{equation}
Now  combining (\ref{Nu}), (\ref{epsu2}) and (\ref{epsuh}), the heat and momentum exchanges are given by
\begin{subeqnarray}
\rm{Re}&\sim&\rm{Ra}^{2/5}\rm{Pr}^{-3/5}, \\
\rm{Nu}&\sim&\rm{Ra}^{1/5}\rm{Pr}^{1/5},
\label{NuRe1/5h}
\end{subeqnarray}
which is denoted as $II_u$ and was first observed in the experiments of Mullarney {\it et al.}\cite{mullarney2004convection} and Wang \& Huang\cite{wang2005experimental}  and later confirmed in the direct numerical simulations of Gayen {\it et al.}\cite{Gayen14}.\\

\subsection{Shishkina \& Wagner (2016) laminar regime $I^*_l$}

At low $\rm{Ra}$ and for large $\rm{Pr}$ and/or large aspect ratio $\Gamma$, the BL thickness $\lambda_u$ saturates and reaches the depth of the domain which gives $\lambda_u\approx h=H$ and eq. (\ref{epsu1}) becomes equivalent to the dissipation in a pressured-driven laminar flow in a channel which writes 
\begin{equation}
\overbar{\epsilon_{u}}_{SW} \sim\nu\frac{U^2}{H^2}= \nu^{3}H^{-4}\rm{Re}^{2}.
\label{epsu_Shishkina}
\end{equation}
Combining (\ref{Nu}), (\ref{epsu2}) and (\ref{epsu_Shishkina}), one obtains the laminar scaling derived in Shishkina \& Wagner \cite{ShishkinaW16}
\begin{subeqnarray}
\rm{Re}&\sim&\rm{Ra}^{1/2}\rm{Pr}^{-1}, \\
\rm{Nu}&\sim&\rm{Ra}^{1/4}\rm{Pr}^{0},
\label{NuRe1/4}
\end{subeqnarray} 
denoted as $I^*_l$ and first observed by Beardsley \& Festa \cite{beardsley1972numerical} in their numerical simulations. Note that Rossby \cite{Rossby98} also observed a steeper scaling than $\rm Nu\sim {\rm Ra}^{1/5}$ in his numerical simulations for low $\rm Ra$ (see page 248 in \cite{Rossby98}). More recent numerical simulations performed by Ramme \& Hansen \cite{ramme2019transition} at infinite Prandtl numbers suggests a similar scaling. However, it should be noted that the observation of this scaling is hardly reported across a decade of Rayleigh numbers in both studies.
It is important to stress that in this regime, The circulation is assumed to span the entire box and this particular aspect will be further explored in the present and companion paper \cite{Passaggia2019LimitigB}.\\ 

\subsection{Turbulent regimes and associated bounds}

Most of the existing work on HC highlighted laminar-type flows, dominated by the behaviour of the boundary layer, at the exception of an analogue of HC  Griffiths \& Gayen (2015)\cite{griffiths2015turbulent} and Rosevear {\it et al.}\cite{rosevear2017turbulent}. In a recent study, they considered a spatially periodic forcing at the conducting boundary with a short wavelength compared to the depth of the domain. In this particular setup, Rosevear {\it et al.} were able to show that $Nu\sim Ra^{1/4}$ with a turbulent-core driven by inertia. They were also the first to report turbulent boundary layers with a log-type layer developing along the conducting wall. This feature is important since it is a necessary condition for turbulent convection scaling to arise in Rayleigh-B\'enard convection\cite{GL00,GrossmannL11}.\\ 

Siggers {\it et al.}'s (2004)\cite{siggers2004bounds} theorised a fully turbulent regime based on the assumption that boundary layers and thus the Nusselt number scales as $\rm Nu\sim Ra^{1/3}$. Recent work by Rocha {\it et al.} refined their original results and showed that short oscillations with large amplitudes of the forcing boundary where favourable to observe a such regime when $\rm Ra\rightarrow \infty$.
However, a such regime is unlikely to exist for a forcing such as step-like boundary condition or a linear profile where the stably stratified layer remains undisturbed over a sufficiently long span,
as suggested by Passaggia {\it et al.}\cite{passaggia2016global} who recast the PY inequality with eq. (\ref{PYbound}) to obtain a bound on the Richardson number in the stable layer underneath the conducting boundary, that is 
%
%\begin{equation}
%Ri \equiv \left(\frac{N}{\partial_z U}\right)^2 \sim \frac{\Delta}{\lambda_b}\left(\frac{\lambda_u}{U}\right)^2 
%\sim \Delta Nu Re^{-2n} U^{-1}
%\sim \Delta \nu^{-1} L Re^{2n+m-1} . 
%\quad \mbox{where} \quad \lambda_u\sim Re^n 
%\quad \mbox{and} \quad Nu\sim Re^m.
%\end{equation}
%

%Siggers {\it et al.}\cite{siggers2004bounds} show that at best $Nu\sim Re$, which translates into $m=1$ and thus $n<0$. This is however impossible as the kinetic boundary layer thickness must grow with increasing $Re$ and shows that non-rotating HC can not achieve the ultimate regime of turbulent convection. 

\begin{equation}
{\rm Ri} \equiv \left(\frac{N}{\partial_z U}\right)^2 \sim \frac{\Delta}{\lambda_b}\left(\frac{\lambda_u}{U}\right)^2 
\sim \Delta L^{-1}\frac{L}{H}\frac{H}{\lambda_b}L^2\frac{\lambda_u^2}{L^2}\frac{\nu^2}{L^2U^2}{L^2}{\nu^{-2}}
\sim \Gamma\,{\rm Ra\,Pr^{-1}\,Nu\,Re^{-2}}\frac{\lambda_u^2}{L^2}.
\end{equation}
For a fully turbulent regime the boundary layers must scale as\cite{siggers2004bounds} $\rm Ra^{-1/3}$, thus $\rm Nu \sim Ra^{1/3}$ (see also Rocha {\it et al.} \cite{rocha2020improved} for an improved estimate). Thus
\begin{equation}
    {\rm Ri}\sim \Gamma\,\rm Pr^{-1}\,Ra^{2/3}\,Re^{-2}.
\end{equation}
In order for $\rm Ri$ to become asymptotically small (at fixed $\rm Pr$), we must have $\rm Re\sim Ra^\alpha$, with $\alpha>1/3$. However, under these assumptions, the normalised dissipation $\overbar{\epsilon_u} L/U^3<\rm Ra^{1-3\alpha}$ (see eq.(\ref{epsu2})) would decay asymptotically, violating the hypothesis of fully turbulent flow.  
What this means is that the stability of the shear layer is not controlled by the Rayleigh number. Note that a such behaviour was observed for instance by Whitehead \& Wang \cite{whitehead2008laboratory} in the their laboratory experiments where a forcing was added. As a conclusion, the Richardson number will only increase with increasing $\rm{Ra}$.

We finally take this moment to underline that this condition is only valid for non-rotating horizontal convection and would no hold true in the case where rotation is taken into account \cite{barkan2013rotating,vreugdenhil2017geostrophic}.

\section{Numerical calculations}\label{sec:numerics}

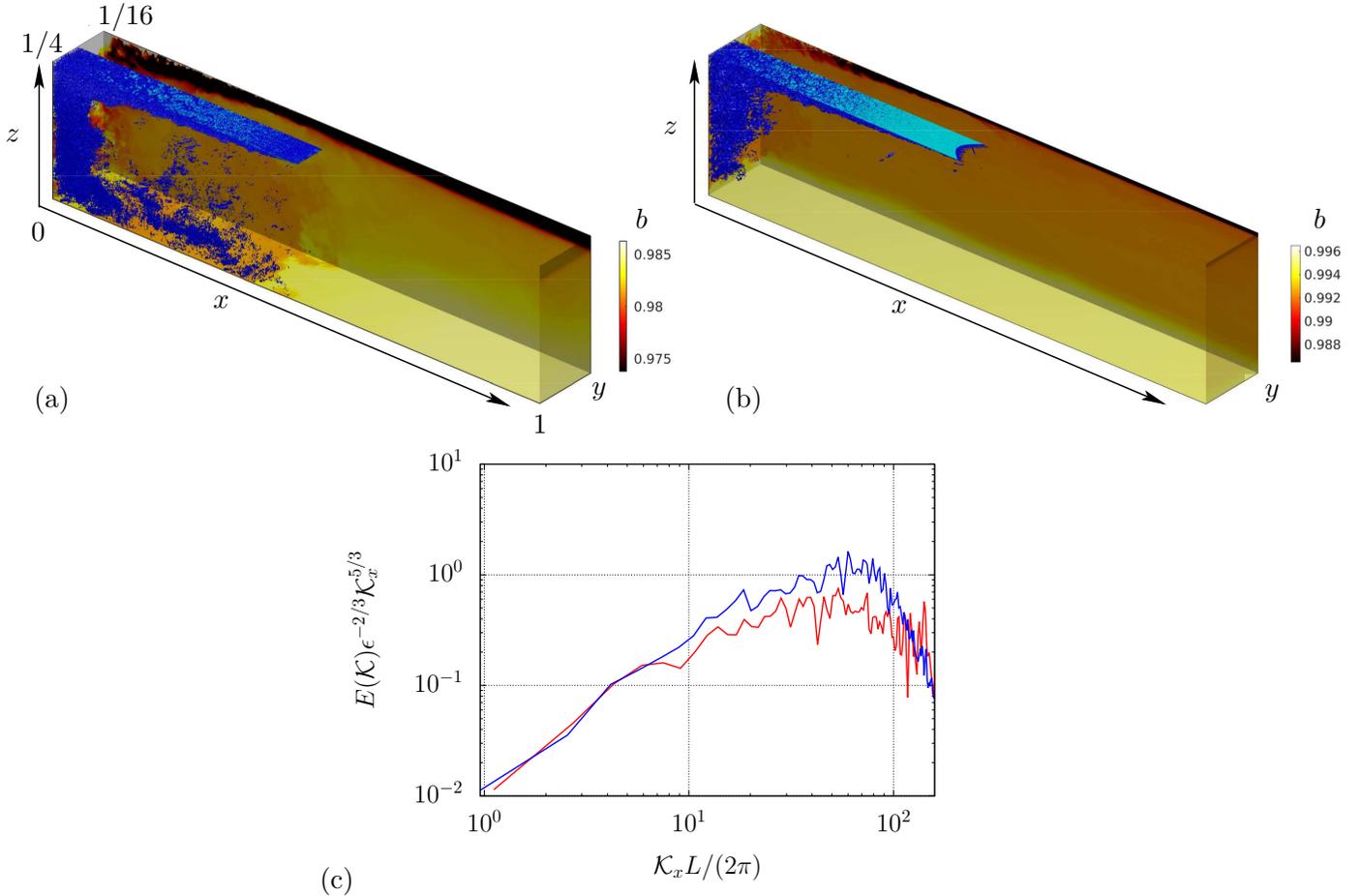
\begin{figure*}[t!]
\centering
\hspace{-0mm}
\hspace{-2mm}\scalebox{0.2}{\Huge\input{snapshot_flow_field.tex}}%
\put(-530,139){$1/4$}\put(-500,150){$1/16$}\put(-526,65){$0$}\put(-330,-10){$1$}
\put(-536,104){$z$}\put(-455,39){$x$}\put(-307,5){$y$}
\put(-279,107){$z$}\put(-190,37){$x$}\put(-44,4){$y$}
\put(-525,0){(a)}\put(-255,0){(b)}
\put(-290,70){$b$}\put(-25,70){$b$}\\
%\put(0,0){\color{black}(b)}
(c)\hspace{-2mm}\scalebox{0.7}{\Large\input{spectra_2}}%
\vspace{-3mm}
\caption{Snapshot of the iso-contours of $\Lambda_2=-\rm{Pr}^{-1}/8$ criteria (blue) and buoyancy $b$ (background) at (a) $\rm Ra=6.4\times 10^{13}$, $\rm Pr=0.01$ showing the new regime ($II_l$) and (b) $\rm{Ra}=6.4\times 10^{13}$, $Pr=1$ corresponding to the Hughes' regime \cite{Hughes07} ($II_u$). (c) Turbulent spectra $E(\mathcal{K}_x)$ at $\rm Ra=6.4\times 10^{13}$ and $\rm{Pr}=0.05$ (red), and $E(\mathcal{K}_x)/2$ at $\rm{Pr}=1$ (blue) for the same $\rm{Ra}$.}
\label{3D}
\end{figure*}
%
% \begin{figure}[ht]
% \hspace{-2mm}\scalebox{1}{\input{spectra}}%
% \vspace{-5mm}
% \caption{Turbulent spectra $\Psi(k)$ at $Ra=6.4\times 10^{13}$ at $\rm{Pr}=0.05$ (red) and $\Psi(k)/2$ at $\rm{Pr}=1$ (blue) for the same $\rm{Ra}$. The black line shows the $k^{-5/3}$ turbulence cascade whereas the green vertical line shows the wavenumber of the forcing.}
% \label{spectra}
% \end{figure}
%

The Navier-Stokes equations are solved numerically on a Cartesian grid, stretched near the upper boundary, using a standard second-order in space and time projection method.  Since we are interested in turbulence dominated regimes where the scaling is not determined by the buoyancy forcing profile, nor the aspect ratio \cite{ShishkinaGL16,Sheard2011}, nor the type of boundary condition \cite{Rossby98,chiu2008very,beardsley1972numerical}, free-slip boundary conditions are used for the velocity on the upper, lower and end walls at $x=\pm L/2$ (see ref.\cite{ScottiW11}) while the domain is assumed periodic in the transverse direction $y$. This is in contrast with Shishkina \& Wagner\cite{ShishkinaW16} where they used no-slip boundary conditions and end walls in the transverse direction. Our approach avoids the numerical difficulties involved with resolving the no-slip BL and a finite domain in the transverse direction and allows us to efficiently explore  Rayleigh numbers as high as  $1.92\times10^{15}$ for a wide range of Prandtl numbers. Snapshots of the flow are shown in figure \ref{3D}(a-b) and by mean of the $\Lambda_2$ criteria defined by the second largest eigenvalue of the matrix $S^2+\Omega^2$ where $S$ and $\Omega$ are the symmetric and anti-symmetric parts of the velocity gradient respectively.

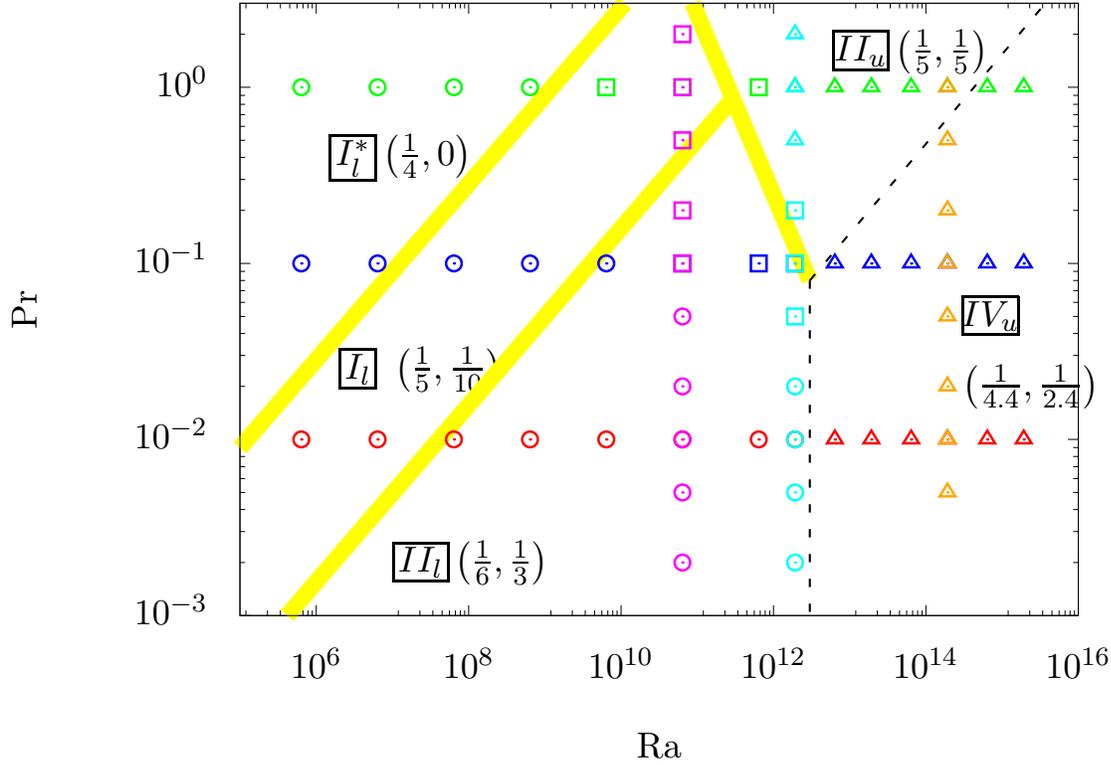
\begin{figure*}[t!]
\hspace{-2mm}\scalebox{1.25}{\input{Ra_Pr_All}}%
\caption{Sketch of the phase diagram in the $(\rm{Ra}, \rm{Pr})$ plane for the laminar regimes $I_l$ and $I^*_l$ together with the turbulent scaling $II_{l}$ with the conducted DNS. The yellow stripes shows the transition from $I^*_l$ to $I_{l}$, and $I_l$ to $II_{l}$, with a slope $\rm Pr\approx Ra^{1/2}$. The transition from $II_l$ to $II_u$ with a slope $\rm Pr\approx Ra^{-1}$.  Symbols reflect the computational meshes in $(x,y,z)$, used in the DNS: $512\times256\times256$ (circle), $1024\times384\times128$ (squares), and $2048\times256\times256$ (triangles). The values ($\alpha,\beta$) in each region provide the exponents $\rm{Nu}\sim \rm{Ra}^{\alpha}\rm{Pr}^{\beta}$ measured in the DNS and derived in the theory.
The transition between regimes is given as follows: 
from $I^*_l$ to $I_l$ is given by $\rm Pr \sim 3.\,10^{-5} Ra^{1/2}$, 
from $I_l$ to $II_l$ by $\rm Pr \sim 3.\,10^{-6} Ra^{1/2}$,  
from $II_u$ to $IV_u$ by $\rm Pr \sim 3.\,10^{-8} Ra^{1/2}$, and
from $II_l$ to $II_u$ by $\rm Pr \sim 3.\,10^{11} Ra^{-1}$.
The black dashed lines show when the boundary layer becomes turbulent and follows from $II_u$ to $IV_u$ by $\rm Pr \sim 3.\,10^{-8} Ra^{1/2}$ and
from $II_l$ to $IV_u$ by $\rm Ra \approx 3.\,10^{12}$.}
\label{Ra_Pr}
\end{figure*}

The turbulent scaling for momentum and buoyancy transport are computed using Direct Numerical Simulations (DNS) in the range $\rm Ra=[6.4\times10^5,1.92\times10^{15}]$ and $0.002\leq \rm Pr\leq 2$.
For $\rm Ra<10^8$ and $0.5\leq \rm Pr\leq 2$, the HC flows are steady \cite{ShishkinaW16,PassaggiaSW17}. With increasing $\rm{Ra}$ and/or decreasing values of $\rm{Pr}$, HC flows become increasingly unsteady, leading to turbulence (as shown in figure \ref{3D}(c)) and the mesh size is decreased in order to resolve the Kolmogorov length scale
(see ref.\cite{ScottiW11} for details about turbulent HC).
In the case of homogeneous turbulence, the Kolmogorov length scale is given by $\eta\approx(\nu^3/\epsilon_u)^{1/4}$. Using the PY inequality, an approximation yield $\eta/L\approx Pr^{1/2}/(\Gamma B Ra)^{1/4} \gtrsim 10^{-4}$ for the largest value of $\rm{Ra}$ and smallest $\rm{Pr}$ considered in this work. These estimates are valid for homogeneous turbulence. In the present case, a substantial amount of the dissipation is located in the boundary layer where the mesh is refined up to $\Delta z/L = 10^{-4}$ in the vertical direction and $\Delta y/L = 10^{-3}$ in the horizontal direction and should ensure that most scales are captured throughout our DNS.
Mesh sizes are reported in Fig.\ref{Ra_Pr}(b) in the $(\rm{Ra},\rm{Pr})$ plane along with the different regimes reported later in this manuscript. Note that turbulence in HC for moderate values of $\rm{Pr}$ is confined to a narrow region located under the cooling/heavy boundary consisting of the plume and the BL where the fluid is statically unstable (cf. Fig.\ref{3D}) \cite{Gayen14,ScottiW11}. Decreasing values of $\rm{Pr}$ increases the volume of fluid subject to turbulence (see Fig.\ref{3D}) and decreases the depth of the circulation.\\
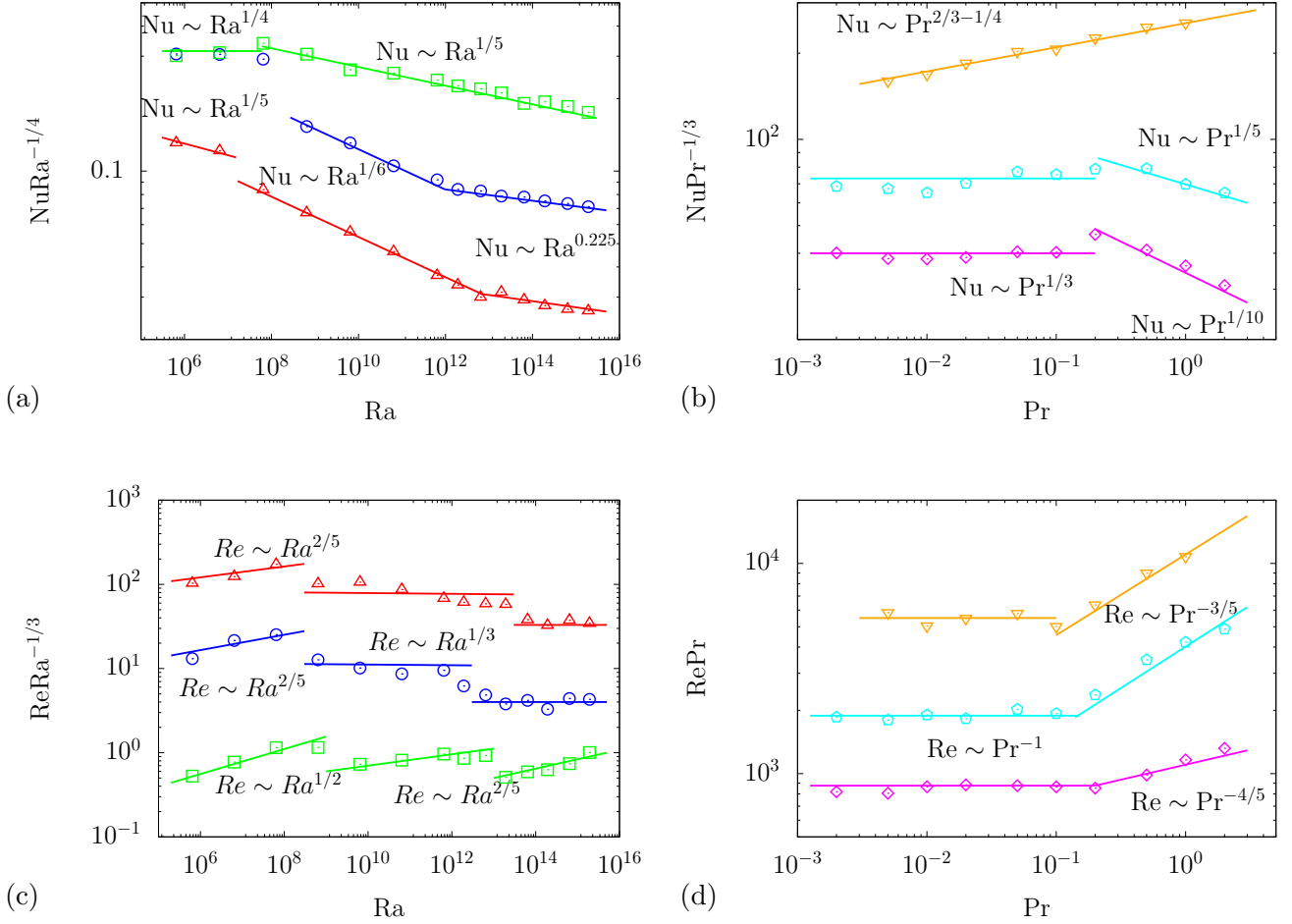
\begin{figure*}[t!]
\begin{minipage}[b]{\linewidth}
\scalebox{0.7}{\Large\input{NuRa1e025_Ra_L}}%\scalebox{0.65}{\input{NuPr1e01_Pr}}
\scalebox{0.7}{\Large\input{NuPr-03_Pr_L}}
\put(-510,10){(a)}\put(-250,10){(b)}\\

\scalebox{0.7}{\Large\input{ReRa-03_Ra_L}}%\scalebox{0.65}{\input{NuPr1e01_Pr}}
\scalebox{0.7}{\Large\input{RePr06_Pr_L}}
%\scalebox{0.55}{\input{ReRa-03_Ra}}%\scalebox{0.65}{\input{RePr_Pr}}
%\scalebox{0.55}{\input{RePr06_Pr}}
\put(-510,10){(c)}\put(-250,10){(d)}
  \end{minipage}
\vspace{-7mm}
\caption{(a),(c) $\rm{Ra}$ dependencies and (b),(d) $\rm{Pr}$ dependencies of (a),(b) the Nusselt number and (c),(d) the Reynolds number, as obtained in the DNS for (a),(c) $\rm{Pr}=1$ (squares), $\rm{Pr}=0.1$ (circles), $\rm{Pr}=0.01$ (triangles) and for (b),(d) $\rm{Ra}=6.4\times10^{10}$  (diamonds) and $\rm{Ra}=1.92\times10^{12}$ (pentagons). $\rm{Pr}$ dependence of $\bar{\epsilon_u}$ with $\rm{Re}$ (d).
The DNS results support the scaling in the regime $I_l$ (solid lines) [Eqs. (\ref{lam_scal}a) and (\ref{lam_scal}b)], transition to $II_l$ (dotted lines) [Eqs. (\ref{NuRe1/6}a),(\ref{NuRe1/6}b) and \ref{NuRe1/6corr}),(\ref{NuRe1/6corr})], transition to $II_u$ (dotted lines) [Eqs. (\ref{NuRe1/5h}a) and (\ref{NuRe1/5h}b)].}
\label{Nu_Re}
\end{figure*}

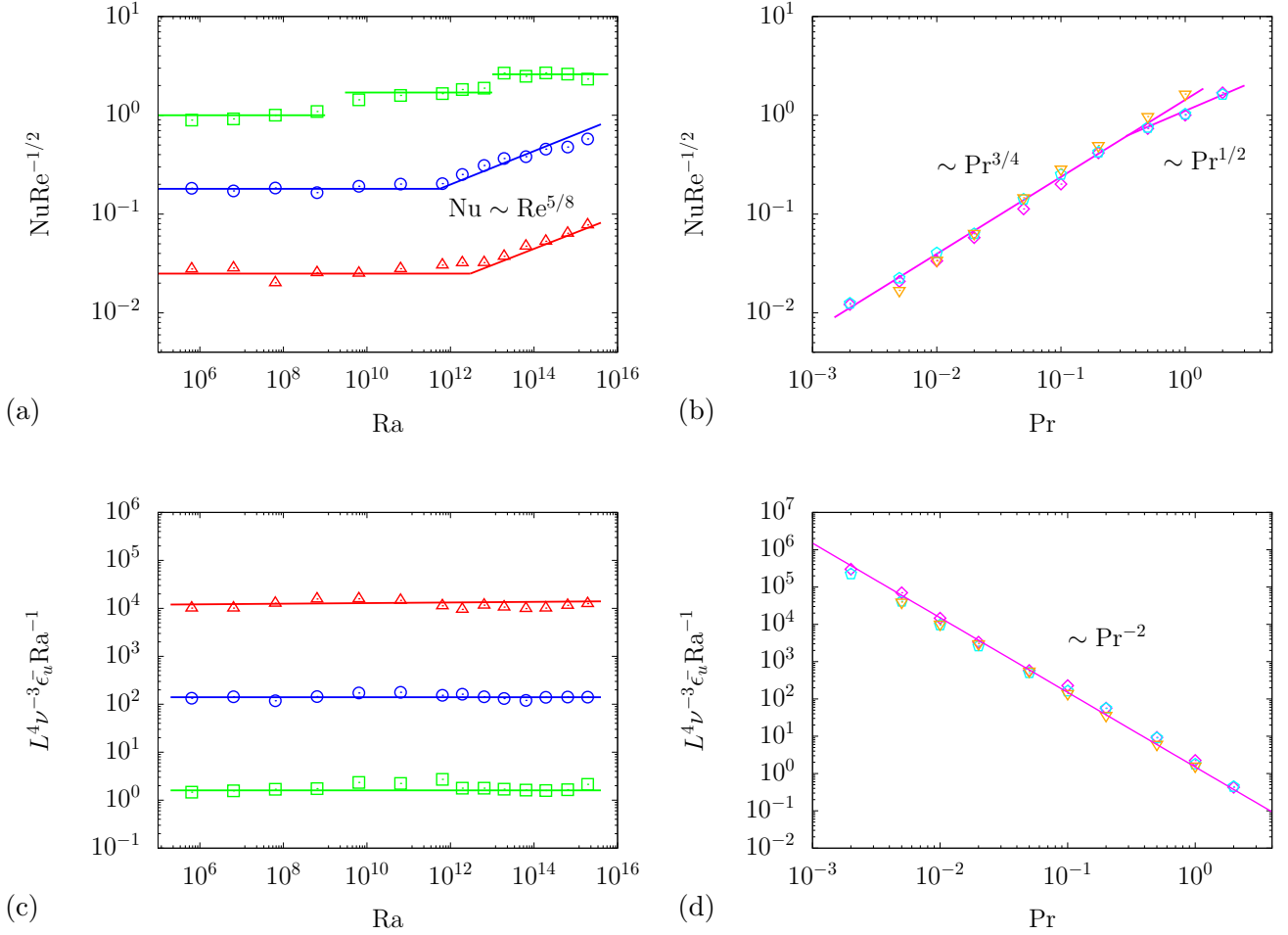
\begin{figure*}[t!]
%\hspace{-13mm}
%\begin{minipage}[b]{0.77\linewidth}
\scalebox{0.7}{\Large\input{NuRe05_Ra_L}}%
\scalebox{0.7}{\Large\input{NuRe05_Pr_L}}%
\put(-510,10){(a)}\put(-250,10){(b)}\\

\scalebox{0.7}{\Large\input{L4nu-3epsilonRa-1_Ra_L}}%
\scalebox{0.7}{\Large\input{L4nu-3epsilonRa-1_Pr_L}}%
\put(-510,10){(c)}\put(-250,10){(d)}
\caption{(a),(c) $\rm{Ra}$ dependencies and (b),(d) $\rm{Pr}$ dependencies of
(a),(b) $\rm{Nu}\rm{Re}^{-1/2}$ and (c),(d) $L^4\nu^{-3}\bar{\epsilon_u}\rm{Ra}^{-1}$, 
as obtained in the DNS for (a),(c) $\rm{Pr}=1$ (squares), $\rm{Pr}=0.1$ (circles), 
$\rm{Pr}=0.01$ (triangles) and for (b),(d) $Ra=10^9$ 
(diamonds) and $\rm{Ra}=2\times10^{10}$ (pentagons). The upper
figures support the estimates in eq. (\ref{Nu}) and eq. (\ref{NuRe1/6corr}b), while the lower figures illustrate eq. (\ref{PYbound}).}
\label{Re_eps}
\end{figure*}

\section{Results and scaling analysis}\label{ssec:low_Pr}

The regimes observed in our numerical simulations are summarised in Fig. \ref{Ra_Pr} together with the exponents computed fitting power laws to the data  (fig.\ref{Nu_Re}(a-d)).
The colours shown in Fig. \ref{Ra_Pr} correspond to the colours shown in Fig.\ref{Nu_Re}(a-d) and the two new regimes, labelled $II_l$ in brown in the laminar case, $II_l$ and $IV_u$ are shown in purple in Fig. \ref{Ra_Pr} for the turbulent scaling laws.\\

The dependence of $\rm{Nu}$ and $\rm{Re}$ with respect to $\rm{Ra}$ and $\rm{Pr}$ are summarised in Fig. \ref{Nu_Re}(a-d). The Nusselt number obeys a scaling law $\rm{Nu}\sim Ra^{\alpha}$ [see Fig. \ref{Nu_Re}(a)] with the exponent $\alpha$ depending on $Ra$ and $Pr$ as follows:
\begin{itemize}
\item $\alpha=1/4$ ,the enhanced laminar scaling, for low $\rm{Ra}$ and higher $\rm{Pr}$,
\item $\alpha=1/5$ , the classical laminar scaling, for small $\rm{Ra}$, 
\item $\alpha=1/6$, for small $\rm{Pr}$, 
\item $\alpha=1/5$, the entrainment-type regime,  at high $\rm{Ra}$ and not too small $Pr$,
\item $\alpha\approx 1/4.4$ for large $\rm{Ra}$ and small $Pr$.
\end{itemize}

We observe the laminar scaling $Re\sim \rm{Ra}^{\gamma}$ with $\gamma=1/2$ (see ref. \cite{ShishkinaW16}) and $\gamma=2/5$ (see ref.\cite{Rossby65}). At higher $\rm{Ra}$, the new scaling $\gamma=1/3$ is also observed and changes back to $\gamma=2/5$ (see ref.\cite{Hughes07}) [Fig. \ref{Nu_Re}(c)].
Similarly, when  $\rm{Pr}<1$ and $\rm Ra$ is fixed, we observe a scaling relationship $\rm{Nu}\sim Pr^{\beta}$ with:
\begin{itemize}
\item $\beta=0$ for higher $\rm{Pr}$ and low $\rm{Ra}$ (see ref. \cite{ShishkinaW16}),
\item $\beta=1/10$ for $\rm{Ra}<10^{11}$ see (see ref.\cite{Rossby65}),  
\item $\beta=1/3$ at low $\rm{Pr}$, 
\item $\beta=1/5$ for $\rm{Ra}>5\times10^{11}$ see (see ref.\cite{Hughes07}). 
\item $\beta \approx 2/3-1/4$ at large $\rm{Ra}$ and low $\rm{Pr}$.
\end{itemize}

The Reynolds number dependence $\rm{Re} \sim \rm{Pr}^{\delta}$ with $\delta=-2/3$ for the smaller value of $\rm{Pr}$, then $\delta=-1$ for $10^{-2}\lesssim\rm{Pr}\lesssim0.2$
changes to $\delta=-4/5$ for increasing $\rm{Ra}$ at all $\rm{Pr}$ and increases at high $\rm{Ra}$ to the HGM scaling $\delta=-3/5$ for the larger values of $\rm{Ra}$ [Fig.\ref{Nu_Re}(d)].\\

The fact that our simulations recover the scaling of the 
SGL\cite{shishkina2017scaling} and HG\cite{hughes2008horizontal} theories validates 
them, and give confidence in the
 scaling regimes occurring in the lower part of parameter space, with $\alpha=\beta=1/6, $ and $\alpha\approx 1/4.4,\,\beta\approx 1/2.4$ which, new to this study, are the focus of the following section. 

\subsection{The low-Prandtl core-driven flow $II_l$}

\subsubsection{The limiting regime}

In the low-Prandtl number regime,  the flow transitions from the Rossby $I_l$ regime to the $II_l$ regime as $\rm Ra$ increases. A snapshot of the flow (Fig. \ref{3D}(a)) shows that,  unlike the Rossby regime, the flow here is clearly turbulent in the core. In this low-Prandtl number regime, the buoyancy flux provided through the boundary is large but the thermal and kinetic boundary layers remain thick, hence laminar, and eq. (\ref{Nu}) still holds. With decreasing $\rm{Pr}$ and/or increasing $\rm{Ra}$, the bulk dynamics hence dominates dissipation with a large-scale overturning flow occupying the entire domain and whose horizontal length scale is $L$. In this case, it is the large-scale velocity $U$ which drives the dissipation of kinetic energy and the latter is given by
%\begin{subeqnarray}
%\overbar{\epsilon_u} &\sim& \nu^{3}L^{-4}\rm{Re}^3, \\
%\overbar{\epsilon_b} &\sim& \kappa\Delta L^{-2}\rm{Re}\rm{Pr}\rm{Nu}^{-1}.
%\label{epsub}
%\end{subeqnarray}
\begin{equation}
\overbar{\epsilon_u} \sim \nu^{3}L^{-4}\rm{Re}^3.
\label{epsub}
\end{equation}
From (\ref{Nu}), (\ref{epsu2}) and (\ref{epsub}), it follows that low-$\rm{Pr}$ HC exhibits dependencies of the form
\begin{subeqnarray}
\rm{Re}&\sim&\rm{Ra}^{1/3}\rm{Pr}^{-2/3}, \\
\rm{Nu}&\sim&\rm{Ra}^{1/6}\rm{Pr}^{1/6},
\label{NuRe1/6}
\end{subeqnarray}
where this scaling regime is denoted as $II_l$ [see Fig. \ref{Ra_Pr}(b) and ref.\cite{ShishkinaGL16}].
Note that these scaling are only observed for the Rayleigh-number dependence but the Prandtl-number dependence is clearly underestimated for both $\rm Nu$ and $\rm Re$. 

The boundary-layer scaling observed thus far was consistent with
$$
\rm{Nu}\sim \rm{Pe}^{1/2}.
$$
In the following, we show that the balance in the boundary has to be modified in order to take into account either core-size modifications or turbulent boundary-layer effects.

\subsubsection{Modification induced by the variable turbulent depth $h$ for the $II_l$ regime}

In low Prandtl number regimes for $\rm{Pr} < 10^{-1}$, turbulence is confined between the plume and the left part of the domain, under the statically unstable boundary layer whose depth is denoted by $h<H$. The PY inequality provides a bound for the the dissipation which can be used to relate the depth $h$ occupied by the turbulent core with the Reynolds number $\rm{Re}$. However, dissipation in horizontal convection is bounded by the dissipation in the laminar kinetic boundary layer, located under the warming (statically stable) boundary. This unbalance between the dissipation in the bulk and the boundary layer is the first occurrence of a turbulent regime subject to two regions with different dissipation rates at low Prandtl numbers. At low values of ${\rm Pr}$, the thermal boundary layer providing the available energy drives the dynamics near the forcing boundary and its dissipation rate is given in eq. (\ref{epsu_lam}). Dissipation in the core supplies a higher dissipation rate, given by eq. (\ref{epsub}) which therefore dissipates the %available 
energy faster than it is created from the forcing boundary. As a consequence, the bulk size $h$ must decrease with respect to the full depth of the domain $H$, in order to take into account for this effect. \\

Similar observation was drawn by Chiu-Webster et al.\cite{chiu2008very} in the case of infinite Prandtl numbers where the core dissipates less than the boundary layer, leaving the Nusselt number scaling independent of the core dynamics.

Here we demonstrate that the turbulent core modifies the Prandtl number dependence by relating the dissipation and buoyancy variance in both the boundary layer and the bulk. The idea is to allow for the ratio $h/H$ to appear and provide the Rayleigh and Prandtl number dependencies and correct for the above estimate.\\

\begin{figure*}[t!]
\centering
\hspace{-0mm}
\includegraphics[width=0.8\textwidth]{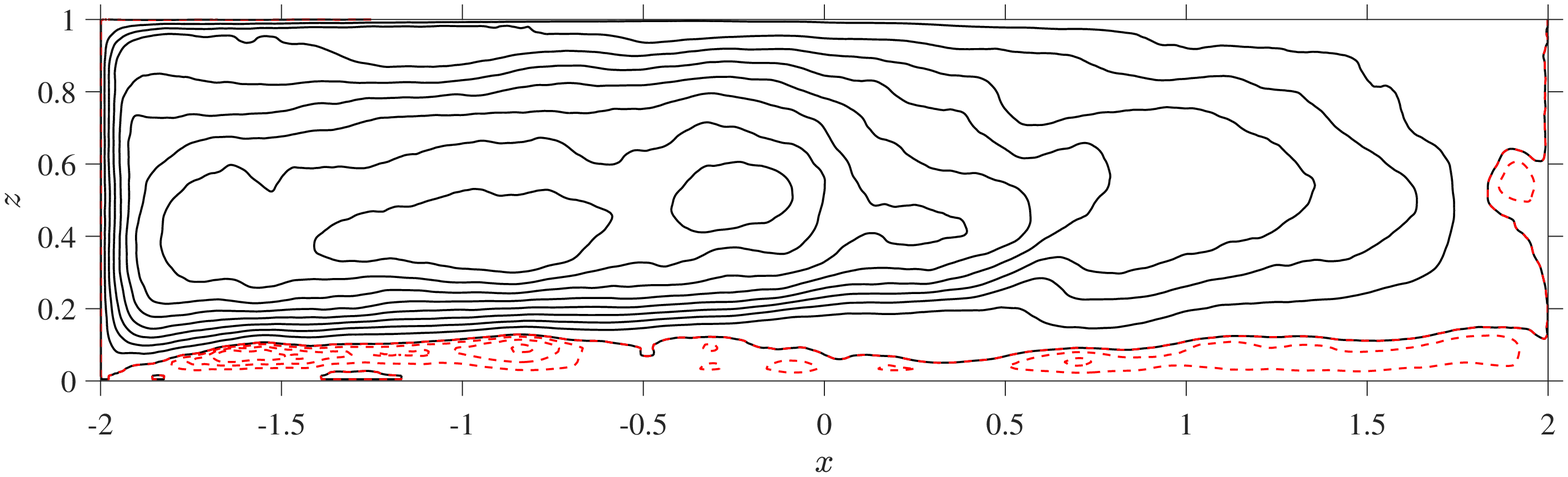}\put(-390,10){(a)}\put(-184,0){\scriptsize$\Gamma$}\put(-360,60){\rotatebox{90}{\scriptsize$\Gamma$}}\\
\includegraphics[width=0.8\textwidth]{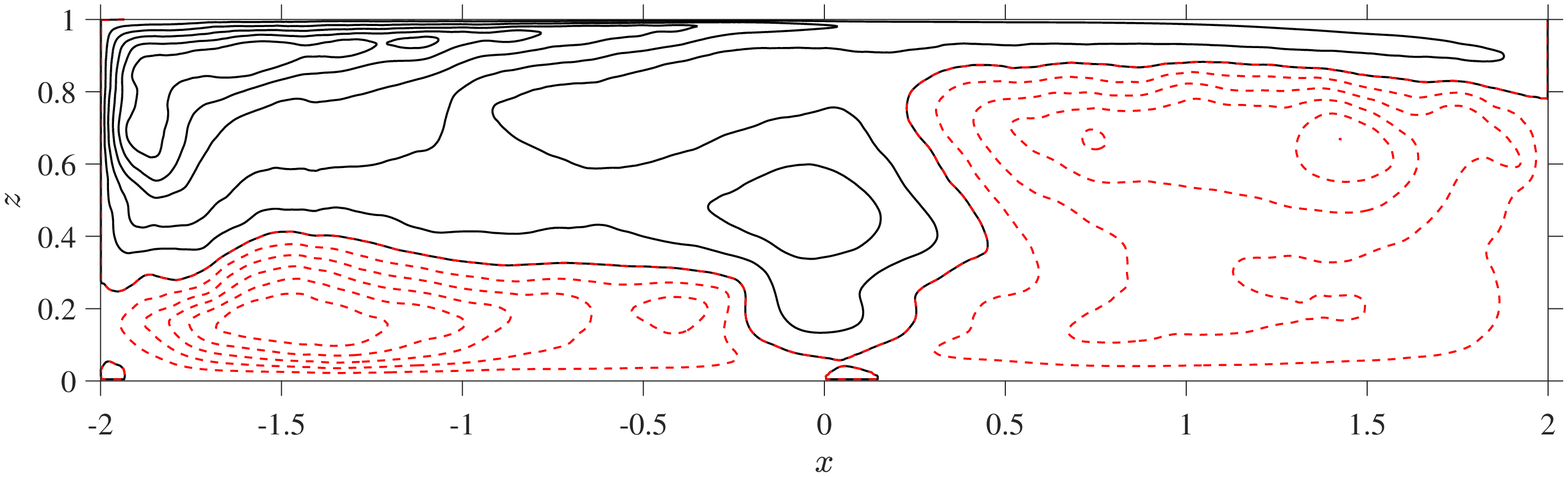}\put(-390,10){(b)}\put(-184,0){\scriptsize$\Gamma$}\put(-360,60){\rotatebox{90}{\scriptsize$\Gamma$}}\\
\includegraphics[width=0.8\textwidth]{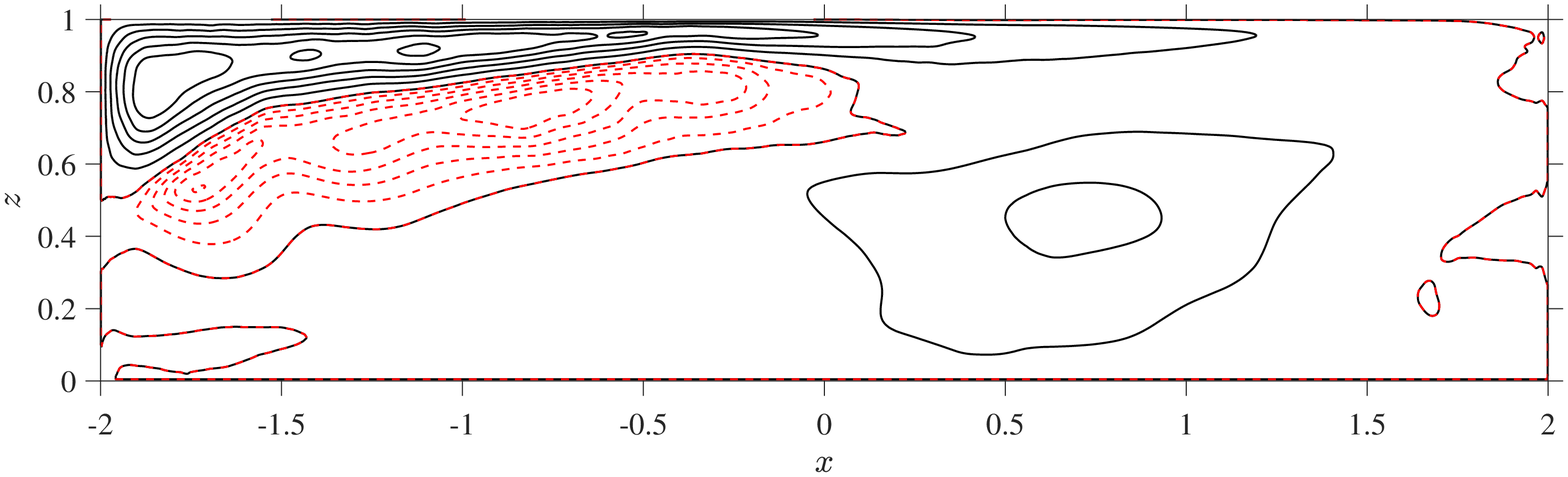}
\put(-390,10){(c)}\put(-184,0){\scriptsize$\Gamma$}\put(-360,60){\rotatebox{90}{\scriptsize$\Gamma$}}\\
\caption{Time-averaged iso-contours of the streamfunction $\psi=[-0.1, -0.075, -0.05, -0.025, 0, 2, $ $ 4, 6, 8,  10, 12, 14, 16]/8\times 10^{-4}$  for $\rm{Ra}=1.92\,10^{15}$; 
(a) $\rm{Pr}=1$ where
(b) $\rm{Pr}=0.1$,
and (c) $\rm{Pr}=0.01$ 
showing the narrowing of the circulation as $\rm{Pr}$ decreases and the circulation within the core  narrowing beneath the deferentially heated surface at $z\Gamma=1$.}
\label{streamlines}
\end{figure*}

The bulk size is assumed to balance the ratio between the thermal dissipation in the BL and in the bulk, which has to equal to the ratio between the kinetic energy dissipation in the BL and in the bulk. The same idea writes
\begin{equation}
\left(\frac{\overbar{\epsilon_{b,bulk}}}{\overbar{{\epsilon}_{b,BL}}}\right) \sim \left(\frac{\overbar{\epsilon_{u,bulk}}}{\overbar{{\epsilon}_{u,BL}}}\right).
\label{eq:ratios_eps}
\end{equation}
We now substitute the expressions for dissipation and the buoyancy variance in regions. For $\overbar{\epsilon_{u,bulk}}$, $\overbar{{\epsilon}_{u,BL}}$ and $\overbar{{\epsilon}_{b,BL}}$, the values are expressed in terms of the Reynolds and the Prandtl number in eqs. (\ref{epsub}), (\ref{epsu_lam}), and (\ref{epsbbl_lam}) respectively. The  buoyancy variance in the bulk is 
\begin{equation}
\overbar{\epsilon_{b,bulk}} \sim \frac{U \Delta^2}{h}\frac{h-\lambda_b}{h} \sim \frac{U \Delta^2}{h} \sim \frac{\kappa\Delta^2}{h^2} Pe.
\end{equation}
Rearranging the terms in eq. (\ref{eq:ratios_eps}), the ratio of bulk-to-domain depth appears naturally and provides the following scaling
\begin{equation}
\frac h H \sim \left({\rm Re^{1/2}}{\rm Pe^{-1/2}}\right)^{-1/2} \sim {\rm Pr}^{1/4}.
\label{bulk_decrease}
\end{equation}
%This scaling is shown in Fig. \ref{}(a) for the Rayleigh number dependence and in Fig. \ref{}(b) for the Prandtl number dependence. 
This decrease of bulk size is depicted in Fig. \ref{streamlines}(a-c) where the plume depth behaves according to the above scaling.
The reduction in $h$ also implies that in this transition regime between $I_l$ and $II_l$ or $II_u$ and $II_l$, eq.(\ref{epsub}) has to take into account the fact that now $\rm Re$ is not based on $H$ but $h$, the overturning depth. With this respect, the overturning may
be rescaled such that
\begin{equation}
\overbar{\epsilon_u} \sim \frac{\nu^{3}}{H^{4}}{\rm Re}^3\left(\frac h H\right)^3,
\label{epsub2}
\end{equation}
in order to take into account the reduction of the bulk size $h$ as $\rm{Pr}$ decreases.
The above argument can also be expressed through the heat transport in the laminar boundary-layer where the bulk modification, similarly to eq. (\ref{epsub2}) is now tied to the amount of heat transport such that
\begin{equation}
U \frac \Delta L \sim \frac{\kappa\Delta}{\lambda_b^2}\left(\frac H h\right).
\label{Nu2}
\end{equation}
Substituting eq. (\ref{bulk_decrease}) into eq. (\ref{epsub2}), the modified dissipation leads to a new Prandtl number dependence for the $II_l$ regime such that
\begin{subeqnarray}
%\overbar{\epsilon_u} &\sim& \nu^{3}L^{-4}{\rm Re}^3\left(\frac{H}{h}\right)^{4}\sim\nu^{3}A^{-4}h^{-4}{\rm Re}^3,\\
{\rm Re} &\sim&  {\rm Ra}^{1/3}{\rm Pr}^{-2/3}\left(\frac{h}{H}\right)^{-1} \sim {\rm Ra}^{1/3}{\rm Pr}^{-11/12},\\
{\rm Nu} &\sim& {\rm Re}^{1/2} {\rm Pr}^{1/2}\left(\frac{h}{H}\right) \quad \sim   {\rm Re}^{1/2} {\rm Pr}^{3/4}
\label{epsub2mod}
\end{subeqnarray}
which is verified empirically in our DNS [see Fig. \ref{Nu_Re}(d) and Fig. \ref{Re_eps}(b)].
Combining eq. (\ref{epsub2}a), and eq. (\ref{epsub2}b) provides a correction for this $\rm{Pr}$ transition in the $II_l$ regime
\begin{subeqnarray}
\rm{Re}&\sim&\rm{Ra}^{1/3}\rm{Pr}^{-11/12}, \\
\rm{Nu}&\sim&\rm{Ra}^{1/6}\rm{Pr}^{7/24},
\label{NuRe1/6corr}
\end{subeqnarray}
found for $\rm{Pr}\lesssim 0.2$ [see Fig. \ref{Nu_Re}(b,d)].
Whence, the advection-diffusion balance in the boundary layer eq. (\ref{Nu}) is modified according to
$$
\rm{Nu} \sim \rm{Pe}^{1/2}\rm{Pr}^{1/4}  \sim \rm{Re}^{1/2}\rm{Pr}^{3/4} .
$$
which now takes into account variable depth effects $h$, decoupled from the domain's depth $H$.
This particular scaling is the last controlled by the laminar boundary layer as ${\rm Ra}$ increases. As noted by Shishkina {\it et al.}\cite{shishkina2017scaling} further increasing the Rayleigh number may eventually lead to core-driven dynamics, as previously observed by Griffith \& Gayen \cite{griffiths2015turbulent} for a somewhat different boundary condition. In the following subsection, we report the transition to a similar regime, which marks the transition to the limiting regime of horizontal turbulent convection for large $\rm{Ra}$. In particular, we leverage on the same scaling analysis for the variable size of the turbulent core with depth $h$ to recover Prandtl-number dependencies.

\subsection{The limiting turbulent boundary-layer regime at large Rayleigh numbers $IV_u$}

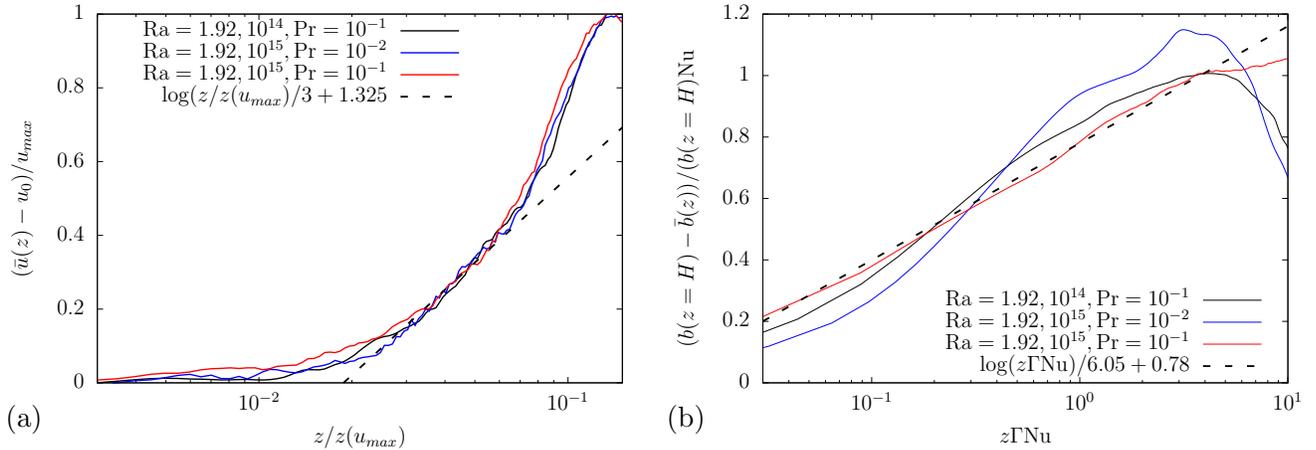
\begin{figure*}[t!]
\scalebox{0.69}{\large\input{turb_BL.tex}}
\scalebox{0.69}{\large\input{turb_BL_b.tex}}
\put(-500,10){(a)}\put(-250,10){(b)}
\caption{(a) Mean turbulent boundary layer profiles and (b) mean turbulent buoyancy profiles measured at $x=-0.325$, rescaled using the slip velocity at the wall $u_0=u(x=-0.75,z=H)$ and the maximum velocity at this particular $x$ location. Note that we are not in the presence of a free-slip-type turbulent boundary layer but we recover a log-type zone as shown by the black line in the above for our largest values of $\rm{Ra}$, small $\rm{Pr}$, and thus largest $\rm{Re}$. The origin of this log-type boundary layer is further discussed in the text.}
\label{fig:log-BL}
\end{figure*}

The modification of circulation depth at higher Rayleigh numbers was first investigated by Griffiths \& Gayen \cite{griffiths2015turbulent,rosevear2017turbulent} who considered a periodic forcing at the surface and very small aspect ratios $\Gamma$. They showed that a laminar-type scaling for the dissipation in the bulk was responsible for the transition to a core dominated turbulence regime. The scaling obtained from the vorticity equation is linked to the previous analysis and considers a balance between the dissipation in the core (or the interior) scaling as $\rm Re\sim Ra^{1/2}Pr^{-1/2}$ and balances the dissipation in the boundary layer, obtained from (\ref{Nu}) boundary layer. This is in contrast with the regime observed in the previous regime ($II_l$). This is shown in Fig. (\ref{Nu_Re}) and Fig. (\ref{Re_eps}) where both $\rm Ra$ and $\rm Pr$ dependencies do not support the above scaling. There is a clear departure from the laminar scaling obtained in eq. (\ref{Nu}) which suggest that turbulent boundary layers, characterised by a log-type profile and modified heat transfer coefficients \cite{GrossmannL11} may be expected.
Note that such boundary layer profiles were already observed in Rosevear {\it et al.}\cite{rosevear2017turbulent} but the latter did not affect the heat- nor the momentum-transfer scaling obtained in their analysis. Here we report different results and show that log-type profiles do influence heat and momentum transfers, in a similar way to results obtained in Rayleigh-B\'enard convection \cite{GrossmannL11,van2015logarithmic}.\\

\subsubsection{Incompatibility with a laminar boundary-layer scaling}

Low-Prandtl number flows are particularly interesting with respect to the study of the transition to the limiting regime in HC since transition to a turbulent-dominated flow is first observed in the $II_l$ regime where the kinetic energy dissipation driving the dynamics scales as $\rm \epsilon_u \sim \rm Re^{3}$. Therefore increasing $\rm{Ra}$ may eventually trigger turbulent boundary as well. However, increasing $\rm{Ra}$
leads to thinner boundary layers and a thinner bulk. Following this logic, once turbulence is triggered in the boundary layer, the thermal boundary layer becomes embedded into the kinetic one and the boundary layer profiles exhibit log-type profiles. 
As recently observed in Reiter \& Shishkina\cite{reiter2020classical}, the velocity of the flow, which carries the temperature in the bulk, reduces from $U$ to $U(\lambda_{b}/\lambda_u)$ and the buoyancy variance dissipation rate\cite{ShishkinaGL16} becomes
\begin{equation}
\overbar{\epsilon_{b,bulk}}\sim(\Gamma/2)\kappa\Delta^2 h^{-2}{\rm Pr\; Re^{3/2}\; Nu^{-1}}.
\label{epsb}
\end{equation}
whereas the total volume $V$ has a buoyancy variance of 
\begin{equation}
\overbar{\epsilon_{b,V}}=(\Gamma/2) \kappa\Delta^2 L^{-2}{\rm Nu}.
\label{epsbV}
\end{equation}
Equating eqs. (\ref{epsb}) and (\ref{epsbV}) gives the expression for the Nusselt number such that 
\begin{equation}
{\rm Nu}\sim {\rm Re}^{3/4}{\rm Pr}^{1/2}\left(\frac h H\right).
\label{NuRe075}
\end{equation}
With increasing $\rm{Ra}$, the bulk dynamics is driven by the large-scale overturning flow whose length scale is $h$ as confirmed by our simulations in the previous subsection. In this case, the dissipation of kinetic energy in the bulk is essentially dependent on the large-scale velocity $U$ and the Reynolds number is again modified from ${\rm Re}$ based on $H$ to ${\rm Re}$ based on $h$ such that
\begin{equation}
\overbar{\epsilon_{u,bulk}} \sim \frac{\nu^{3}}{H^{4}}{\rm Re}^3\left(\frac h H\right)^3,
\label{epsub4u}
\end{equation}
while the Rayleigh and Prandtl number dependencies of the bulk now yield 
\begin{equation}
\left(\frac{\overbar{\epsilon_{b,bulk}}}{\overbar{{\epsilon}_{b,BL}}}\right) \sim \left(\frac{\overbar{\epsilon_{u,bulk}}}{\overbar{{\epsilon}_{u,BL}}}\right) 
\equiv \frac h H \sim {\rm Re}^{-1/8} {\rm Pr}^{1/4}.
\label{eq:bulk_reduce_high_Ra}
\end{equation}
In the above, the buoyancy variance in the bulk still follows $\overbar{\epsilon_{b,bulk}}\sim h^{-2}{\rm Pe}$, but in the boundary layer, the buoyancy variance follows eq. (\ref{NuRe075}) and  $\overbar{{\epsilon}_{b,BL}}\sim ^{-2}{\rm Re}^{3/4}{\rm Pr}^{1/2}$ (see ref.\cite{reiter2020classical}). The turbulent kinetic energy dissipation is given by eq. (\ref{epsub4u}) and reads $\overbar{\epsilon_{u,bulk}}\sim L^{-4}{\rm Re}^3$ while in the boundary layer, dissipation is bounded by the stably stratified layer $\overbar{\epsilon_{u,BL}}\sim L^{-4}{\rm Re}^{5/2}$.
Combining (\ref{epsb}) and (\ref{epsub4u}), together with the bulk reduction effect $\left(h/H\right)$ for the Nusselt number as in eq. (\ref{epsub2mod}b), the relation for the Nusselt number reads
\begin{equation}
{\rm Nu}\sim {\rm Re}^{3/4}{\rm Pr}^{1/2}\left(\frac h H\right) \sim {\rm Re}^{5/8}{\rm Pr}^{3/4},
\label{NuRe075_h}
\end{equation}
which agrees with the results shown in Fig. \ref{Re_eps}(a,b). Combining (\ref{epsb}), (\ref{epsu2}), (\ref{epsub4u}a) and (\ref{epsub4u}b), one obtains
\begin{subeqnarray}
\rm{Re}&\sim&\rm{Ra}^{8/21}\rm{Pr}^{-22/21},\\
\rm{Nu}&\sim&\rm{Ra}^{5/21}\rm{Pr}^{17/96},
\label{NuRe1/4_turb}
\end{subeqnarray}
which slightly over estimates the ${\rm Ra}$ number dependence for the Nusselt number with respect to $\rm Ra$ (i.e. $\approx 0.225$ from the DNS vs. $\approx 0.238$ from the theory) but clearly under estimates the Prandtl number dependence ($\approx 0.41$ for the DNS vs. $\approx 0.17$ for the theory)
in that particular region of the $(\rm{Ra},\rm{Pr})$ plane and is denoted as $IV_u$ in Fig. \ref{Ra_Pr}(a,b) and \ref{Re_eps}(a,b).\\

This scaling analysis shows that the effect of turbulence must play a role. In particular, the dissipation scaling has to take into account the turbulent boundary layer characteristics and log-type corrections have to be eventually reintroduced in order to predict accurately both the Prandtl and Reynolds number dependencies observed from the simulations.

\subsubsection{A fully turbulent boundary-layer scaling}

The above scaling overestimates the heat flux, since the observed exponent $\rm Nu\sim {\rm Ra}^{0.225}$ is close to the exponent ${\rm Nu}\sim{\rm Ra}^{5/21}\sim{\rm Ra}^{0.238}$ but smaller and the turbulent ${\rm Nu}\sim{\rm Ra}^{1/4}$ scaling (Fig. \ref{Nu_Re}(a)) and suggests a new correction for the dissipation in what may appear as turbulent thermal and the kinetic boundary layers. In addition, both Prandtl-number dependencies found in the numerical simulations are not predicted by eq. (\ref{NuRe1/4_turb}a) and (\ref{NuRe1/4_turb}b) which suggests that turbulence plays a non trivial role in the above scaling. 
For small $\rm Pr$ and large $\rm Ra$, the statically unstable boundary layers becomes indeed turbulent and for decreasing ${\rm Pr}$, the Reynolds number increases [Fig.\ref{Nu_Re}(c)] which causes the boundary layer to transition to turbulence. The dissipation of kinetic energy may be split between a viscous sub-layer $\overbar{\epsilon_{vs}}$ 
and a log layer $\overbar{\epsilon_{ll}}$ such that $\overbar{\epsilon_{u,BL}}=\overbar{\epsilon_{vs}}+\overbar{\epsilon_{ll}}$. \cite{GrossmannL11} 
In the log layer, the dissipation writes
\begin{equation}
{\epsilon_{u,ll}}(z) = \frac{u_*}{C_\kappa z}
\end{equation}
which may also be rearranged as
\begin{equation}
{\epsilon_{u,ll}}(z) = \nu^3 L^{-4} \frac{\rm{Re}^3}{C_\kappa z} \left(\frac{u_*}{U}\right)^3.
\end{equation}
The mean kinetic energy dissipation in the log layer can therefore be obtained by integrating the above such that
\begin{equation}
\overbar{\epsilon_{u,ll}}  = \int_{z^*}^{L_2}\epsilon_{u,ll}(z) \mbox{d}z,
\end{equation}
where $z_* = \nu/u_*$ and $L_2$ corresponds to the edge of the logarithmic zone. The dissipation in the log layer thus acts as a buffer to heat exchanges and induces a log-type correction denoted as ${\cal L}(\cdot)$, which is dependent on the Reynolds number such that
\begin{equation}
\overbar{\epsilon_{u,ll}} := \nu^{3}L^{-4}{\rm Re}^3\left(\frac{u_*}{U}\right)^3 \frac{2}{C_\kappa}\log\left(Re\frac{u_*}{U}\frac{1}{2}\right),
\label{logll}
\end{equation}
where the length scale is now $L$ (i.e. the length of the domain), $C_\kappa\approx0.4$ is the von K\'arm\'an constant, and $u_*=\overline{u'w'}$ is the typical velocity fluctuation scale\cite{GrossmannL11}.
Indeed in the present simulations, friction directly at the wall is null because of the free-slip boundary condition and the friction velocity $u_*$ does not refer to the wall shear stress but the turbulent shear stresses, induced by turbulent fluctuations of the buoyancy flux $\overline{w'b'}$, originating from the statically unstable buoyancy profile in this near-wall region. The log profiles for both the velocity and the buoyancy are shown in Fig. \ref{fig:log-BL}(a) and Fig. \ref{fig:log-BL}(b). For all turbulent profiles, a logarithmic region can be observed for the velocity profile. In addition the profile seems to be self-similar, at least for the three profiles reported in the figure. The source of turbulent stresses here is suggested by the large log profile, measured for two decades in Fig. \ref{fig:log-BL}(b) for the buoyancy profile at $\rm Ra = 1.92\, 10^{15}$ and $\rm Pr=10^{-2}$.
\begin{figure*}[t!]
\begin{minipage}[b]{\linewidth}
\scalebox{0.7}{\Large\input{logBLRe_Ra}}
\scalebox{0.7}{\Large\input{logBLRe_Pr}}
\put(-510,10){(a)}\put(-250,10){(b)}
\end{minipage}
\vspace{-7mm}
\caption{Dependencies of $\rm{Nu}$ (red) and $\rm{Re}$ (blue) with respect to $\rm Ra$ (a) and $\rm Pr$ (b) showing the modification and weak variations of both $\rm{Nu}$ and $\rm{Re}$ in the regime considered here. The continuous lines show the predictions from eqs. \ref{NuRe1/3ll}(a,b) while the dashed line shows the measurements from Fig. \ref{Nu_Re}(a-d).}
\label{fig:log-scaling}
\end{figure*}
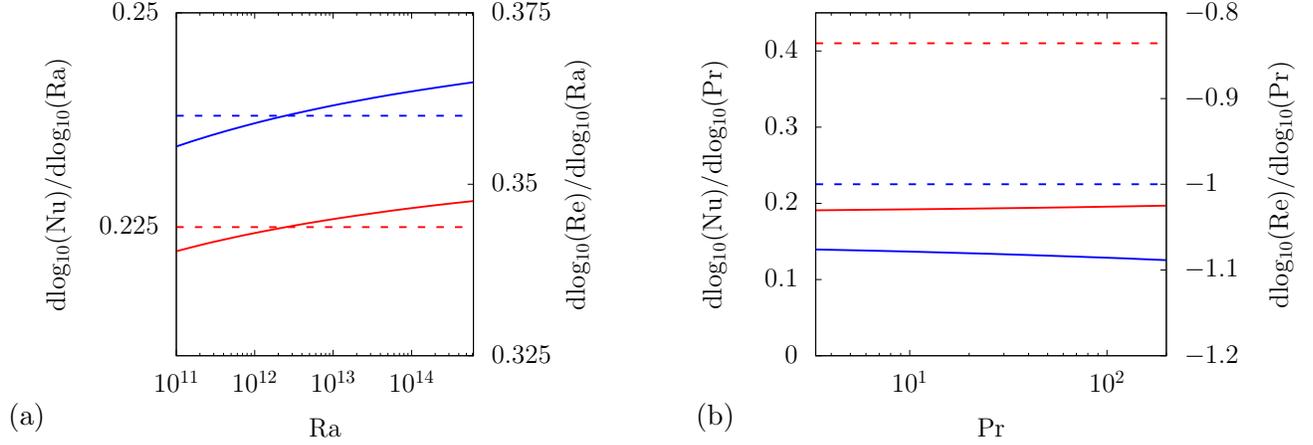
The buoyancy variance in the boundary layer can also expressed using a similar analogy. As shown in Fig. \ref{fig:log-BL}(b), the thermal layer displays a log-type layer which is a common feature of turbulent statically stable and unstable boundary layers  
\begin{equation}
{\epsilon_{b,ll}}(z) = \frac{\kappa b_*^2}{C_\kappa z^2},
\end{equation}
where $b_*= \overbar{w'b'}/u_*$. The fluctuating velocity $u_*$ can be connected to the outer velocity $U$ or $\rm Re$, and equivalently for the buoyancy variance  by
\begin{equation}
\frac{u_*}{U} = \frac{c_\kappa}{\log\left({\rm Re}\frac{u_*}{U}\frac{1}{\theta}\right)} \quad \mbox{and} \quad \frac{b_*}{\Delta} = \frac{c_\kappa}{\log\left({\rm Re}\frac{u_*}{U}\frac{1}{\theta}\right)}.
%TODO Should it be Peclet in the second expression in the above??????
\label{eq:epsilon_exp}
\end{equation}
In the above, we assumed that the flux Richardson number $R_f=\overline{w'b'}/(\overline{u'w'} \partial u/\partial z)$ is constant, which should hold provided that the structure of the turbulent boundary layer remains self-similar with $\rm Ra$ and $\rm Pr$ (see Fig. \ref{fig:log-scaling}(a,b)). The empirical constant $b$ depends on the system geometry, along a plate $\theta$ is empirically found to be equal to $0.13$ (see\cite{GrossmannL11}). Again, the mean buoyancy variance can be integrated such that
\begin{equation}
\overbar{\epsilon_{b,ll}}  = \int_{z^*}^{L/2}\epsilon_{b,ll}(z) \mbox{d}z,
\end{equation}
which reduces to
\begin{equation}
\overbar{\epsilon_{b,ll}}  = \frac{2 \kappa b^2_*}{C_\kappa L^2} {\rm Re}\frac{u_*}{U},
\end{equation}
and using eq. (\ref{eq:epsilon_exp}), the expression becomes
\begin{equation}
\overbar{\epsilon_{b,ll}}  = \kappa \Delta^2 L^{-2} {\rm Re}\left(\frac{u_*}{U}\right)\frac{2}{C_\kappa }{\log\left({\rm Re}\frac{u_*}{U}\frac{1}{\theta}\right)^{-2}}.
\end{equation}
%In this region, the  buoyancy profile is weakly statically unstable and the flux Richardson number $R_f=\overline{w'b'}/(\overline{u'w'} \partial u/\partial z)$ provides a measure for the importance of the buoyancy fluctuations in this region.\\
%
In the above expressions, the unknown ratio $u_*/U$ can be computed using Lambert's W-function where $u_*/U=\bar{\kappa}/W({\rm Re}\bar{\kappa}/\theta)$. The dissipation in the turbulent regime is thus modified from eq. (\ref{epsub4u}a) and becomes
\begin{equation}
\overbar{\epsilon_{u,ll}} \sim \nu^{3}L^{-4}\rm{Re}^3{{\cal L}(Re)}
\label{logu}
\end{equation}
where ${{\cal L}(Re)}$ is given by eq. (\ref{logll}). The bulk however still dissipates at a rate expressed in eq. (\ref{NuRe1/4_turb}a) above  (see Figs. \ref{Nu_Re}(c,d)). One can think of this correction as a decrease in heat transfer through the boundary layer which is responsible for an even faster modification of the relative depth of the recirculation region $(h/H)$. The later can is estimated using eq. (\ref{eq:bulk_reduce_high_Ra}) which writes
$$
\left(\frac{\overbar{\epsilon_{b,bulk}}}{\overbar{{\epsilon}_{b,BL}}}\right) \sim \left(\frac{\overbar{\epsilon_{u,bulk}}}{\overbar{{\epsilon}_{u,BL}}}\right) 
\equiv \frac h H \sim {\rm Re}^{-1/8} {\rm Pr}^{1/4}.
$$
Thus the scaling for the Nusselt number becomes
\begin{equation}
{\rm Nu}={\rm Re}^{3/4}{\rm Pr}^{1/2}\left(\frac h H\right)\sim\rm{Re}^{5/8}\rm{Pr}^{3/4}{\cal L}(Re)^{1/2},
\label{NuRe075_turb}
\end{equation}
%
%In the case where the thermal boundary layer also becomes turbulent,
%the dissipation of buoyancy variance becomes
%\begin{equation}
%\epsilon_{b,ll} := \kappa\Delta L^{-2}\rm{Re}\rm{Pr}{{\cal L}(Re)},
%\label{logb}
%\end{equation}
% 
and from (\ref{epsub4u}b) (\ref{epsb}), (\ref{epsu2}), and (\ref{logu}) it follows that 
\begin{subeqnarray}
\rm{Re}&\sim&\rm{Ra}^{8/21}\rm{Pr}^{-22/21}{\cal L}(Re)^{-8/21}, \\
\rm{Nu}&\sim&\rm{Ra}^{5/21}\rm{Pr}^{17/96}{\cal L}(Re)^{-5/27}.
\label{NuRe1/3ll}
\end{subeqnarray}

These scaling are verified in Fig. \ref{fig:log-scaling}(a,b) for both the Nusselt number and the Reynolds number with respect to both $\rm Ra$. The Prandtl number dependence prediction is also improved compared with eq. \ref{NuRe1/4_turb}(a,b). The exponent of the Prandtl number for the Reynolds number is found at $\rm Re \sim Pr^{-1}$ for the DNS while the log-corrected exponent is $\rm Re \sim Pr^{-1.075}$. The Nusselt-number dependence is $\rm Nu \sim Pr^{0.41}$ in the DNS while the log-corrected scaling provides $\rm Nu \sim Pr^{0.2}$ which hints at a possible Prandtl-number dependence due to the plume dynamics, as observed in Rayleigh-B\'enard convection \cite{GrossmannL11,ni2011local}.

To the best of our knowledge, it is the first time that such log-type corrections are applied and verified for the Prandtl number dependence. The log region of time-averaged turbulent boundary layers are shown in Fig. \ref{fig:log-BL} for two different Prandtl numbers and two different Rayleigh numbers in the turbulent regime in support of our assumption and analysis.
These scaling laws thus support the evidence of a new limiting regime in horizontal convection which can be considered as the limiting regime in horizontal convection. It is worth pointing that as $\rm{Ra}$ further increases, the $\alpha$ exponent progressively reaches the value of $5/21$ where log-corrections become less significant. Note that this exponent is also in very good agreement with the recent study of Reiter \& Shishkina\cite{reiter2020classical}.
The transition from the $II_l$ regime to the $IV_u$ regime in Fig. \ref{Ra_Pr} is marked by the vertical black dotted line and is obtained by matching the Reynolds number between each region. Equating eq. (\ref{NuRe1/3ll}a) with eq. (\ref{NuRe1/6corr}), the transition is found very close to a constant at $\rm Ra \approx 3\,10^{12}$.
\\

The next subsection investigates whether this last transition can be thought in terms of turbulent regimes. As one expects to see a transition to the limiting regime of convection, the question therefore arises whether turbulence in the present flow is dependent or not on viscosity and if we are reaching the asymptotic regime known as the ultimate regime where dissipation is not dependent on viscosity. Note that a viscous-independent turbulent regime would make the present result relevant for large-scale geophysical applications. 

\section{"Hard" or "Soft" turbulence?}\label{sec:Ko}

The picture drawn in the previous section provides an understanding on the heat and momentum transition leading to the ultimate regime of horizontal convection. Although the above picture is rather convincing with successive scaling transitions in agreement with what was previously observed in Rayleigh-B\'enard convection, it remains complex, with two control parameters and at least five different regimes spanning ten orders of magnitude in $\rm{Ra}$.

We propose to reanalyse our data in the framework of Kolmogorov turbulence which frames the above analysis differently than the Grossman \& Lohse\cite{GL00} picture and provide a valuable tool for diagnosing the state of turbulence observed and how these regimes and its transitions may be further analysed.\\

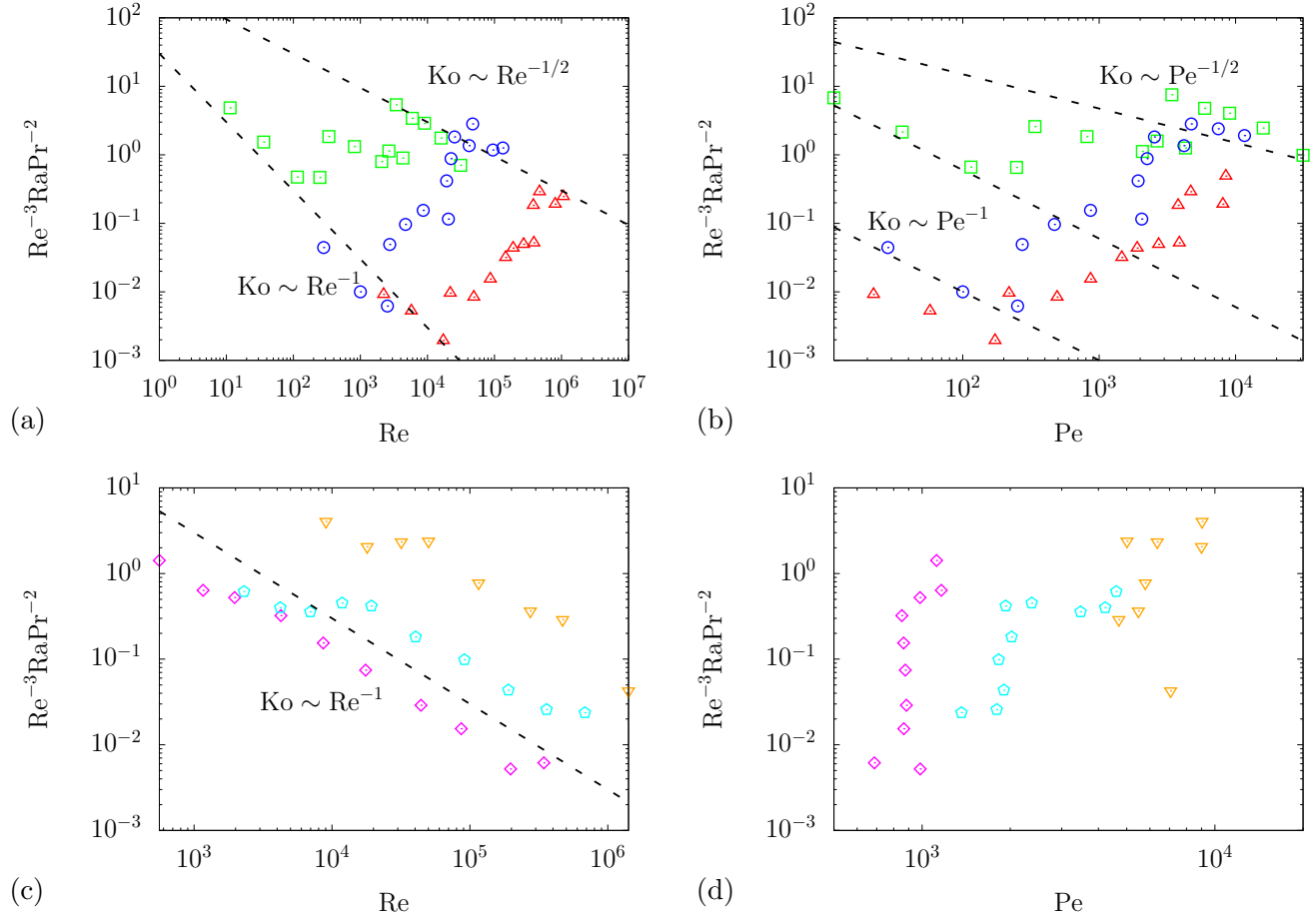
\begin{figure*}[t!]
\begin{minipage}[b]{\linewidth}
\scalebox{0.7}{\Large\input{Ko_Re}}
\scalebox{0.7}{\Large\input{Ko_Pe}}
\put(-510,10){(a)}\put(-250,10){(b)}\\
\scalebox{0.7}{\Large\input{Ko_Pr_Re}}
\scalebox{0.7}{\Large\input{Ko_Pr_Pe}}
\put(-510,10){(c)}\put(-250,10){(d)}
\end{minipage}
\vspace{-7mm}
\caption{(a) $\rm{Re}$ dependencies and (b) $\rm{Pe}$ dependencies of the Kolmogorov number $\rm{Ko}$ for variations with respect to $\rm{Ra}$. Same for (c) and (d) but for variations with respect to $\rm{Pr}$ (refer to figure \ref{Ra_Pr} for colour code).}
\label{fig:Ko}
\end{figure*}

We first define what we  denote as the Kolmogorov number, which is obtained rescaling the Paparella \& Young constraint on dissipation using  $U$ and $L$, giving
\begin{equation}
\rm{Ko} = {\rm Re}^{-3} {\rm Ra} {\rm Pr}^{-2}.
\end{equation}
The scaling law relating $\rm{Ko}$ with respect to $\rm{Re}$ and $\rm{Pe}$ provides a new way to analyse whether the flow is laminar, transitional or driven by "soft" or "hard" turbulence. In the laminar case, the Kolmogorov number and hence dissipation is solely caused by the vertical gradient of velocity where $\rm{Ko}\sim\rm{Re}^{-1}$. At the contrary, hard turbulence achieves complete similarity with respect to parameters that contain viscosity\cite{vassilicos2015dissipation}, that is we should expect $\rm{Ko}=Cst$. An intermediate regime $\rm{Ko}\sim\rm{Re}^{-1/2}$ may also be expected if boundary layers dominate dissipation, since the latter may not achieve complete similarity.

The evolution of $\rm{Ko}$ with respect to both $\rm{Re}$ and $\rm{Pe}$ is shown in  Fig. \ref{fig:Ko} where for laminar flows we obtain for all $\rm{Pr}$, $\rm{Ko}\sim\rm{Re}^{-1}$. Across the $I_l$ and $I^*_l$ regime, the dependency of $\rm{Ko}$ exhibits a $\rm{Re}^{1}$ transition, were dissipation is enhanced throughout this core-driven mixing regime. A $\rm{Ko}\sim\rm{Re}^{-1/2}$ type-regime is then recovered for all values of $\rm{Pr}$ in the $IV_u$ regime (see Fig. \ref{fig:Ko}(a)). The same observation can be done for the scaling with respect to $\rm{Pe}$ where the same conclusions arise (see Fig. \ref{fig:Ko}(b)). 
Variations with respect to the Prandtl number follow the same rationale. Transitions between regimes varying the Prandtl number is found at constant $\rm{Ko}$ (see Fig. \ref{fig:Ko}(c)) whereas in the $II_l$ and $II^*_l$ regimes, variations with respect to $\rm{Pr}$ occurs for $\rm{Ko}\sim \rm{Re}^{-1}$. At higher $\rm{Ra}$ and in the $IV_u$ regime, conclusions are yet hard to draw but we may expect a $\rm{Ko}\sim \rm{Re}^{-1/2}$.

It is straightforward to conclude that despite we have reached the limiting regime of horizontal convection at large $\rm{Ra}$,  the transition to "hard" turbulence or the "ultimate" regime of turbulent convection in natural horizontal convection does not seem attainable using our numerical simulations. In addition, this may not arise, even at extremely high $\rm{Ra}$. This agrees with Sandstr\"om inference which  may be explained by the bound on the Richardson number which holds for all regimes reported in this paper and the scaling laws available in the literature.
As shown in ref.\cite{toppaladoddi2017roughness}, such a regime may be achievable by introducing appropriate roughness elements along the statically unstable boundary, optimising the distribution of the forcing boundary\cite{rocha2020improved}, adding active forcing in this same region, or for instance through radiative heat transfer \cite{lepot2018radiative}. A such strategy may also allow for observing the $IV_l$ regime predicted by Siggers {\it et al.}\cite{siggers2004bounds,rocha2020improved}.

\section{Conclusions}\label{sec:conclusion}

In conclusion, we report evidences of two new turbulent regimes in horizontal convection based on scaling arguments at low Prandtl numbers. More precisely we first highlight regimes that are known as limiting regimes. For asymptotically small Prandtl numbers, we highlight a regime where the core is driven by turbulence but where the boundary layer remain laminar and name this regime $II_l$ following the nomenclature of Shishkina, Grossmann \& Lohse \cite{ShishkinaGL16}. The second regime is characterised  by both a turbulent core and turbulent boundary layers. It is also found to be a limiting regime for asymptotically large Rayleigh numbers called $IV_u$ following SGL's nomenclature.

%The transition occurs from $\rm{Re} \sim \rm{Ra}^{2/5}\rm{Pr}^{-4/5}, \rm{Nu} \sim \rm{Ra}^{1/5}\rm{Pr}^{1/10}$ to  $\rm{Nu}\sim \rm{Ra}^{1/6}\rm{Pr}^{1/6}, \rm{Re}\sim \rm{Ra}^{1/3}\rm{Pr}^{-2/3}$ at low $\rm{Pr}$ and sufficiently large $\rm{Ra}$ and then $\rm{Re} \sim \rm{Ra}^{2/5}\rm{Pr}^{-3/5}, \rm{Nu} \sim \rm{Ra}^{1/5}\rm{Pr}^{1/5}$.

Our results, support and integrate previous evidence from Shishkina \& Wagner \cite{ShishkinaW16} and the model of Hughes {\it et al.} (see ref. \cite{Hughes07}) in the SGL theory of HC (see ref. \cite{GL00,ShishkinaGL16}).\\

In the $II_l$ regime, we observe a new scaling where the modification of the turbulent bulk size modifies the Prandtl number dependence for both the Reynolds and the Nusselt numbers scaling. This reduction of the bulk size is found to be essentially Prandtl-number dependent where the bulk decreases in size when $\rm{Pr}$ decreases.\\

The transition to the turbulent limiting regime denoted as $IV_u$ is also observed. In this particular regime, the flow becomes turbulent, that is both the boundary layer and the core follow turbulent-type scaling laws. Similarly to the study of Rosevear {\it et al.}\cite{rosevear2017turbulent}, the turbulent flow is essentially located beneath the horizontal forcing and progressively clusters underneath as both $\rm{Ra}$ increases and $\rm{Pr}$ decreases. According to Shishkina {\it et al.}\cite{ShishkinaGL16}, this last regime marks the final transition at large $\rm{Ra}$. The log-corrections allow for recovering the correct $\rm{Nu}$ and $\rm{Re}$ dependencies with respect $\rm{Ra}$ and improved estimates for $\rm{Pr}$. however further work need to be dedicated to the exact Nusselt number dependence to the Prandtl number, which may be attributed to plume dynamics\cite{GrossmannL11,ni2011local}.\\

We also propose a new analysis, based on the Kolmgorov number $\rm Ko$, a rescaled dissipation rate, defined in such a way so that  $\rm{Ko}\sim{Re}^{-3}\rm{Ra}\rm{Pr}^{-2}$. The analysis confirms that the flow transitions from laminar to soft turbulence but also shows that the flow never transitions to hard turbulence which would be akin to the $IV_l$ regime of Siggers {\it et al.}\cite{siggers2004bounds,ShishkinaGL16,rocha2020improved}.

The ultimate regime $IV_l$, if it exists,  thus has yet to be observed (see ref. \cite{ShishkinaGL16,rocha2020improved}). It is therefore of particular interest to study new types of horizontal convection where the turbulence can be strong enough to get rid of the effect of boundary layers and trigger purely inertial, core-driven, turbulent horizontal convection regimes. A such regime would be of particular importance for geophysical applications such as the overturning circulation.\\

The companion paper Part II\cite{Passaggia2019LimitigB}, gathers the results from this study together with an experimental study at Large Prandtl number. In particular, a regime diagram is provided and highlights all known limiting regimes of horizontal convection.\\

The authors acknowledge the support of the National Science Foundation Grant Number OCE--1155558 and OCE--1736989.

\bibliographystyle{plain}

\bibliography{bib}

\end{document}

%% file: schematics.tex
\begin{picture}(0,0)%
\includegraphics{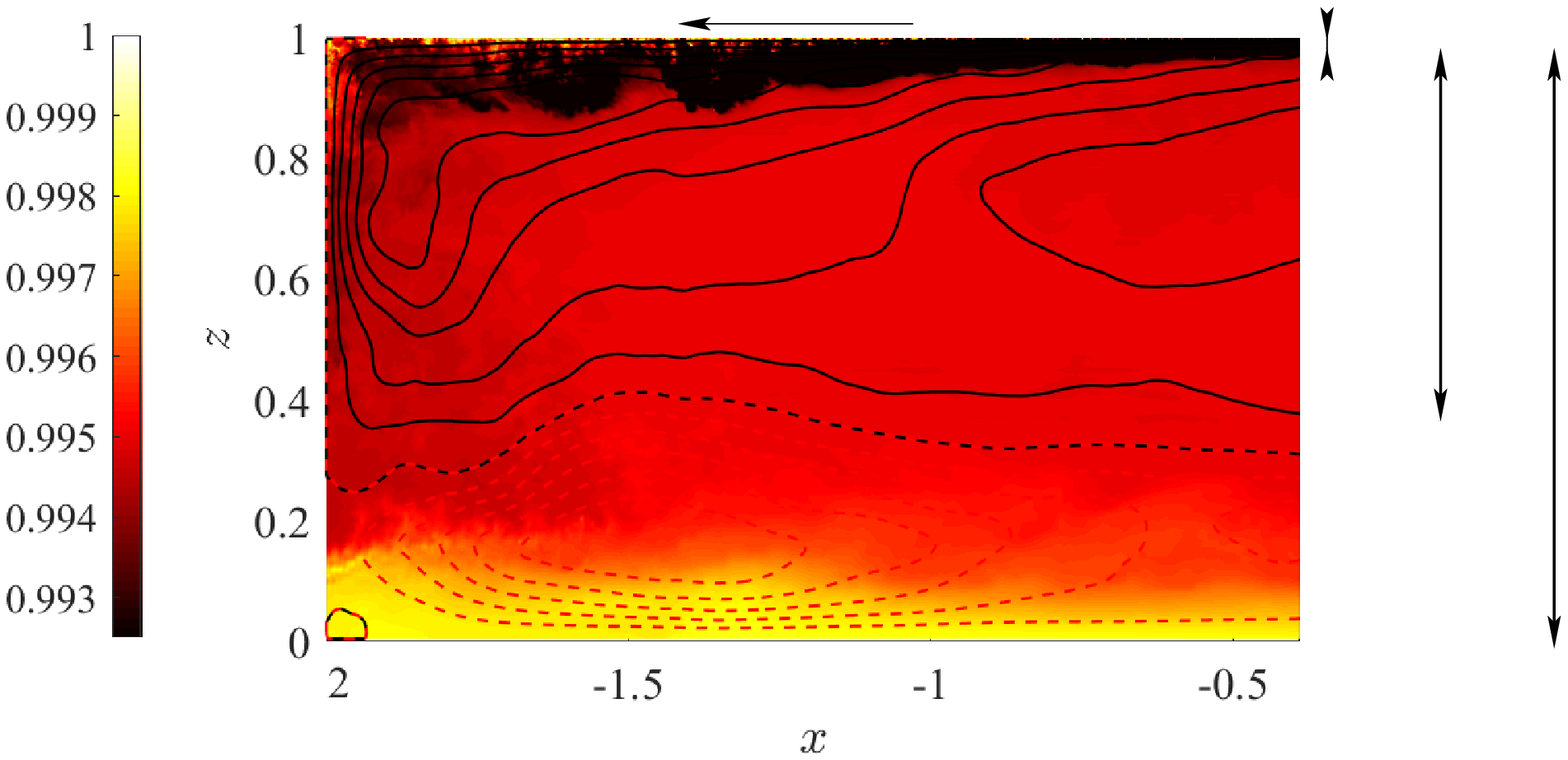}%
\end{picture}%
\setlength{\unitlength}{4144sp}%
\begingroup\makeatletter\ifx\SetFigFont\undefined%
\gdef\SetFigFont#1#2#3#4#5{%
  \reset@font\fontsize{#1}{#2pt}%
  \fontfamily{#3}\fontseries{#4}\fontshape{#5}%
  \selectfont}%
\fi\endgroup%
\begin{picture}(8938,4413)(1261,-4654)
\put(10126,-2266){\makebox(0,0)[lb]{\smash{{\SetFigFont{12}{14.4}{\rmdefault}{\mddefault}{\updefault}{\color[rgb]{0,0,0}$H$}%
}}}}
\put(9451,-1681){\makebox(0,0)[lb]{\smash{{\SetFigFont{12}{14.4}{\rmdefault}{\mddefault}{\updefault}{\color[rgb]{0,0,0}$h$}%
}}}}
\put(8821,-691){\makebox(0,0)[lb]{\smash{{\SetFigFont{12}{14.4}{\rmdefault}{\mddefault}{\updefault}{\color[rgb]{0,0,0}$\lambda_{u,b}$}%
}}}}
\put(5716,-421){\makebox(0,0)[lb]{\smash{{\SetFigFont{12}{14.4}{\rmdefault}{\mddefault}{\updefault}{\color[rgb]{0,0,0}$U$}%
}}}}
\put(1846,-4201){\makebox(0,0)[lb]{\smash{{\SetFigFont{12}{14.4}{\rmdefault}{\mddefault}{\updefault}{\color[rgb]{0,0,0}$b$}%
}}}}
\end{picture}%

%% file: snapshot_flow_field.tex
\begin{picture}(0,0)%
\includegraphics{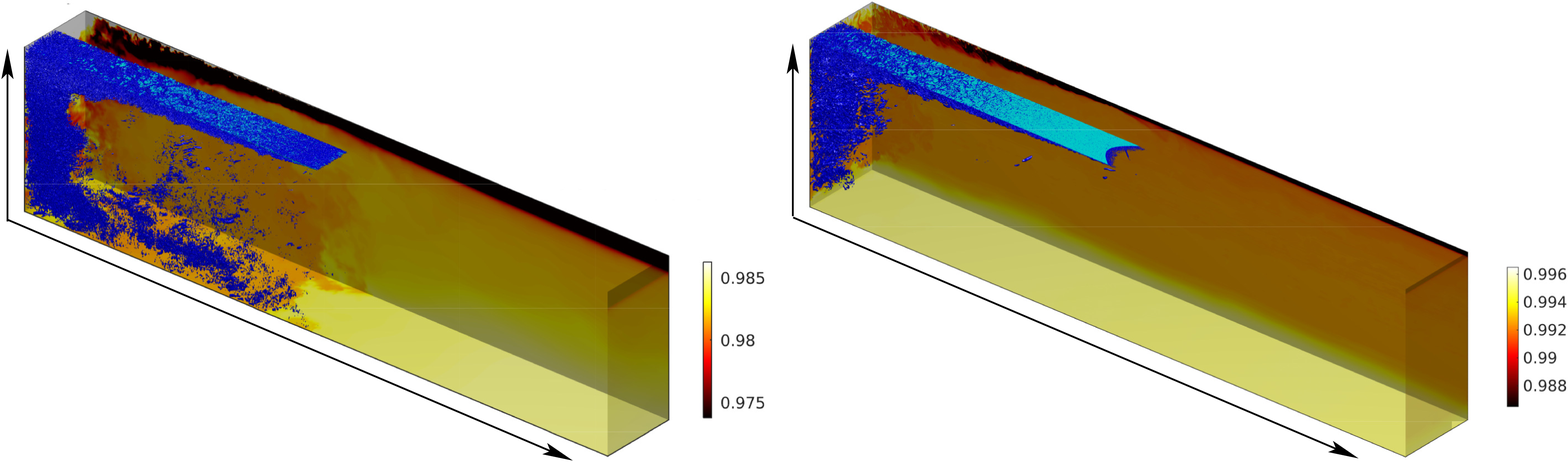}%
\end{picture}%
\setlength{\unitlength}{4144sp}%
\begingroup\makeatletter\ifx\SetFigFont\undefined%
\gdef\SetFigFont#1#2#3#4#5{%
  \reset@font\fontsize{#1}{#2pt}%
  \fontfamily{#3}\fontseries{#4}\fontshape{#5}%
  \selectfont}%
\fi\endgroup%
\begin{picture}(41519,12149)(2206,-13380)
%\put(2251,-5056){\makebox(0,0)[lb]{\smash{{\SetFigFont{248}{257.6}{\rmdefault}{\mddefault}{\updefault}{\color[rgb]{0,0,0}$z$}%
%}}}}
%\put(8281,-10231){\makebox(0,0)[lb]{\smash{{\SetFigFont{248}{257.6}{\rmdefault}{\mddefault}{\updefault}{\color[rgb]{0,0,0}$x$}%
%}}}}
%\put(19576,-12976){\makebox(0,0)[lb]{\smash{{\SetFigFont{248}{257.6}{\rmdefault}{\mddefault}{\updefault}{\color[rgb]{0,0,0}$y$}%
%}}}}
%\put(29431,-10366){\makebox(0,0)[lb]{\smash{{\SetFigFont{248}{257}{\rmdefault}{\mddefault}{\updefault}{\color[rgb]{0,0,0}$x$}%
%}}}}
%\put(22591,-4786){\makebox(0,0)[lb]{\smash{{\SetFigFont{248}{257.6}{\rmdefault}{\mddefault}{\updefault}{\color[rgb]{0,0,0}$z$}%
%}}}}
%\put(2296,-2851){\makebox(0,0)[lb]{\smash{{\SetFigFont{248}{257.6}{\rmdefault}{\mddefault}{\updefault}{\color[rgb]{0,0,0}$1/4$}%
%}}}}
%\put(15571,-13246){\makebox(0,0)[lb]{\smash{{\SetFigFont{248}{257.6}{\rmdefault}{\mddefault}{\updefault}{\color[rgb]{0,0,0}$1$}%
%}}}}
%\put(4501,-1501){\makebox(0,0)[lb]{\smash{{\SetFigFont{248}{257.6}{\rmdefault}{\mddefault}{\updefault}{\color[rgb]{0,0,0}$1/16$}%
%}}}}
%\put(2296,-7171){\makebox(0,0)[lb]{\smash{{\SetFigFont{248}{257.6}{\rmdefault}{\mddefault}{\updefault}{\color[rgb]{0,0,0}$0$}%
%}}}}
%\put(40231,-12931){\makebox(0,0)[lb]{\smash{{\SetFigFont{248}{257.6}{\rmdefault}{\mddefault}{\updefault}{\color[rgb]{0,0,0}$y$}%
%}}}}
\end{picture}%

%% file: spectra_2.tex
% GNUPLOT: LaTeX picture with Postscript
\begingroup
  \fontfamily{Times-Roman}%
  \selectfont
  \makeatletter
  \providecommand\color[2][]{%
    \GenericError{(gnuplot) \space\space\space\@spaces}{%
      Package color not loaded in conjunction with
      terminal option `colourtext'%
    }{See the gnuplot documentation for explanation.%
    }{Either use 'blacktext' in gnuplot or load the package
      color.sty in LaTeX.}%
    \renewcommand\color[2][]{}%
  }%
  \providecommand\includegraphics[2][]{%
    \GenericError{(gnuplot) \space\space\space\@spaces}{%
      Package graphicx or graphics not loaded%
    }{See the gnuplot documentation for explanation.%
    }{The gnuplot epslatex terminal needs graphicx.sty or graphics.sty.}%
    \renewcommand\includegraphics[2][]{}%
  }%
  \providecommand\rotatebox[2]{#2}%
  \@ifundefined{ifGPcolor}{%
    \newif\ifGPcolor
    \GPcolortrue
  }{}%
  \@ifundefined{ifGPblacktext}{%
    \newif\ifGPblacktext
    \GPblacktexttrue
  }{}%
  % define a \g@addto@macro without @ in the name:
  \let\gplgaddtomacro\g@addto@macro
  % define empty templates for all commands taking text:
  \gdef\gplbacktext{}%
  \gdef\gplfronttext{}%
  \makeatother
  \ifGPblacktext
    % no textcolor at all
    \def\colorrgb#1{}%
    \def\colorgray#1{}%
  \else
    % gray or color?
    \ifGPcolor
      \def\colorrgb#1{\color[rgb]{#1}}%
      \def\colorgray#1{\color[gray]{#1}}%
      \expandafter\def\csname LTw\endcsname{\color{white}}%
      \expandafter\def\csname LTb\endcsname{\color{black}}%
      \expandafter\def\csname LTa\endcsname{\color{black}}%
      \expandafter\def\csname LT0\endcsname{\color[rgb]{1,0,0}}%
      \expandafter\def\csname LT1\endcsname{\color[rgb]{0,1,0}}%
      \expandafter\def\csname LT2\endcsname{\color[rgb]{0,0,1}}%
      \expandafter\def\csname LT3\endcsname{\color[rgb]{1,0,1}}%
      \expandafter\def\csname LT4\endcsname{\color[rgb]{0,1,1}}%
      \expandafter\def\csname LT5\endcsname{\color[rgb]{1,1,0}}%
      \expandafter\def\csname LT6\endcsname{\color[rgb]{0,0,0}}%
      \expandafter\def\csname LT7\endcsname{\color[rgb]{1,0.3,0}}%
      \expandafter\def\csname LT8\endcsname{\color[rgb]{0.5,0.5,0.5}}%
    \else
      % gray
      \def\colorrgb#1{\color{black}}%
      \def\colorgray#1{\color[gray]{#1}}%
      \expandafter\def\csname LTw\endcsname{\color{white}}%
      \expandafter\def\csname LTb\endcsname{\color{black}}%
      \expandafter\def\csname LTa\endcsname{\color{black}}%
      \expandafter\def\csname LT0\endcsname{\color{black}}%
      \expandafter\def\csname LT1\endcsname{\color{black}}%
      \expandafter\def\csname LT2\endcsname{\color{black}}%
      \expandafter\def\csname LT3\endcsname{\color{black}}%
      \expandafter\def\csname LT4\endcsname{\color{black}}%
      \expandafter\def\csname LT5\endcsname{\color{black}}%
      \expandafter\def\csname LT6\endcsname{\color{black}}%
      \expandafter\def\csname LT7\endcsname{\color{black}}%
      \expandafter\def\csname LT8\endcsname{\color{black}}%
    \fi
  \fi
    \setlength{\unitlength}{0.0500bp}%
    \ifx\gptboxheight\undefined%
      \newlength{\gptboxheight}%
      \newlength{\gptboxwidth}%
      \newsavebox{\gptboxtext}%
    \fi%
    \setlength{\fboxrule}{0.5pt}%
    \setlength{\fboxsep}{1pt}%
\begin{picture}(7200.00,5040.00)%
    \gplgaddtomacro\gplbacktext{%
      \csname LTb\endcsname%%
      \put(1376,1024){\makebox(0,0)[r]{\strut{}$10^{-2}$}}%
      \csname LTb\endcsname%%
      \put(1376,2256){\makebox(0,0)[r]{\strut{}$10^{-1}$}}%
      \csname LTb\endcsname%%
      \put(1376,3487){\makebox(0,0)[r]{\strut{}$10^{0}$}}%
      \csname LTb\endcsname%%
      \put(1376,4719){\makebox(0,0)[r]{\strut{}$10^{1}$}}%
      \csname LTb\endcsname%%
      \put(1614,704){\makebox(0,0){\strut{}$10^{0}$}}%
      \csname LTb\endcsname%%
      \put(3889,704){\makebox(0,0){\strut{}$10^{1}$}}%
      \csname LTb\endcsname%%
      \put(6165,704){\makebox(0,0){\strut{}$10^{2}$}}%
    }%
    \gplgaddtomacro\gplfronttext{%
      \csname LTb\endcsname%%
      \put(288,2871){\rotatebox{-270}{\makebox(0,0){\strut{}$E(\mathcal{K})\epsilon^{-2/3}\mathcal{K}_x^{5/3}$}}}%
      \put(4095,224){\makebox(0,0){\strut{}$\mathcal{K}_xL/(2\pi)$}}%
    }%
    \gplbacktext
    \put(0,0){\includegraphics{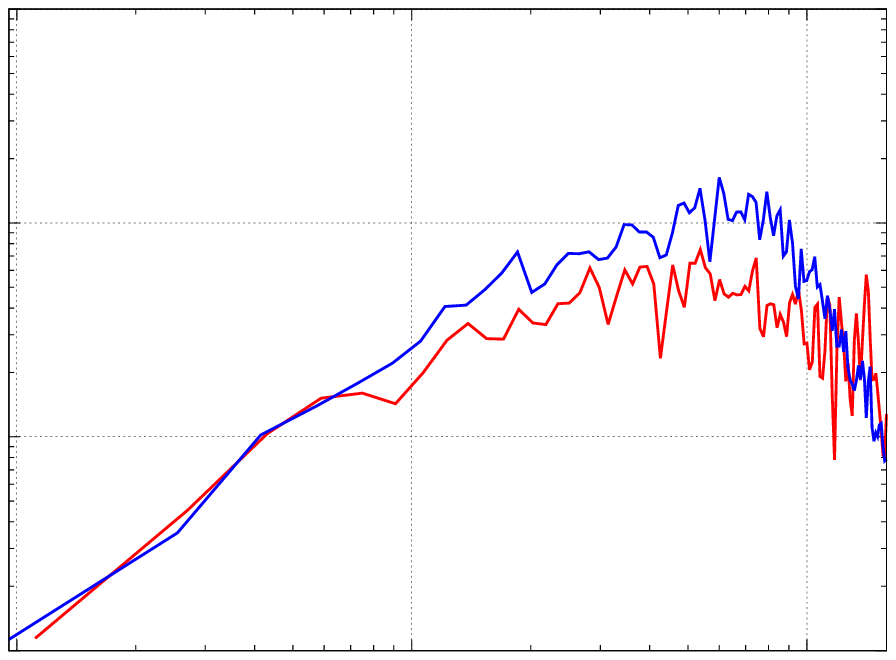}}%
    \gplfronttext
  \end{picture}%
\endgroup

%% file: Ra_Pr_All.tex
% GNUPLOT: LaTeX picture with Postscript
\begingroup
  \fontfamily{Times-Roman}%
  \selectfont
  \makeatletter
  \providecommand\color[2][]{%
    \GenericError{(gnuplot) \space\space\space\@spaces}{%
      Package color not loaded in conjunction with
      terminal option `colourtext'%
    }{See the gnuplot documentation for explanation.%
    }{Either use 'blacktext' in gnuplot or load the package
      color.sty in LaTeX.}%
    \renewcommand\color[2][]{}%
  }%
  \providecommand\includegraphics[2][]{%
    \GenericError{(gnuplot) \space\space\space\@spaces}{%
      Package graphicx or graphics not loaded%
    }{See the gnuplot documentation for explanation.%
    }{The gnuplot epslatex terminal needs graphicx.sty or graphics.sty.}%
    \renewcommand\includegraphics[2][]{}%
  }%
  \providecommand\rotatebox[2]{#2}%
  \@ifundefined{ifGPcolor}{%
    \newif\ifGPcolor
    \GPcolortrue
  }{}%
  \@ifundefined{ifGPblacktext}{%
    \newif\ifGPblacktext
    \GPblacktexttrue
  }{}%
  % define a \g@addto@macro without @ in the name:
  \let\gplgaddtomacro\g@addto@macro
  % define empty templates for all commands taking text:
  \gdef\gplbacktext{}%
  \gdef\gplfronttext{}%
  \makeatother
  \ifGPblacktext
    % no textcolor at all
    \def\colorrgb#1{}%
    \def\colorgray#1{}%
  \else
    % gray or color?
    \ifGPcolor
      \def\colorrgb#1{\color[rgb]{#1}}%
      \def\colorgray#1{\color[gray]{#1}}%
      \expandafter\def\csname LTw\endcsname{\color{white}}%
      \expandafter\def\csname LTb\endcsname{\color{black}}%
      \expandafter\def\csname LTa\endcsname{\color{black}}%
      \expandafter\def\csname LT0\endcsname{\color[rgb]{1,0,0}}%
      \expandafter\def\csname LT1\endcsname{\color[rgb]{0,1,0}}%
      \expandafter\def\csname LT2\endcsname{\color[rgb]{0,0,1}}%
      \expandafter\def\csname LT3\endcsname{\color[rgb]{1,0,1}}%
      \expandafter\def\csname LT4\endcsname{\color[rgb]{0,1,1}}%
      \expandafter\def\csname LT5\endcsname{\color[rgb]{1,1,0}}%
      \expandafter\def\csname LT6\endcsname{\color[rgb]{0,0,0}}%
      \expandafter\def\csname LT7\endcsname{\color[rgb]{1,0.3,0}}%
      \expandafter\def\csname LT8\endcsname{\color[rgb]{0.5,0.5,0.5}}%
    \else
      % gray
      \def\colorrgb#1{\color{black}}%
      \def\colorgray#1{\color[gray]{#1}}%
      \expandafter\def\csname LTw\endcsname{\color{white}}%
      \expandafter\def\csname LTb\endcsname{\color{black}}%
      \expandafter\def\csname LTa\endcsname{\color{black}}%
      \expandafter\def\csname LT0\endcsname{\color{black}}%
      \expandafter\def\csname LT1\endcsname{\color{black}}%
      \expandafter\def\csname LT2\endcsname{\color{black}}%
      \expandafter\def\csname LT3\endcsname{\color{black}}%
      \expandafter\def\csname LT4\endcsname{\color{black}}%
      \expandafter\def\csname LT5\endcsname{\color{black}}%
      \expandafter\def\csname LT6\endcsname{\color{black}}%
      \expandafter\def\csname LT7\endcsname{\color{black}}%
      \expandafter\def\csname LT8\endcsname{\color{black}}%
    \fi
  \fi
    \setlength{\unitlength}{0.0500bp}%
    \ifx\gptboxheight\undefined%
      \newlength{\gptboxheight}%
      \newlength{\gptboxwidth}%
      \newsavebox{\gptboxtext}%
    \fi%
    \setlength{\fboxrule}{0.5pt}%
    \setlength{\fboxsep}{1pt}%
\begin{picture}(7200.00,5040.00)%
    \gplgaddtomacro\gplbacktext{%
      \csname LTb\endcsname%%
      \put(1376,1024){\makebox(0,0)[r]{\strut{}$10^{-3}$}}%
      \put(1376,2087){\makebox(0,0)[r]{\strut{}$10^{-2}$}}%
      \put(1376,3149){\makebox(0,0)[r]{\strut{}$10^{-1}$}}%
      \put(1376,4212){\makebox(0,0)[r]{\strut{}$10^{0}$}}%
      \put(2028,704){\makebox(0,0){\strut{}$10^{6}$}}%
      \put(2947,704){\makebox(0,0){\strut{}$10^{8}$}}%
      \put(3866,704){\makebox(0,0){\strut{}$10^{10}$}}%
      \put(4785,704){\makebox(0,0){\strut{}$10^{12}$}}%
      \put(5704,704){\makebox(0,0){\strut{}$10^{14}$}}%
      \put(6623,704){\makebox(0,0){\strut{}$10^{16}$}}%
      \put(2108,3789){\makebox(0,0)[l]{\strut{}\framebox[5mm]{$I^*_{l}$}\,$\left(\frac{1}{4},0\right)$}}%
      \put(2166,2510){\makebox(0,0)[l]{\strut{}\framebox[4mm]{$I_l$} \,$\left(\frac{1}{5},\frac{1}{10}\right)$}}%
      \put(2487,1344){\makebox(0,0)[l]{\strut{}\framebox[6mm]{$II_l$}\,$\left(\frac{1}{6},\frac{1}{3}\right)$}}%
      \put(5142,4429){\makebox(0,0)[l]{\strut{}\framebox[6mm]{$II_u$}\,$\left(\frac{1}{5},\frac{1}{5}\right)$}}%
      \put(5923,2829){\makebox(0,0)[l]{\strut{}\framebox[6mm]{$IV_u$}}}%
      \put(5923,2407){\makebox(0,0)[l]{\strut{}$\left(\frac{1}{4.4},\frac{1}{2.4}\right)$}}%
    }%
    \gplgaddtomacro\gplfronttext{%
      \csname LTb\endcsname%%
      \put(288,2871){\rotatebox{-270}{\makebox(0,0){\strut{}$\rm{Pr}$}}}%
      \put(4095,224){\makebox(0,0){\strut{}$\rm{Ra}$}}%
    }%
    \gplbacktext
    \put(0,0){\includegraphics{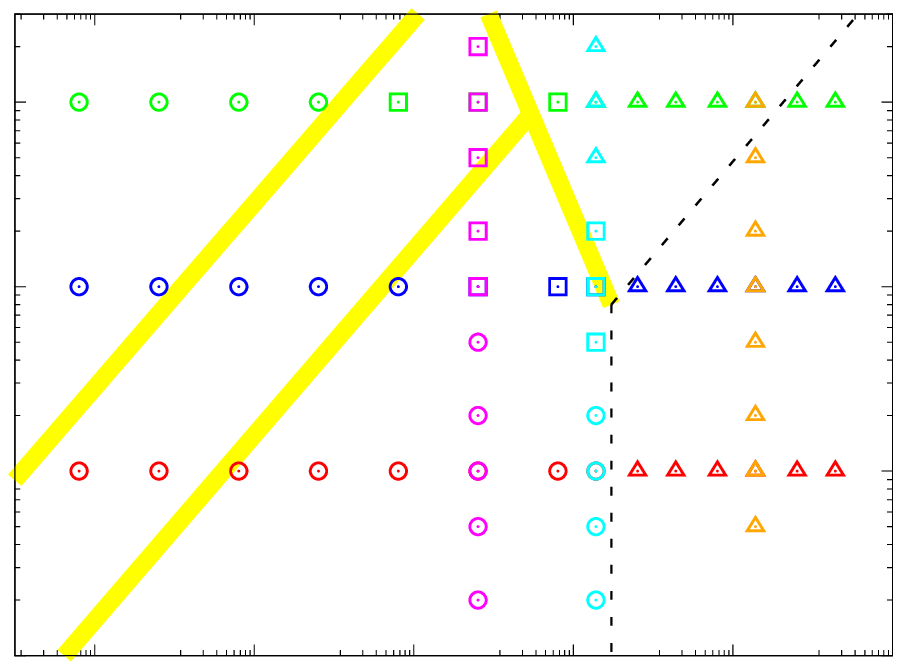}}%
    \gplfronttext
  \end{picture}%
\endgroup

%% file: NuRa1e025_Ra_L.tex
% GNUPLOT: LaTeX picture with Postscript
\begingroup
  \fontfamily{Times-Roman}%
  \selectfont
  \makeatletter
  \providecommand\color[2][]{%
    \GenericError{(gnuplot) \space\space\space\@spaces}{%
      Package color not loaded in conjunction with
      terminal option `colourtext'%
    }{See the gnuplot documentation for explanation.%
    }{Either use 'blacktext' in gnuplot or load the package
      color.sty in LaTeX.}%
    \renewcommand\color[2][]{}%
  }%
  \providecommand\includegraphics[2][]{%
    \GenericError{(gnuplot) \space\space\space\@spaces}{%
      Package graphicx or graphics not loaded%
    }{See the gnuplot documentation for explanation.%
    }{The gnuplot epslatex terminal needs graphicx.sty or graphics.sty.}%
    \renewcommand\includegraphics[2][]{}%
  }%
  \providecommand\rotatebox[2]{#2}%
  \@ifundefined{ifGPcolor}{%
    \newif\ifGPcolor
    \GPcolortrue
  }{}%
  \@ifundefined{ifGPblacktext}{%
    \newif\ifGPblacktext
    \GPblacktexttrue
  }{}%
  % define a \g@addto@macro without @ in the name:
  \let\gplgaddtomacro\g@addto@macro
  % define empty templates for all commands taking text:
  \gdef\gplbacktext{}%
  \gdef\gplfronttext{}%
  \makeatother
  \ifGPblacktext
    % no textcolor at all
    \def\colorrgb#1{}%
    \def\colorgray#1{}%
  \else
    % gray or color?
    \ifGPcolor
      \def\colorrgb#1{\color[rgb]{#1}}%
      \def\colorgray#1{\color[gray]{#1}}%
      \expandafter\def\csname LTw\endcsname{\color{white}}%
      \expandafter\def\csname LTb\endcsname{\color{black}}%
      \expandafter\def\csname LTa\endcsname{\color{black}}%
      \expandafter\def\csname LT0\endcsname{\color[rgb]{1,0,0}}%
      \expandafter\def\csname LT1\endcsname{\color[rgb]{0,1,0}}%
      \expandafter\def\csname LT2\endcsname{\color[rgb]{0,0,1}}%
      \expandafter\def\csname LT3\endcsname{\color[rgb]{1,0,1}}%
      \expandafter\def\csname LT4\endcsname{\color[rgb]{0,1,1}}%
      \expandafter\def\csname LT5\endcsname{\color[rgb]{1,1,0}}%
      \expandafter\def\csname LT6\endcsname{\color[rgb]{0,0,0}}%
      \expandafter\def\csname LT7\endcsname{\color[rgb]{1,0.3,0}}%
      \expandafter\def\csname LT8\endcsname{\color[rgb]{0.5,0.5,0.5}}%
    \else
      % gray
      \def\colorrgb#1{\color{black}}%
      \def\colorgray#1{\color[gray]{#1}}%
      \expandafter\def\csname LTw\endcsname{\color{white}}%
      \expandafter\def\csname LTb\endcsname{\color{black}}%
      \expandafter\def\csname LTa\endcsname{\color{black}}%
      \expandafter\def\csname LT0\endcsname{\color{black}}%
      \expandafter\def\csname LT1\endcsname{\color{black}}%
      \expandafter\def\csname LT2\endcsname{\color{black}}%
      \expandafter\def\csname LT3\endcsname{\color{black}}%
      \expandafter\def\csname LT4\endcsname{\color{black}}%
      \expandafter\def\csname LT5\endcsname{\color{black}}%
      \expandafter\def\csname LT6\endcsname{\color{black}}%
      \expandafter\def\csname LT7\endcsname{\color{black}}%
      \expandafter\def\csname LT8\endcsname{\color{black}}%
    \fi
  \fi
    \setlength{\unitlength}{0.0500bp}%
    \ifx\gptboxheight\undefined%
      \newlength{\gptboxheight}%
      \newlength{\gptboxwidth}%
      \newsavebox{\gptboxtext}%
    \fi%
    \setlength{\fboxrule}{0.5pt}%
    \setlength{\fboxsep}{1pt}%
\begin{picture}(7200.00,5040.00)%
    \gplgaddtomacro\gplbacktext{%
      \csname LTb\endcsname%%
      \put(1184,2872){\makebox(0,0)[r]{\strut{}$0.1$}}%
      \put(1853,704){\makebox(0,0){\strut{}$10^{6}$}}%
      \put(2807,704){\makebox(0,0){\strut{}$10^{8}$}}%
      \put(3761,704){\makebox(0,0){\strut{}$10^{10}$}}%
      \put(4715,704){\makebox(0,0){\strut{}$10^{12}$}}%
      \put(5669,704){\makebox(0,0){\strut{}$10^{14}$}}%
      \put(6623,704){\makebox(0,0){\strut{}$10^{16}$}}%
      \put(2663,2800){\makebox(0,0)[l]{\strut{}$\rm{Nu}\sim \rm{Ra}^{1/6}$}}%
      \put(3951,4170){\makebox(0,0)[l]{\strut{}$\rm{Nu}\sim \rm{Ra}^{1/5}$}}%
      \put(5036,2005){\makebox(0,0)[l]{\strut{}$\rm{Nu}\sim \rm{Ra}^{0.225}$}}%
      \put(1376,4463){\makebox(0,0)[l]{\strut{}$\rm{Nu}\sim \rm{Ra}^{1/4}$}}%
      \put(1376,3546){\makebox(0,0)[l]{\strut{}$\rm{Nu}\sim \rm{Ra}^{1/5}$}}%
    }%
    \gplgaddtomacro\gplfronttext{%
      \csname LTb\endcsname%%
      \put(288,2871){\rotatebox{-270}{\makebox(0,0){\strut{}$\rm NuRa^{-1/4}$}}}%
      \put(3999,224){\makebox(0,0){\strut{}$\rm Ra$}}%
    }%
    \gplbacktext
    \put(0,0){\includegraphics{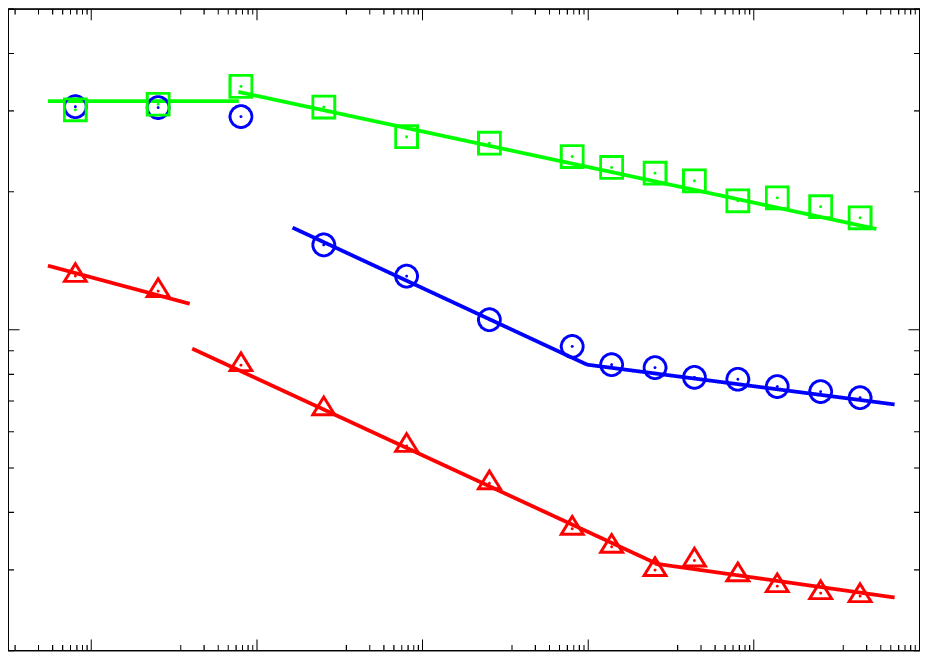}}%
    \gplfronttext
  \end{picture}%
\endgroup

%% file: NuPr-03_Pr_L.tex
% GNUPLOT: LaTeX picture with Postscript
\begingroup
  \fontfamily{Times-Roman}%
  \selectfont
  \makeatletter
  \providecommand\color[2][]{%
    \GenericError{(gnuplot) \space\space\space\@spaces}{%
      Package color not loaded in conjunction with
      terminal option `colourtext'%
    }{See the gnuplot documentation for explanation.%
    }{Either use 'blacktext' in gnuplot or load the package
      color.sty in LaTeX.}%
    \renewcommand\color[2][]{}%
  }%
  \providecommand\includegraphics[2][]{%
    \GenericError{(gnuplot) \space\space\space\@spaces}{%
      Package graphicx or graphics not loaded%
    }{See the gnuplot documentation for explanation.%
    }{The gnuplot epslatex terminal needs graphicx.sty or graphics.sty.}%
    \renewcommand\includegraphics[2][]{}%
  }%
  \providecommand\rotatebox[2]{#2}%
  \@ifundefined{ifGPcolor}{%
    \newif\ifGPcolor
    \GPcolortrue
  }{}%
  \@ifundefined{ifGPblacktext}{%
    \newif\ifGPblacktext
    \GPblacktexttrue
  }{}%
  % define a \g@addto@macro without @ in the name:
  \let\gplgaddtomacro\g@addto@macro
  % define empty templates for all commands taking text:
  \gdef\gplbacktext{}%
  \gdef\gplfronttext{}%
  \makeatother
  \ifGPblacktext
    % no textcolor at all
    \def\colorrgb#1{}%
    \def\colorgray#1{}%
  \else
    % gray or color?
    \ifGPcolor
      \def\colorrgb#1{\color[rgb]{#1}}%
      \def\colorgray#1{\color[gray]{#1}}%
      \expandafter\def\csname LTw\endcsname{\color{white}}%
      \expandafter\def\csname LTb\endcsname{\color{black}}%
      \expandafter\def\csname LTa\endcsname{\color{black}}%
      \expandafter\def\csname LT0\endcsname{\color[rgb]{1,0,0}}%
      \expandafter\def\csname LT1\endcsname{\color[rgb]{0,1,0}}%
      \expandafter\def\csname LT2\endcsname{\color[rgb]{0,0,1}}%
      \expandafter\def\csname LT3\endcsname{\color[rgb]{1,0,1}}%
      \expandafter\def\csname LT4\endcsname{\color[rgb]{0,1,1}}%
      \expandafter\def\csname LT5\endcsname{\color[rgb]{1,1,0}}%
      \expandafter\def\csname LT6\endcsname{\color[rgb]{0,0,0}}%
      \expandafter\def\csname LT7\endcsname{\color[rgb]{1,0.3,0}}%
      \expandafter\def\csname LT8\endcsname{\color[rgb]{0.5,0.5,0.5}}%
    \else
      % gray
      \def\colorrgb#1{\color{black}}%
      \def\colorgray#1{\color[gray]{#1}}%
      \expandafter\def\csname LTw\endcsname{\color{white}}%
      \expandafter\def\csname LTb\endcsname{\color{black}}%
      \expandafter\def\csname LTa\endcsname{\color{black}}%
      \expandafter\def\csname LT0\endcsname{\color{black}}%
      \expandafter\def\csname LT1\endcsname{\color{black}}%
      \expandafter\def\csname LT2\endcsname{\color{black}}%
      \expandafter\def\csname LT3\endcsname{\color{black}}%
      \expandafter\def\csname LT4\endcsname{\color{black}}%
      \expandafter\def\csname LT5\endcsname{\color{black}}%
      \expandafter\def\csname LT6\endcsname{\color{black}}%
      \expandafter\def\csname LT7\endcsname{\color{black}}%
      \expandafter\def\csname LT8\endcsname{\color{black}}%
    \fi
  \fi
    \setlength{\unitlength}{0.0500bp}%
    \ifx\gptboxheight\undefined%
      \newlength{\gptboxheight}%
      \newlength{\gptboxwidth}%
      \newsavebox{\gptboxtext}%
    \fi%
    \setlength{\fboxrule}{0.5pt}%
    \setlength{\fboxsep}{1pt}%
\begin{picture}(7200.00,5040.00)%
    \gplgaddtomacro\gplbacktext{%
      \csname LTb\endcsname%%
      \put(1184,3220){\makebox(0,0)[r]{\strut{}$10^{2}$}}%
      \put(1376,704){\makebox(0,0){\strut{}$10^{-3}$}}%
      \put(2795,704){\makebox(0,0){\strut{}$10^{-2}$}}%
      \put(4213,704){\makebox(0,0){\strut{}$10^{-1}$}}%
      \put(5632,704){\makebox(0,0){\strut{}$10^{0}$}}%
      \put(1803,4470){\makebox(0,0)[l]{\strut{}$\rm Nu\sim Pr^{2/3-1/4}$}}%
      \put(5104,3220){\makebox(0,0)[l]{\strut{}$\rm Nu\sim Pr^{1/5}$}}%
      \put(5032,1170){\makebox(0,0)[l]{\strut{}$\rm Nu\sim Pr^{1/10}$}}%
      \put(3044,1577){\makebox(0,0)[l]{\strut{}$\rm Nu\sim Pr^{1/3}$}}%
    }%
    \gplgaddtomacro\gplfronttext{%
      \csname LTb\endcsname%%
      \put(288,2871){\rotatebox{-270}{\makebox(0,0){\strut{}$\rm NuPr^{-1/3}$}}}%
      \put(3999,224){\makebox(0,0){\strut{}$\rm Pr$}}%
    }%
    \gplbacktext
    \put(0,0){\includegraphics{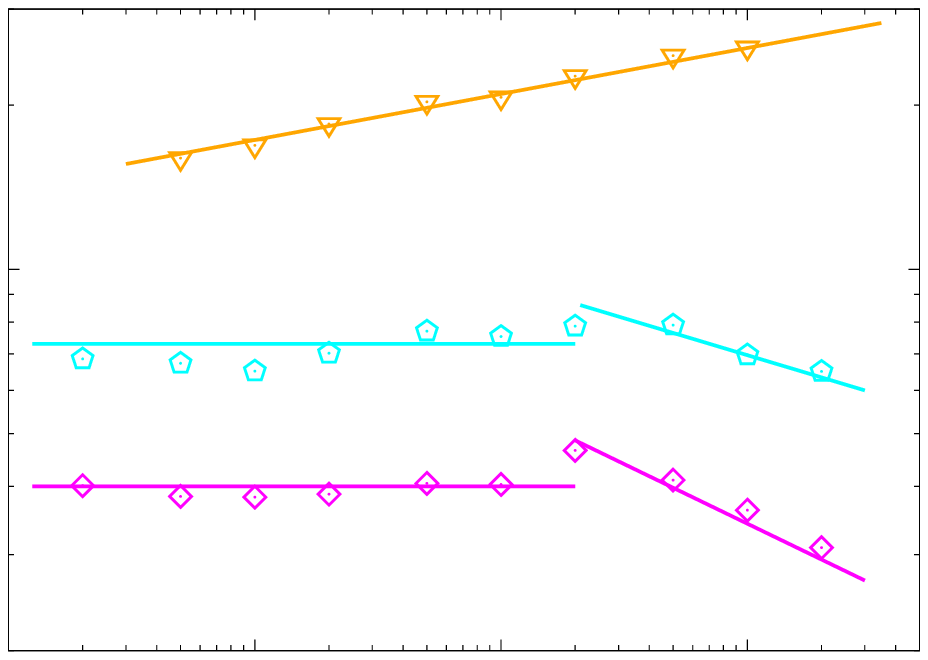}}%
    \gplfronttext
  \end{picture}%
\endgroup

%% file: ReRa-03_Ra_L.tex
% GNUPLOT: LaTeX picture with Postscript
\begingroup
  \fontfamily{Times-Roman}%
  \selectfont
  \makeatletter
  \providecommand\color[2][]{%
    \GenericError{(gnuplot) \space\space\space\@spaces}{%
      Package color not loaded in conjunction with
      terminal option `colourtext'%
    }{See the gnuplot documentation for explanation.%
    }{Either use 'blacktext' in gnuplot or load the package
      color.sty in LaTeX.}%
    \renewcommand\color[2][]{}%
  }%
  \providecommand\includegraphics[2][]{%
    \GenericError{(gnuplot) \space\space\space\@spaces}{%
      Package graphicx or graphics not loaded%
    }{See the gnuplot documentation for explanation.%
    }{The gnuplot epslatex terminal needs graphicx.sty or graphics.sty.}%
    \renewcommand\includegraphics[2][]{}%
  }%
  \providecommand\rotatebox[2]{#2}%
  \@ifundefined{ifGPcolor}{%
    \newif\ifGPcolor
    \GPcolortrue
  }{}%
  \@ifundefined{ifGPblacktext}{%
    \newif\ifGPblacktext
    \GPblacktexttrue
  }{}%
  % define a \g@addto@macro without @ in the name:
  \let\gplgaddtomacro\g@addto@macro
  % define empty templates for all commands taking text:
  \gdef\gplbacktext{}%
  \gdef\gplfronttext{}%
  \makeatother
  \ifGPblacktext
    % no textcolor at all
    \def\colorrgb#1{}%
    \def\colorgray#1{}%
  \else
    % gray or color?
    \ifGPcolor
      \def\colorrgb#1{\color[rgb]{#1}}%
      \def\colorgray#1{\color[gray]{#1}}%
      \expandafter\def\csname LTw\endcsname{\color{white}}%
      \expandafter\def\csname LTb\endcsname{\color{black}}%
      \expandafter\def\csname LTa\endcsname{\color{black}}%
      \expandafter\def\csname LT0\endcsname{\color[rgb]{1,0,0}}%
      \expandafter\def\csname LT1\endcsname{\color[rgb]{0,1,0}}%
      \expandafter\def\csname LT2\endcsname{\color[rgb]{0,0,1}}%
      \expandafter\def\csname LT3\endcsname{\color[rgb]{1,0,1}}%
      \expandafter\def\csname LT4\endcsname{\color[rgb]{0,1,1}}%
      \expandafter\def\csname LT5\endcsname{\color[rgb]{1,1,0}}%
      \expandafter\def\csname LT6\endcsname{\color[rgb]{0,0,0}}%
      \expandafter\def\csname LT7\endcsname{\color[rgb]{1,0.3,0}}%
      \expandafter\def\csname LT8\endcsname{\color[rgb]{0.5,0.5,0.5}}%
    \else
      % gray
      \def\colorrgb#1{\color{black}}%
      \def\colorgray#1{\color[gray]{#1}}%
      \expandafter\def\csname LTw\endcsname{\color{white}}%
      \expandafter\def\csname LTb\endcsname{\color{black}}%
      \expandafter\def\csname LTa\endcsname{\color{black}}%
      \expandafter\def\csname LT0\endcsname{\color{black}}%
      \expandafter\def\csname LT1\endcsname{\color{black}}%
      \expandafter\def\csname LT2\endcsname{\color{black}}%
      \expandafter\def\csname LT3\endcsname{\color{black}}%
      \expandafter\def\csname LT4\endcsname{\color{black}}%
      \expandafter\def\csname LT5\endcsname{\color{black}}%
      \expandafter\def\csname LT6\endcsname{\color{black}}%
      \expandafter\def\csname LT7\endcsname{\color{black}}%
      \expandafter\def\csname LT8\endcsname{\color{black}}%
    \fi
  \fi
    \setlength{\unitlength}{0.0500bp}%
    \ifx\gptboxheight\undefined%
      \newlength{\gptboxheight}%
      \newlength{\gptboxwidth}%
      \newsavebox{\gptboxtext}%
    \fi%
    \setlength{\fboxrule}{0.5pt}%
    \setlength{\fboxsep}{1pt}%
\begin{picture}(7200.00,5040.00)%
    \gplgaddtomacro\gplbacktext{%
      \csname LTb\endcsname%%
      \put(1376,1024){\makebox(0,0)[r]{\strut{}$10^{-1}$}}%
      \put(1376,1948){\makebox(0,0)[r]{\strut{}$10^{0}$}}%
      \put(1376,2872){\makebox(0,0)[r]{\strut{}$10^{1}$}}%
      \put(1376,3795){\makebox(0,0)[r]{\strut{}$10^{2}$}}%
      \put(1376,4719){\makebox(0,0)[r]{\strut{}$10^{3}$}}%
      \put(2028,704){\makebox(0,0){\strut{}$10^{6}$}}%
      \put(2947,704){\makebox(0,0){\strut{}$10^{8}$}}%
      \put(3866,704){\makebox(0,0){\strut{}$10^{10}$}}%
      \put(4785,704){\makebox(0,0){\strut{}$10^{12}$}}%
      \put(5704,704){\makebox(0,0){\strut{}$10^{14}$}}%
      \put(6623,704){\makebox(0,0){\strut{}$10^{16}$}}%
      \put(2180,1570){\makebox(0,0)[l]{\strut{}$Re\sim Ra^{1/2}$}}%
      \put(1787,2639){\makebox(0,0)[l]{\strut{}$Re\sim Ra^{2/5}$}}%
      \put(2166,4171){\makebox(0,0)[l]{\strut{}$Re\sim Ra^{2/5}$}}%
      \put(4142,1465){\makebox(0,0)[l]{\strut{}$Re\sim Ra^{2/5}$}}%
      \put(3866,3150){\makebox(0,0)[l]{\strut{}$Re\sim Ra^{1/3}$}}%
    }%
    \gplgaddtomacro\gplfronttext{%
      \csname LTb\endcsname%%
      \put(288,2871){\rotatebox{-270}{\makebox(0,0){\strut{}$\rm Re Ra^{-1/3}$}}}%
      \put(4095,224){\makebox(0,0){\strut{}$\rm Ra$}}%
    }%
    \gplbacktext
    \put(0,0){\includegraphics{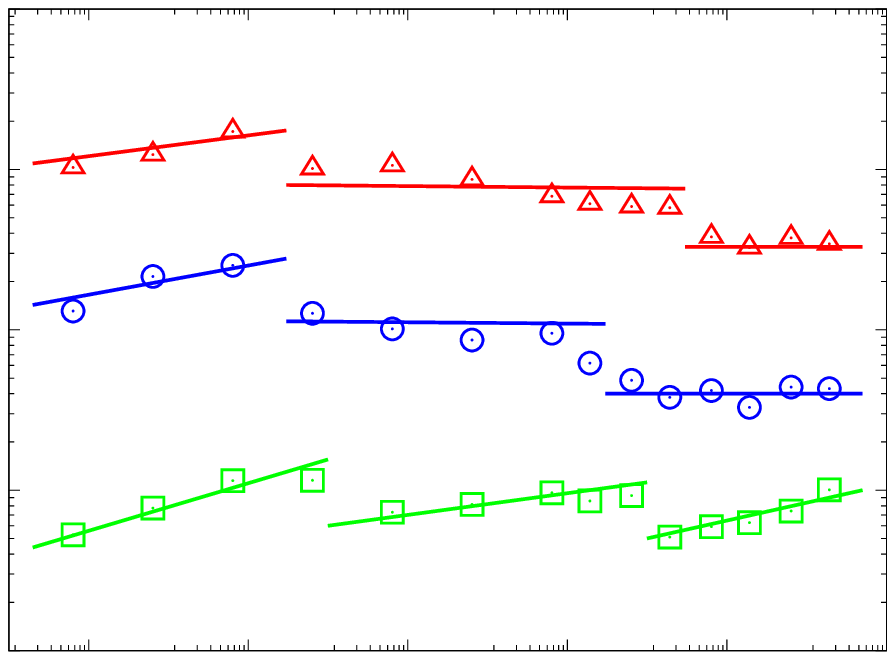}}%
    \gplfronttext
  \end{picture}%
\endgroup

%% file: RePr06_Pr_L.tex
% GNUPLOT: LaTeX picture with Postscript
\begingroup
  \fontfamily{Times-Roman}%
  \selectfont
  \makeatletter
  \providecommand\color[2][]{%
    \GenericError{(gnuplot) \space\space\space\@spaces}{%
      Package color not loaded in conjunction with
      terminal option `colourtext'%
    }{See the gnuplot documentation for explanation.%
    }{Either use 'blacktext' in gnuplot or load the package
      color.sty in LaTeX.}%
    \renewcommand\color[2][]{}%
  }%
  \providecommand\includegraphics[2][]{%
    \GenericError{(gnuplot) \space\space\space\@spaces}{%
      Package graphicx or graphics not loaded%
    }{See the gnuplot documentation for explanation.%
    }{The gnuplot epslatex terminal needs graphicx.sty or graphics.sty.}%
    \renewcommand\includegraphics[2][]{}%
  }%
  \providecommand\rotatebox[2]{#2}%
  \@ifundefined{ifGPcolor}{%
    \newif\ifGPcolor
    \GPcolortrue
  }{}%
  \@ifundefined{ifGPblacktext}{%
    \newif\ifGPblacktext
    \GPblacktexttrue
  }{}%
  % define a \g@addto@macro without @ in the name:
  \let\gplgaddtomacro\g@addto@macro
  % define empty templates for all commands taking text:
  \gdef\gplbacktext{}%
  \gdef\gplfronttext{}%
  \makeatother
  \ifGPblacktext
    % no textcolor at all
    \def\colorrgb#1{}%
    \def\colorgray#1{}%
  \else
    % gray or color?
    \ifGPcolor
      \def\colorrgb#1{\color[rgb]{#1}}%
      \def\colorgray#1{\color[gray]{#1}}%
      \expandafter\def\csname LTw\endcsname{\color{white}}%
      \expandafter\def\csname LTb\endcsname{\color{black}}%
      \expandafter\def\csname LTa\endcsname{\color{black}}%
      \expandafter\def\csname LT0\endcsname{\color[rgb]{1,0,0}}%
      \expandafter\def\csname LT1\endcsname{\color[rgb]{0,1,0}}%
      \expandafter\def\csname LT2\endcsname{\color[rgb]{0,0,1}}%
      \expandafter\def\csname LT3\endcsname{\color[rgb]{1,0,1}}%
      \expandafter\def\csname LT4\endcsname{\color[rgb]{0,1,1}}%
      \expandafter\def\csname LT5\endcsname{\color[rgb]{1,1,0}}%
      \expandafter\def\csname LT6\endcsname{\color[rgb]{0,0,0}}%
      \expandafter\def\csname LT7\endcsname{\color[rgb]{1,0.3,0}}%
      \expandafter\def\csname LT8\endcsname{\color[rgb]{0.5,0.5,0.5}}%
    \else
      % gray
      \def\colorrgb#1{\color{black}}%
      \def\colorgray#1{\color[gray]{#1}}%
      \expandafter\def\csname LTw\endcsname{\color{white}}%
      \expandafter\def\csname LTb\endcsname{\color{black}}%
      \expandafter\def\csname LTa\endcsname{\color{black}}%
      \expandafter\def\csname LT0\endcsname{\color{black}}%
      \expandafter\def\csname LT1\endcsname{\color{black}}%
      \expandafter\def\csname LT2\endcsname{\color{black}}%
      \expandafter\def\csname LT3\endcsname{\color{black}}%
      \expandafter\def\csname LT4\endcsname{\color{black}}%
      \expandafter\def\csname LT5\endcsname{\color{black}}%
      \expandafter\def\csname LT6\endcsname{\color{black}}%
      \expandafter\def\csname LT7\endcsname{\color{black}}%
      \expandafter\def\csname LT8\endcsname{\color{black}}%
    \fi
  \fi
    \setlength{\unitlength}{0.0500bp}%
    \ifx\gptboxheight\undefined%
      \newlength{\gptboxheight}%
      \newlength{\gptboxwidth}%
      \newsavebox{\gptboxtext}%
    \fi%
    \setlength{\fboxrule}{0.5pt}%
    \setlength{\fboxsep}{1pt}%
\begin{picture}(7200.00,5040.00)%
    \gplgaddtomacro\gplbacktext{%
      \csname LTb\endcsname%%
      \put(1184,1718){\makebox(0,0)[r]{\strut{}$10^{3}$}}%
      \put(1184,4025){\makebox(0,0)[r]{\strut{}$10^{4}$}}%
      \put(1376,704){\makebox(0,0){\strut{}$10^{-3}$}}%
      \put(2795,704){\makebox(0,0){\strut{}$10^{-2}$}}%
      \put(4213,704){\makebox(0,0){\strut{}$10^{-1}$}}%
      \put(5632,704){\makebox(0,0){\strut{}$10^{0}$}}%
      \put(5032,1395){\makebox(0,0)[l]{\strut{}$\rm Re\sim Pr^{-4/5}$}}%
      \put(4732,3444){\makebox(0,0)[l]{\strut{}$\rm Re\sim Pr^{-3/5}$}}%
      \put(2795,1966){\makebox(0,0)[l]{\strut{}$\rm Re\sim Pr^{-1}$}}%
    }%
    \gplgaddtomacro\gplfronttext{%
      \csname LTb\endcsname%%
      \put(288,2871){\rotatebox{-270}{\makebox(0,0){\strut{}$\rm Re Pr$}}}%
      \put(3999,224){\makebox(0,0){\strut{}$\rm Pr$}}%
    }%
    \gplbacktext
    \put(0,0){\includegraphics{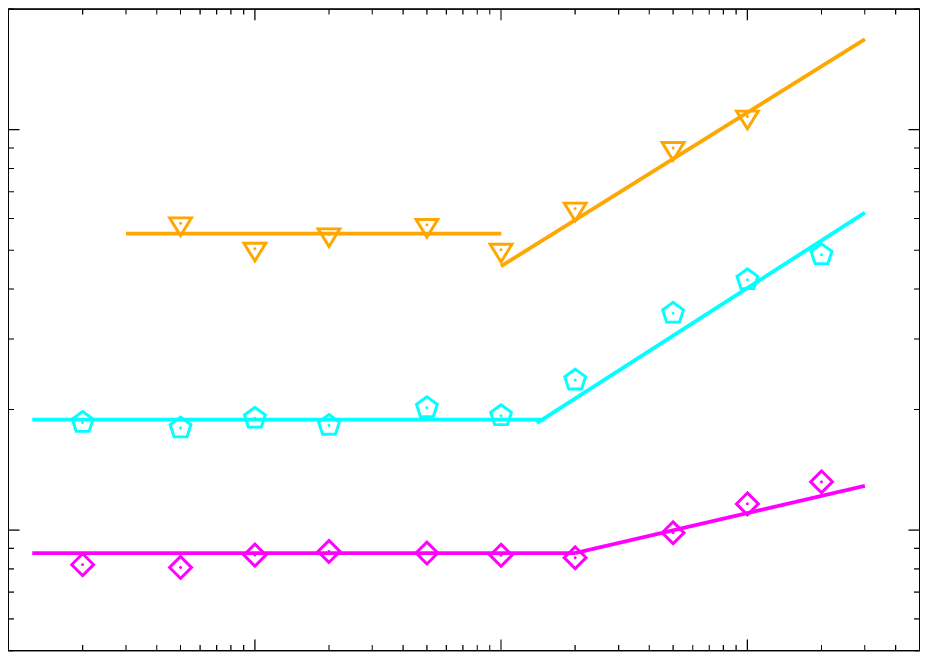}}%
    \gplfronttext
  \end{picture}%
\endgroup

%% file: NuRe05_Ra_L.tex
% GNUPLOT: LaTeX picture with Postscript
\begingroup
  \fontfamily{Times-Roman}%
  \selectfont
  \makeatletter
  \providecommand\color[2][]{%
    \GenericError{(gnuplot) \space\space\space\@spaces}{%
      Package color not loaded in conjunction with
      terminal option `colourtext'%
    }{See the gnuplot documentation for explanation.%
    }{Either use 'blacktext' in gnuplot or load the package
      color.sty in LaTeX.}%
    \renewcommand\color[2][]{}%
  }%
  \providecommand\includegraphics[2][]{%
    \GenericError{(gnuplot) \space\space\space\@spaces}{%
      Package graphicx or graphics not loaded%
    }{See the gnuplot documentation for explanation.%
    }{The gnuplot epslatex terminal needs graphicx.sty or graphics.sty.}%
    \renewcommand\includegraphics[2][]{}%
  }%
  \providecommand\rotatebox[2]{#2}%
  \@ifundefined{ifGPcolor}{%
    \newif\ifGPcolor
    \GPcolortrue
  }{}%
  \@ifundefined{ifGPblacktext}{%
    \newif\ifGPblacktext
    \GPblacktexttrue
  }{}%
  % define a \g@addto@macro without @ in the name:
  \let\gplgaddtomacro\g@addto@macro
  % define empty templates for all commands taking text:
  \gdef\gplbacktext{}%
  \gdef\gplfronttext{}%
  \makeatother
  \ifGPblacktext
    % no textcolor at all
    \def\colorrgb#1{}%
    \def\colorgray#1{}%
  \else
    % gray or color?
    \ifGPcolor
      \def\colorrgb#1{\color[rgb]{#1}}%
      \def\colorgray#1{\color[gray]{#1}}%
      \expandafter\def\csname LTw\endcsname{\color{white}}%
      \expandafter\def\csname LTb\endcsname{\color{black}}%
      \expandafter\def\csname LTa\endcsname{\color{black}}%
      \expandafter\def\csname LT0\endcsname{\color[rgb]{1,0,0}}%
      \expandafter\def\csname LT1\endcsname{\color[rgb]{0,1,0}}%
      \expandafter\def\csname LT2\endcsname{\color[rgb]{0,0,1}}%
      \expandafter\def\csname LT3\endcsname{\color[rgb]{1,0,1}}%
      \expandafter\def\csname LT4\endcsname{\color[rgb]{0,1,1}}%
      \expandafter\def\csname LT5\endcsname{\color[rgb]{1,1,0}}%
      \expandafter\def\csname LT6\endcsname{\color[rgb]{0,0,0}}%
      \expandafter\def\csname LT7\endcsname{\color[rgb]{1,0.3,0}}%
      \expandafter\def\csname LT8\endcsname{\color[rgb]{0.5,0.5,0.5}}%
    \else
      % gray
      \def\colorrgb#1{\color{black}}%
      \def\colorgray#1{\color[gray]{#1}}%
      \expandafter\def\csname LTw\endcsname{\color{white}}%
      \expandafter\def\csname LTb\endcsname{\color{black}}%
      \expandafter\def\csname LTa\endcsname{\color{black}}%
      \expandafter\def\csname LT0\endcsname{\color{black}}%
      \expandafter\def\csname LT1\endcsname{\color{black}}%
      \expandafter\def\csname LT2\endcsname{\color{black}}%
      \expandafter\def\csname LT3\endcsname{\color{black}}%
      \expandafter\def\csname LT4\endcsname{\color{black}}%
      \expandafter\def\csname LT5\endcsname{\color{black}}%
      \expandafter\def\csname LT6\endcsname{\color{black}}%
      \expandafter\def\csname LT7\endcsname{\color{black}}%
      \expandafter\def\csname LT8\endcsname{\color{black}}%
    \fi
  \fi
    \setlength{\unitlength}{0.0500bp}%
    \ifx\gptboxheight\undefined%
      \newlength{\gptboxheight}%
      \newlength{\gptboxwidth}%
      \newsavebox{\gptboxtext}%
    \fi%
    \setlength{\fboxrule}{0.5pt}%
    \setlength{\fboxsep}{1pt}%
\begin{picture}(7200.00,5040.00)%
    \gplgaddtomacro\gplbacktext{%
      \csname LTb\endcsname%%
      \put(1376,1457){\makebox(0,0)[r]{\strut{}$10^{-2}$}}%
      \put(1376,2544){\makebox(0,0)[r]{\strut{}$10^{-1}$}}%
      \put(1376,3632){\makebox(0,0)[r]{\strut{}$10^{0}$}}%
      \put(1376,4719){\makebox(0,0)[r]{\strut{}$10^{1}$}}%
      \put(2028,704){\makebox(0,0){\strut{}$10^{6}$}}%
      \put(2947,704){\makebox(0,0){\strut{}$10^{8}$}}%
      \put(3866,704){\makebox(0,0){\strut{}$10^{10}$}}%
      \put(4785,704){\makebox(0,0){\strut{}$10^{12}$}}%
      \put(5704,704){\makebox(0,0){\strut{}$10^{14}$}}%
      \put(6623,704){\makebox(0,0){\strut{}$10^{16}$}}%
      \put(4764,2589){\makebox(0,0)[l]{\strut{}$\rm  Nu\sim Re^{5/8}$}}%
    }%
    \gplgaddtomacro\gplfronttext{%
      \csname LTb\endcsname%%
      \put(288,2871){\rotatebox{-270}{\makebox(0,0){\strut{}$\rm NuRe^{-1/2}$}}}%
      \put(4095,224){\makebox(0,0){\strut{}$\rm Ra$}}%
    }%
    \gplbacktext
    \put(0,0){\includegraphics{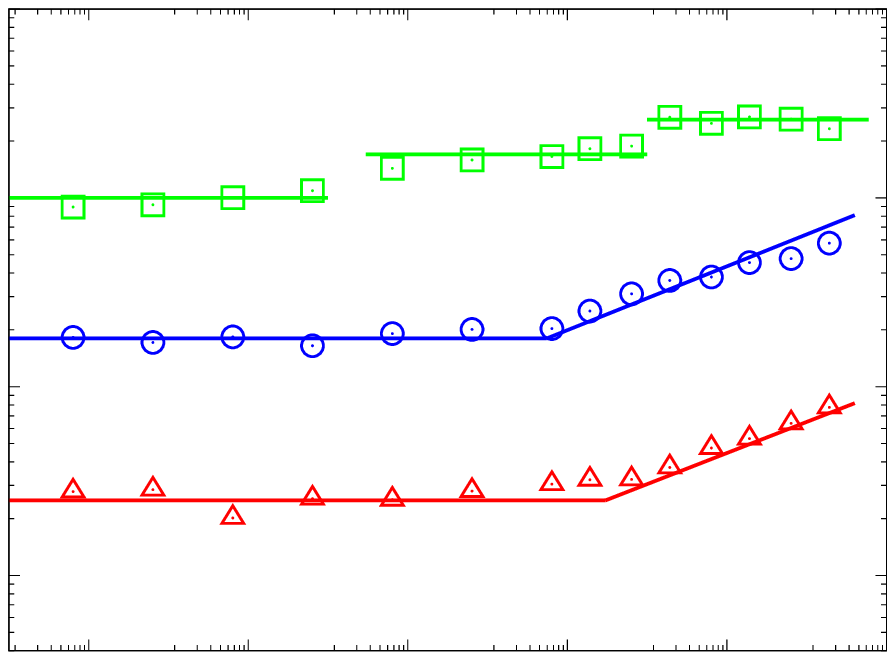}}%
    \gplfronttext
  \end{picture}%
\endgroup

%% file: NuRe05_Pr_L.tex
% GNUPLOT: LaTeX picture with Postscript
\begingroup
  \fontfamily{Times-Roman}%
  \selectfont
  \makeatletter
  \providecommand\color[2][]{%
    \GenericError{(gnuplot) \space\space\space\@spaces}{%
      Package color not loaded in conjunction with
      terminal option `colourtext'%
    }{See the gnuplot documentation for explanation.%
    }{Either use 'blacktext' in gnuplot or load the package
      color.sty in LaTeX.}%
    \renewcommand\color[2][]{}%
  }%
  \providecommand\includegraphics[2][]{%
    \GenericError{(gnuplot) \space\space\space\@spaces}{%
      Package graphicx or graphics not loaded%
    }{See the gnuplot documentation for explanation.%
    }{The gnuplot epslatex terminal needs graphicx.sty or graphics.sty.}%
    \renewcommand\includegraphics[2][]{}%
  }%
  \providecommand\rotatebox[2]{#2}%
  \@ifundefined{ifGPcolor}{%
    \newif\ifGPcolor
    \GPcolortrue
  }{}%
  \@ifundefined{ifGPblacktext}{%
    \newif\ifGPblacktext
    \GPblacktexttrue
  }{}%
  % define a \g@addto@macro without @ in the name:
  \let\gplgaddtomacro\g@addto@macro
  % define empty templates for all commands taking text:
  \gdef\gplbacktext{}%
  \gdef\gplfronttext{}%
  \makeatother
  \ifGPblacktext
    % no textcolor at all
    \def\colorrgb#1{}%
    \def\colorgray#1{}%
  \else
    % gray or color?
    \ifGPcolor
      \def\colorrgb#1{\color[rgb]{#1}}%
      \def\colorgray#1{\color[gray]{#1}}%
      \expandafter\def\csname LTw\endcsname{\color{white}}%
      \expandafter\def\csname LTb\endcsname{\color{black}}%
      \expandafter\def\csname LTa\endcsname{\color{black}}%
      \expandafter\def\csname LT0\endcsname{\color[rgb]{1,0,0}}%
      \expandafter\def\csname LT1\endcsname{\color[rgb]{0,1,0}}%
      \expandafter\def\csname LT2\endcsname{\color[rgb]{0,0,1}}%
      \expandafter\def\csname LT3\endcsname{\color[rgb]{1,0,1}}%
      \expandafter\def\csname LT4\endcsname{\color[rgb]{0,1,1}}%
      \expandafter\def\csname LT5\endcsname{\color[rgb]{1,1,0}}%
      \expandafter\def\csname LT6\endcsname{\color[rgb]{0,0,0}}%
      \expandafter\def\csname LT7\endcsname{\color[rgb]{1,0.3,0}}%
      \expandafter\def\csname LT8\endcsname{\color[rgb]{0.5,0.5,0.5}}%
    \else
      % gray
      \def\colorrgb#1{\color{black}}%
      \def\colorgray#1{\color[gray]{#1}}%
      \expandafter\def\csname LTw\endcsname{\color{white}}%
      \expandafter\def\csname LTb\endcsname{\color{black}}%
      \expandafter\def\csname LTa\endcsname{\color{black}}%
      \expandafter\def\csname LT0\endcsname{\color{black}}%
      \expandafter\def\csname LT1\endcsname{\color{black}}%
      \expandafter\def\csname LT2\endcsname{\color{black}}%
      \expandafter\def\csname LT3\endcsname{\color{black}}%
      \expandafter\def\csname LT4\endcsname{\color{black}}%
      \expandafter\def\csname LT5\endcsname{\color{black}}%
      \expandafter\def\csname LT6\endcsname{\color{black}}%
      \expandafter\def\csname LT7\endcsname{\color{black}}%
      \expandafter\def\csname LT8\endcsname{\color{black}}%
    \fi
  \fi
    \setlength{\unitlength}{0.0500bp}%
    \ifx\gptboxheight\undefined%
      \newlength{\gptboxheight}%
      \newlength{\gptboxwidth}%
      \newsavebox{\gptboxtext}%
    \fi%
    \setlength{\fboxrule}{0.5pt}%
    \setlength{\fboxsep}{1pt}%
\begin{picture}(7200.00,5040.00)%
    \gplgaddtomacro\gplbacktext{%
      \csname LTb\endcsname%%
      \put(1376,1457){\makebox(0,0)[r]{\strut{}$10^{-2}$}}%
      \put(1376,2544){\makebox(0,0)[r]{\strut{}$10^{-1}$}}%
      \put(1376,3632){\makebox(0,0)[r]{\strut{}$10^{0}$}}%
      \put(1376,4719){\makebox(0,0)[r]{\strut{}$10^{1}$}}%
      \put(1568,704){\makebox(0,0){\strut{}$10^{-3}$}}%
      \put(2935,704){\makebox(0,0){\strut{}$10^{-2}$}}%
      \put(4301,704){\makebox(0,0){\strut{}$10^{-1}$}}%
      \put(5668,704){\makebox(0,0){\strut{}$10^{0}$}}%
      \put(5427,3113){\makebox(0,0)[l]{\strut{}$\sim \rm Pr^{1/2}$}}%
      \put(2935,3063){\makebox(0,0)[l]{\strut{}$\sim \rm Pr^{3/4}$}}%
    }%
    \gplgaddtomacro\gplfronttext{%
      \csname LTb\endcsname%%
      \put(288,2871){\rotatebox{-270}{\makebox(0,0){\strut{}$\rm NuRe^{-1/2}$}}}%
      \put(4095,224){\makebox(0,0){\strut{}$\rm Pr$}}%
    }%
    \gplbacktext
    \put(0,0){\includegraphics{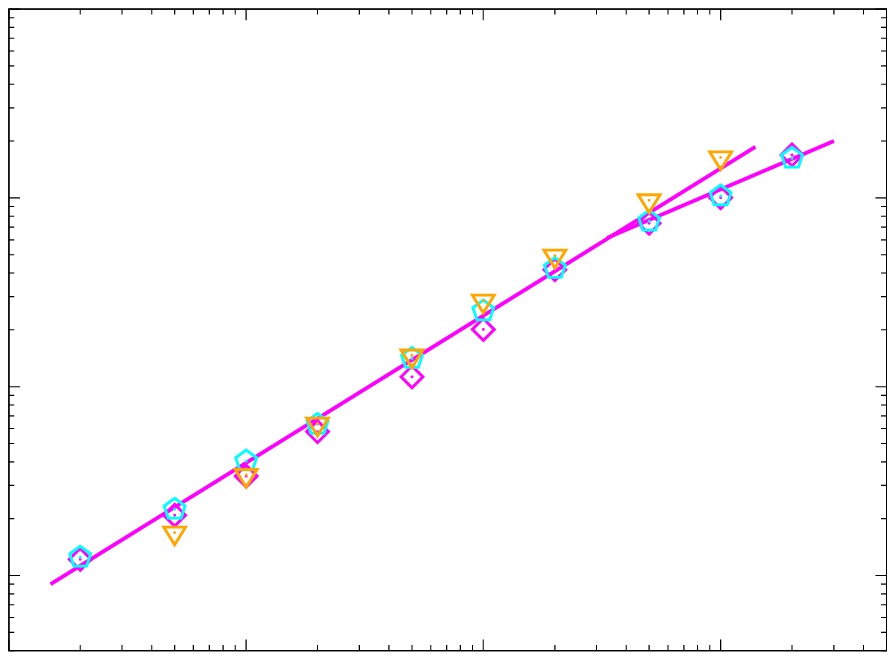}}%
    \gplfronttext
  \end{picture}%
\endgroup

%% file: L4nu-3epsilonRa-1_Ra_L.tex
% GNUPLOT: LaTeX picture with Postscript
\begingroup
  \fontfamily{Times-Roman}%
  \selectfont
  \makeatletter
  \providecommand\color[2][]{%
    \GenericError{(gnuplot) \space\space\space\@spaces}{%
      Package color not loaded in conjunction with
      terminal option `colourtext'%
    }{See the gnuplot documentation for explanation.%
    }{Either use 'blacktext' in gnuplot or load the package
      color.sty in LaTeX.}%
    \renewcommand\color[2][]{}%
  }%
  \providecommand\includegraphics[2][]{%
    \GenericError{(gnuplot) \space\space\space\@spaces}{%
      Package graphicx or graphics not loaded%
    }{See the gnuplot documentation for explanation.%
    }{The gnuplot epslatex terminal needs graphicx.sty or graphics.sty.}%
    \renewcommand\includegraphics[2][]{}%
  }%
  \providecommand\rotatebox[2]{#2}%
  \@ifundefined{ifGPcolor}{%
    \newif\ifGPcolor
    \GPcolortrue
  }{}%
  \@ifundefined{ifGPblacktext}{%
    \newif\ifGPblacktext
    \GPblacktexttrue
  }{}%
  % define a \g@addto@macro without @ in the name:
  \let\gplgaddtomacro\g@addto@macro
  % define empty templates for all commands taking text:
  \gdef\gplbacktext{}%
  \gdef\gplfronttext{}%
  \makeatother
  \ifGPblacktext
    % no textcolor at all
    \def\colorrgb#1{}%
    \def\colorgray#1{}%
  \else
    % gray or color?
    \ifGPcolor
      \def\colorrgb#1{\color[rgb]{#1}}%
      \def\colorgray#1{\color[gray]{#1}}%
      \expandafter\def\csname LTw\endcsname{\color{white}}%
      \expandafter\def\csname LTb\endcsname{\color{black}}%
      \expandafter\def\csname LTa\endcsname{\color{black}}%
      \expandafter\def\csname LT0\endcsname{\color[rgb]{1,0,0}}%
      \expandafter\def\csname LT1\endcsname{\color[rgb]{0,1,0}}%
      \expandafter\def\csname LT2\endcsname{\color[rgb]{0,0,1}}%
      \expandafter\def\csname LT3\endcsname{\color[rgb]{1,0,1}}%
      \expandafter\def\csname LT4\endcsname{\color[rgb]{0,1,1}}%
      \expandafter\def\csname LT5\endcsname{\color[rgb]{1,1,0}}%
      \expandafter\def\csname LT6\endcsname{\color[rgb]{0,0,0}}%
      \expandafter\def\csname LT7\endcsname{\color[rgb]{1,0.3,0}}%
      \expandafter\def\csname LT8\endcsname{\color[rgb]{0.5,0.5,0.5}}%
    \else
      % gray
      \def\colorrgb#1{\color{black}}%
      \def\colorgray#1{\color[gray]{#1}}%
      \expandafter\def\csname LTw\endcsname{\color{white}}%
      \expandafter\def\csname LTb\endcsname{\color{black}}%
      \expandafter\def\csname LTa\endcsname{\color{black}}%
      \expandafter\def\csname LT0\endcsname{\color{black}}%
      \expandafter\def\csname LT1\endcsname{\color{black}}%
      \expandafter\def\csname LT2\endcsname{\color{black}}%
      \expandafter\def\csname LT3\endcsname{\color{black}}%
      \expandafter\def\csname LT4\endcsname{\color{black}}%
      \expandafter\def\csname LT5\endcsname{\color{black}}%
      \expandafter\def\csname LT6\endcsname{\color{black}}%
      \expandafter\def\csname LT7\endcsname{\color{black}}%
      \expandafter\def\csname LT8\endcsname{\color{black}}%
    \fi
  \fi
    \setlength{\unitlength}{0.0500bp}%
    \ifx\gptboxheight\undefined%
      \newlength{\gptboxheight}%
      \newlength{\gptboxwidth}%
      \newsavebox{\gptboxtext}%
    \fi%
    \setlength{\fboxrule}{0.5pt}%
    \setlength{\fboxsep}{1pt}%
\begin{picture}(7200.00,5040.00)%
    \gplgaddtomacro\gplbacktext{%
      \csname LTb\endcsname%%
      \put(1376,1024){\makebox(0,0)[r]{\strut{}$10^{-1}$}}%
      \put(1376,1552){\makebox(0,0)[r]{\strut{}$10^{0}$}}%
      \put(1376,2080){\makebox(0,0)[r]{\strut{}$10^{1}$}}%
      \put(1376,2608){\makebox(0,0)[r]{\strut{}$10^{2}$}}%
      \put(1376,3135){\makebox(0,0)[r]{\strut{}$10^{3}$}}%
      \put(1376,3663){\makebox(0,0)[r]{\strut{}$10^{4}$}}%
      \put(1376,4191){\makebox(0,0)[r]{\strut{}$10^{5}$}}%
      \put(1376,4719){\makebox(0,0)[r]{\strut{}$10^{6}$}}%
      \put(2028,704){\makebox(0,0){\strut{}$10^{6}$}}%
      \put(2947,704){\makebox(0,0){\strut{}$10^{8}$}}%
      \put(3866,704){\makebox(0,0){\strut{}$10^{10}$}}%
      \put(4785,704){\makebox(0,0){\strut{}$10^{12}$}}%
      \put(5704,704){\makebox(0,0){\strut{}$10^{14}$}}%
      \put(6623,704){\makebox(0,0){\strut{}$10^{16}$}}%
    }%
    \gplgaddtomacro\gplfronttext{%
      \csname LTb\endcsname%%
      \put(288,2871){\rotatebox{-270}{\makebox(0,0){\strut{}$ L^4\nu^{-3}\bar{\epsilon_u}{\rm Ra}^{-1}$}}}%
      \put(4095,224){\makebox(0,0){\strut{}$\rm Ra$}}%
    }%
    \gplbacktext
    \put(0,0){\includegraphics{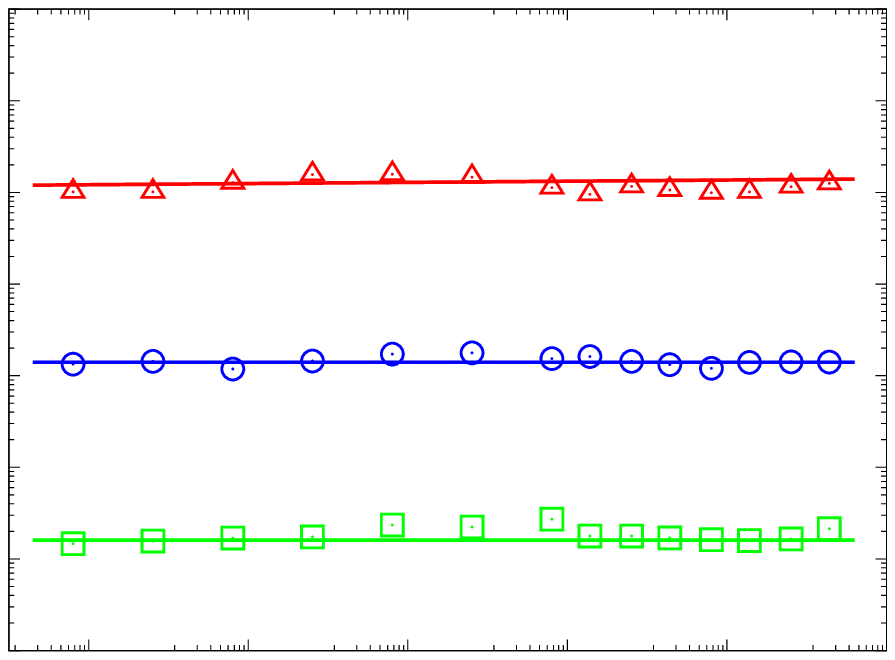}}%
    \gplfronttext
  \end{picture}%
\endgroup

%% file: L4nu-3epsilonRa-1_Pr_L.tex
% GNUPLOT: LaTeX picture with Postscript
\begingroup
  \fontfamily{Times-Roman}%
  \selectfont
  \makeatletter
  \providecommand\color[2][]{%
    \GenericError{(gnuplot) \space\space\space\@spaces}{%
      Package color not loaded in conjunction with
      terminal option `colourtext'%
    }{See the gnuplot documentation for explanation.%
    }{Either use 'blacktext' in gnuplot or load the package
      color.sty in LaTeX.}%
    \renewcommand\color[2][]{}%
  }%
  \providecommand\includegraphics[2][]{%
    \GenericError{(gnuplot) \space\space\space\@spaces}{%
      Package graphicx or graphics not loaded%
    }{See the gnuplot documentation for explanation.%
    }{The gnuplot epslatex terminal needs graphicx.sty or graphics.sty.}%
    \renewcommand\includegraphics[2][]{}%
  }%
  \providecommand\rotatebox[2]{#2}%
  \@ifundefined{ifGPcolor}{%
    \newif\ifGPcolor
    \GPcolortrue
  }{}%
  \@ifundefined{ifGPblacktext}{%
    \newif\ifGPblacktext
    \GPblacktexttrue
  }{}%
  % define a \g@addto@macro without @ in the name:
  \let\gplgaddtomacro\g@addto@macro
  % define empty templates for all commands taking text:
  \gdef\gplbacktext{}%
  \gdef\gplfronttext{}%
  \makeatother
  \ifGPblacktext
    % no textcolor at all
    \def\colorrgb#1{}%
    \def\colorgray#1{}%
  \else
    % gray or color?
    \ifGPcolor
      \def\colorrgb#1{\color[rgb]{#1}}%
      \def\colorgray#1{\color[gray]{#1}}%
      \expandafter\def\csname LTw\endcsname{\color{white}}%
      \expandafter\def\csname LTb\endcsname{\color{black}}%
      \expandafter\def\csname LTa\endcsname{\color{black}}%
      \expandafter\def\csname LT0\endcsname{\color[rgb]{1,0,0}}%
      \expandafter\def\csname LT1\endcsname{\color[rgb]{0,1,0}}%
      \expandafter\def\csname LT2\endcsname{\color[rgb]{0,0,1}}%
      \expandafter\def\csname LT3\endcsname{\color[rgb]{1,0,1}}%
      \expandafter\def\csname LT4\endcsname{\color[rgb]{0,1,1}}%
      \expandafter\def\csname LT5\endcsname{\color[rgb]{1,1,0}}%
      \expandafter\def\csname LT6\endcsname{\color[rgb]{0,0,0}}%
      \expandafter\def\csname LT7\endcsname{\color[rgb]{1,0.3,0}}%
      \expandafter\def\csname LT8\endcsname{\color[rgb]{0.5,0.5,0.5}}%
    \else
      % gray
      \def\colorrgb#1{\color{black}}%
      \def\colorgray#1{\color[gray]{#1}}%
      \expandafter\def\csname LTw\endcsname{\color{white}}%
      \expandafter\def\csname LTb\endcsname{\color{black}}%
      \expandafter\def\csname LTa\endcsname{\color{black}}%
      \expandafter\def\csname LT0\endcsname{\color{black}}%
      \expandafter\def\csname LT1\endcsname{\color{black}}%
      \expandafter\def\csname LT2\endcsname{\color{black}}%
      \expandafter\def\csname LT3\endcsname{\color{black}}%
      \expandafter\def\csname LT4\endcsname{\color{black}}%
      \expandafter\def\csname LT5\endcsname{\color{black}}%
      \expandafter\def\csname LT6\endcsname{\color{black}}%
      \expandafter\def\csname LT7\endcsname{\color{black}}%
      \expandafter\def\csname LT8\endcsname{\color{black}}%
    \fi
  \fi
    \setlength{\unitlength}{0.0500bp}%
    \ifx\gptboxheight\undefined%
      \newlength{\gptboxheight}%
      \newlength{\gptboxwidth}%
      \newsavebox{\gptboxtext}%
    \fi%
    \setlength{\fboxrule}{0.5pt}%
    \setlength{\fboxsep}{1pt}%
\begin{picture}(7200.00,5040.00)%
    \gplgaddtomacro\gplbacktext{%
      \csname LTb\endcsname%%
      \put(1376,1024){\makebox(0,0)[r]{\strut{}$10^{-2}$}}%
      \put(1376,1435){\makebox(0,0)[r]{\strut{}$10^{-1}$}}%
      \put(1376,1845){\makebox(0,0)[r]{\strut{}$10^{0}$}}%
      \put(1376,2256){\makebox(0,0)[r]{\strut{}$10^{1}$}}%
      \put(1376,2666){\makebox(0,0)[r]{\strut{}$10^{2}$}}%
      \put(1376,3077){\makebox(0,0)[r]{\strut{}$10^{3}$}}%
      \put(1376,3487){\makebox(0,0)[r]{\strut{}$10^{4}$}}%
      \put(1376,3898){\makebox(0,0)[r]{\strut{}$10^{5}$}}%
      \put(1376,4308){\makebox(0,0)[r]{\strut{}$10^{6}$}}%
      \put(1376,4719){\makebox(0,0)[r]{\strut{}$10^{7}$}}%
      \put(1568,704){\makebox(0,0){\strut{}$10^{-3}$}}%
      \put(2971,704){\makebox(0,0){\strut{}$10^{-2}$}}%
      \put(4375,704){\makebox(0,0){\strut{}$10^{-1}$}}%
      \put(5778,704){\makebox(0,0){\strut{}$10^{0}$}}%
      \put(4375,3324){\makebox(0,0)[l]{\strut{}$\sim \rm Pr^{-2}$}}%
    }%
    \gplgaddtomacro\gplfronttext{%
      \csname LTb\endcsname%%
      \put(288,2871){\rotatebox{-270}{\makebox(0,0){\strut{}$L^4\nu^{-3}\bar{\epsilon_u}{\rm Ra}^{-1}$}}}%
      \put(4095,224){\makebox(0,0){\strut{}$\rm Pr$}}%
    }%
    \gplbacktext
    \put(0,0){\includegraphics{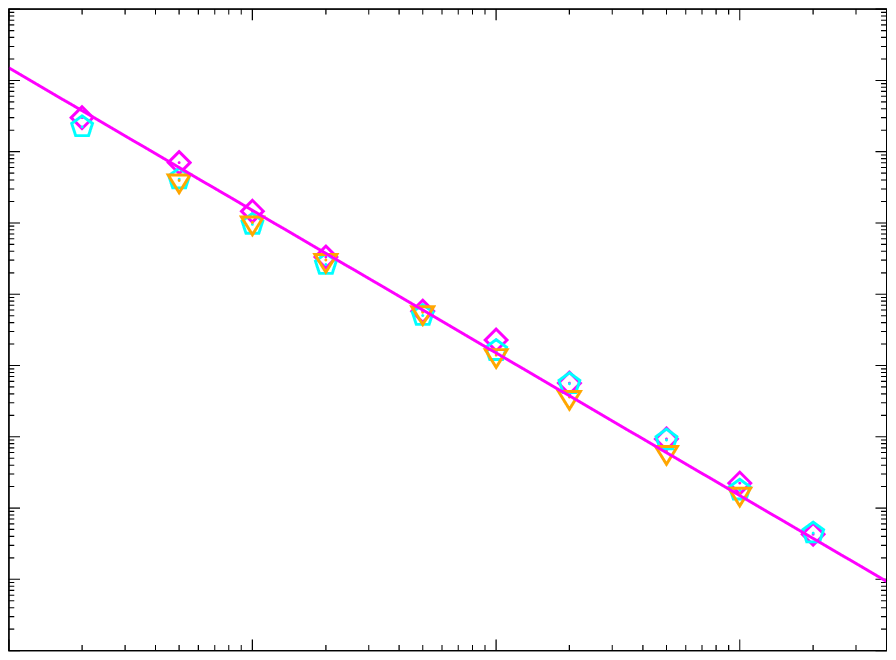}}%
    \gplfronttext
  \end{picture}%
\endgroup

%% file: turb_BL.tex
% GNUPLOT: LaTeX picture with Postscript
\begingroup
  \fontfamily{Times-Roman}%
  \selectfont
  \makeatletter
  \providecommand\color[2][]{%
    \GenericError{(gnuplot) \space\space\space\@spaces}{%
      Package color not loaded in conjunction with
      terminal option `colourtext'%
    }{See the gnuplot documentation for explanation.%
    }{Either use 'blacktext' in gnuplot or load the package
      color.sty in LaTeX.}%
    \renewcommand\color[2][]{}%
  }%
  \providecommand\includegraphics[2][]{%
    \GenericError{(gnuplot) \space\space\space\@spaces}{%
      Package graphicx or graphics not loaded%
    }{See the gnuplot documentation for explanation.%
    }{The gnuplot epslatex terminal needs graphicx.sty or graphics.sty.}%
    \renewcommand\includegraphics[2][]{}%
  }%
  \providecommand\rotatebox[2]{#2}%
  \@ifundefined{ifGPcolor}{%
    \newif\ifGPcolor
    \GPcolortrue
  }{}%
  \@ifundefined{ifGPblacktext}{%
    \newif\ifGPblacktext
    \GPblacktexttrue
  }{}%
  % define a \g@addto@macro without @ in the name:
  \let\gplgaddtomacro\g@addto@macro
  % define empty templates for all commands taking text:
  \gdef\gplbacktext{}%
  \gdef\gplfronttext{}%
  \makeatother
  \ifGPblacktext
    % no textcolor at all
    \def\colorrgb#1{}%
    \def\colorgray#1{}%
  \else
    % gray or color?
    \ifGPcolor
      \def\colorrgb#1{\color[rgb]{#1}}%
      \def\colorgray#1{\color[gray]{#1}}%
      \expandafter\def\csname LTw\endcsname{\color{white}}%
      \expandafter\def\csname LTb\endcsname{\color{black}}%
      \expandafter\def\csname LTa\endcsname{\color{black}}%
      \expandafter\def\csname LT0\endcsname{\color[rgb]{1,0,0}}%
      \expandafter\def\csname LT1\endcsname{\color[rgb]{0,1,0}}%
      \expandafter\def\csname LT2\endcsname{\color[rgb]{0,0,1}}%
      \expandafter\def\csname LT3\endcsname{\color[rgb]{1,0,1}}%
      \expandafter\def\csname LT4\endcsname{\color[rgb]{0,1,1}}%
      \expandafter\def\csname LT5\endcsname{\color[rgb]{1,1,0}}%
      \expandafter\def\csname LT6\endcsname{\color[rgb]{0,0,0}}%
      \expandafter\def\csname LT7\endcsname{\color[rgb]{1,0.3,0}}%
      \expandafter\def\csname LT8\endcsname{\color[rgb]{0.5,0.5,0.5}}%
    \else
      % gray
      \def\colorrgb#1{\color{black}}%
      \def\colorgray#1{\color[gray]{#1}}%
      \expandafter\def\csname LTw\endcsname{\color{white}}%
      \expandafter\def\csname LTb\endcsname{\color{black}}%
      \expandafter\def\csname LTa\endcsname{\color{black}}%
      \expandafter\def\csname LT0\endcsname{\color{black}}%
      \expandafter\def\csname LT1\endcsname{\color{black}}%
      \expandafter\def\csname LT2\endcsname{\color{black}}%
      \expandafter\def\csname LT3\endcsname{\color{black}}%
      \expandafter\def\csname LT4\endcsname{\color{black}}%
      \expandafter\def\csname LT5\endcsname{\color{black}}%
      \expandafter\def\csname LT6\endcsname{\color{black}}%
      \expandafter\def\csname LT7\endcsname{\color{black}}%
      \expandafter\def\csname LT8\endcsname{\color{black}}%
    \fi
  \fi
    \setlength{\unitlength}{0.0500bp}%
    \ifx\gptboxheight\undefined%
      \newlength{\gptboxheight}%
      \newlength{\gptboxwidth}%
      \newsavebox{\gptboxtext}%
    \fi%
    \setlength{\fboxrule}{0.5pt}%
    \setlength{\fboxsep}{1pt}%
\begin{picture}(7200.00,5040.00)%
    \gplgaddtomacro\gplbacktext{%
      \csname LTb\endcsname%%
      \put(888,768){\makebox(0,0)[r]{\strut{}$0$}}%
      \put(888,1574){\makebox(0,0)[r]{\strut{}$0.2$}}%
      \put(888,2380){\makebox(0,0)[r]{\strut{}$0.4$}}%
      \put(888,3187){\makebox(0,0)[r]{\strut{}$0.6$}}%
      \put(888,3993){\makebox(0,0)[r]{\strut{}$0.8$}}%
      \put(888,4799){\makebox(0,0)[r]{\strut{}$1$}}%
      \put(2797,528){\makebox(0,0){\strut{}$10^{-2}$}}%
      \put(6173,528){\makebox(0,0){\strut{}$10^{-1}$}}%
    }%
    \gplgaddtomacro\gplfronttext{%
      \csname LTb\endcsname%%
      \put(216,2783){\rotatebox{-270}{\makebox(0,0){\strut{}$(\bar{u}(z)-u_0)/u_{max}$}}}%
      \put(3899,168){\makebox(0,0){\strut{}$z/z(u_{max})$}}%
      \csname LTb\endcsname%%
      \put(4200,4616){\makebox(0,0)[r]{\strut{}$\rm Ra=1.92,10^{14}, Pr=10^{-1}$}}%
      \csname LTb\endcsname%%
      \put(4200,4376){\makebox(0,0)[r]{\strut{}$\rm Ra=1.92,10^{15}, Pr=10^{-2}$}}%
      \csname LTb\endcsname%%
      \put(4200,4136){\makebox(0,0)[r]{\strut{}$\rm Ra=1.92,10^{15}, Pr=10^{-1}$}}%
      \csname LTb\endcsname%%
      \put(4200,3896){\makebox(0,0)[r]{\strut{}$\log(z/z(u_{max})/3+1.325$}}%
    }%
    \gplbacktext
    \put(0,0){\includegraphics{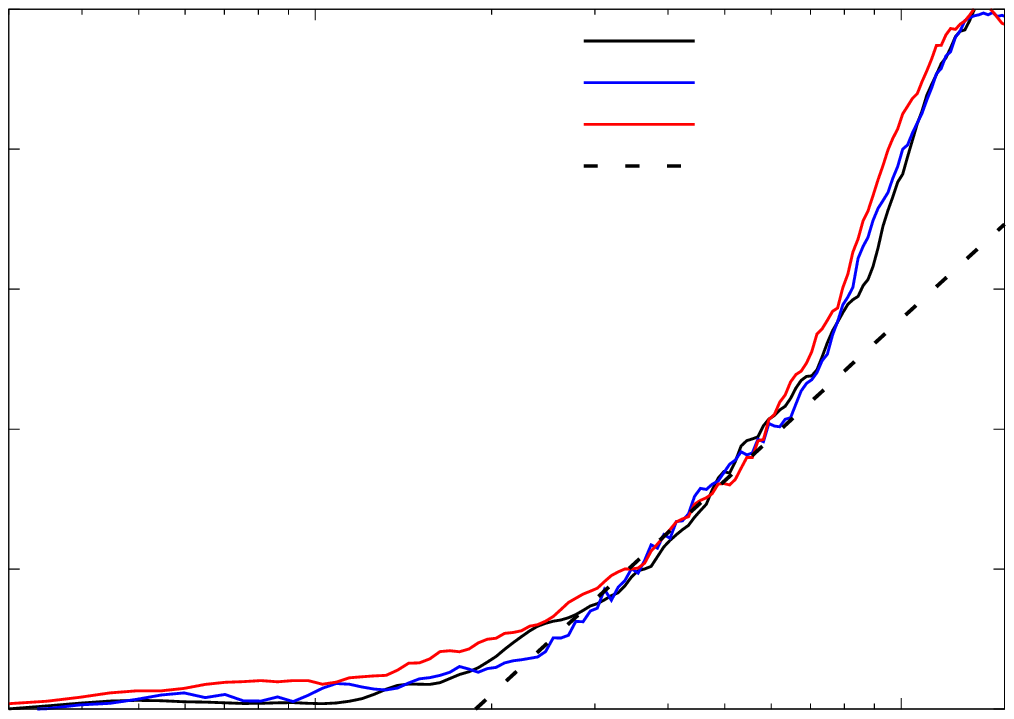}}%
    \gplfronttext
  \end{picture}%
\endgroup

%% file: turb_BL_b.tex
% GNUPLOT: LaTeX picture with Postscript
\begingroup
  \fontfamily{Times-Roman}%
  \selectfont
  \makeatletter
  \providecommand\color[2][]{%
    \GenericError{(gnuplot) \space\space\space\@spaces}{%
      Package color not loaded in conjunction with
      terminal option `colourtext'%
    }{See the gnuplot documentation for explanation.%
    }{Either use 'blacktext' in gnuplot or load the package
      color.sty in LaTeX.}%
    \renewcommand\color[2][]{}%
  }%
  \providecommand\includegraphics[2][]{%
    \GenericError{(gnuplot) \space\space\space\@spaces}{%
      Package graphicx or graphics not loaded%
    }{See the gnuplot documentation for explanation.%
    }{The gnuplot epslatex terminal needs graphicx.sty or graphics.sty.}%
    \renewcommand\includegraphics[2][]{}%
  }%
  \providecommand\rotatebox[2]{#2}%
  \@ifundefined{ifGPcolor}{%
    \newif\ifGPcolor
    \GPcolortrue
  }{}%
  \@ifundefined{ifGPblacktext}{%
    \newif\ifGPblacktext
    \GPblacktexttrue
  }{}%
  % define a \g@addto@macro without @ in the name:
  \let\gplgaddtomacro\g@addto@macro
  % define empty templates for all commands taking text:
  \gdef\gplbacktext{}%
  \gdef\gplfronttext{}%
  \makeatother
  \ifGPblacktext
    % no textcolor at all
    \def\colorrgb#1{}%
    \def\colorgray#1{}%
  \else
    % gray or color?
    \ifGPcolor
      \def\colorrgb#1{\color[rgb]{#1}}%
      \def\colorgray#1{\color[gray]{#1}}%
      \expandafter\def\csname LTw\endcsname{\color{white}}%
      \expandafter\def\csname LTb\endcsname{\color{black}}%
      \expandafter\def\csname LTa\endcsname{\color{black}}%
      \expandafter\def\csname LT0\endcsname{\color[rgb]{1,0,0}}%
      \expandafter\def\csname LT1\endcsname{\color[rgb]{0,1,0}}%
      \expandafter\def\csname LT2\endcsname{\color[rgb]{0,0,1}}%
      \expandafter\def\csname LT3\endcsname{\color[rgb]{1,0,1}}%
      \expandafter\def\csname LT4\endcsname{\color[rgb]{0,1,1}}%
      \expandafter\def\csname LT5\endcsname{\color[rgb]{1,1,0}}%
      \expandafter\def\csname LT6\endcsname{\color[rgb]{0,0,0}}%
      \expandafter\def\csname LT7\endcsname{\color[rgb]{1,0.3,0}}%
      \expandafter\def\csname LT8\endcsname{\color[rgb]{0.5,0.5,0.5}}%
    \else
      % gray
      \def\colorrgb#1{\color{black}}%
      \def\colorgray#1{\color[gray]{#1}}%
      \expandafter\def\csname LTw\endcsname{\color{white}}%
      \expandafter\def\csname LTb\endcsname{\color{black}}%
      \expandafter\def\csname LTa\endcsname{\color{black}}%
      \expandafter\def\csname LT0\endcsname{\color{black}}%
      \expandafter\def\csname LT1\endcsname{\color{black}}%
      \expandafter\def\csname LT2\endcsname{\color{black}}%
      \expandafter\def\csname LT3\endcsname{\color{black}}%
      \expandafter\def\csname LT4\endcsname{\color{black}}%
      \expandafter\def\csname LT5\endcsname{\color{black}}%
      \expandafter\def\csname LT6\endcsname{\color{black}}%
      \expandafter\def\csname LT7\endcsname{\color{black}}%
      \expandafter\def\csname LT8\endcsname{\color{black}}%
    \fi
  \fi
    \setlength{\unitlength}{0.0500bp}%
    \ifx\gptboxheight\undefined%
      \newlength{\gptboxheight}%
      \newlength{\gptboxwidth}%
      \newsavebox{\gptboxtext}%
    \fi%
    \setlength{\fboxrule}{0.5pt}%
    \setlength{\fboxsep}{1pt}%
\begin{picture}(7200.00,5040.00)%
    \gplgaddtomacro\gplbacktext{%
      \csname LTb\endcsname%%
      \put(888,768){\makebox(0,0)[r]{\strut{}$0$}}%
      \put(888,1440){\makebox(0,0)[r]{\strut{}$0.2$}}%
      \put(888,2112){\makebox(0,0)[r]{\strut{}$0.4$}}%
      \put(888,2784){\makebox(0,0)[r]{\strut{}$0.6$}}%
      \put(888,3455){\makebox(0,0)[r]{\strut{}$0.8$}}%
      \put(888,4127){\makebox(0,0)[r]{\strut{}$1$}}%
      \put(888,4799){\makebox(0,0)[r]{\strut{}$1.2$}}%
      \put(2221,528){\makebox(0,0){\strut{}$10^{-1}$}}%
      \put(4494,528){\makebox(0,0){\strut{}$10^{0}$}}%
      \put(6767,528){\makebox(0,0){\strut{}$10^{1}$}}%
    }%
    \gplgaddtomacro\gplfronttext{%
      \csname LTb\endcsname%%
      \put(216,2783){\rotatebox{-270}{\makebox(0,0){\strut{}$(b(z=H)-\bar{b}(z))/(b(z=H){\rm Nu}$}}}%
      \put(3899,168){\makebox(0,0){\strut{}$z\Gamma{\rm Nu}$}}%
      \csname LTb\endcsname%%
      \put(5696,1671){\makebox(0,0)[r]{\strut{}$\rm Ra=1.92,10^{14}, Pr=10^{-1}$}}%
      \csname LTb\endcsname%%
      \put(5696,1431){\makebox(0,0)[r]{\strut{}$\rm Ra=1.92,10^{15}, Pr=10^{-2}$}}%
      \csname LTb\endcsname%%
      \put(5696,1191){\makebox(0,0)[r]{\strut{}$\rm Ra=1.92,10^{15}, Pr=10^{-1}$}}%
      \csname LTb\endcsname%%
      \put(5696,951){\makebox(0,0)[r]{\strut{}$\log(z\Gamma{\rm Nu})/6.05+0.78$}}%
    }%
    \gplbacktext
    \put(0,0){\includegraphics{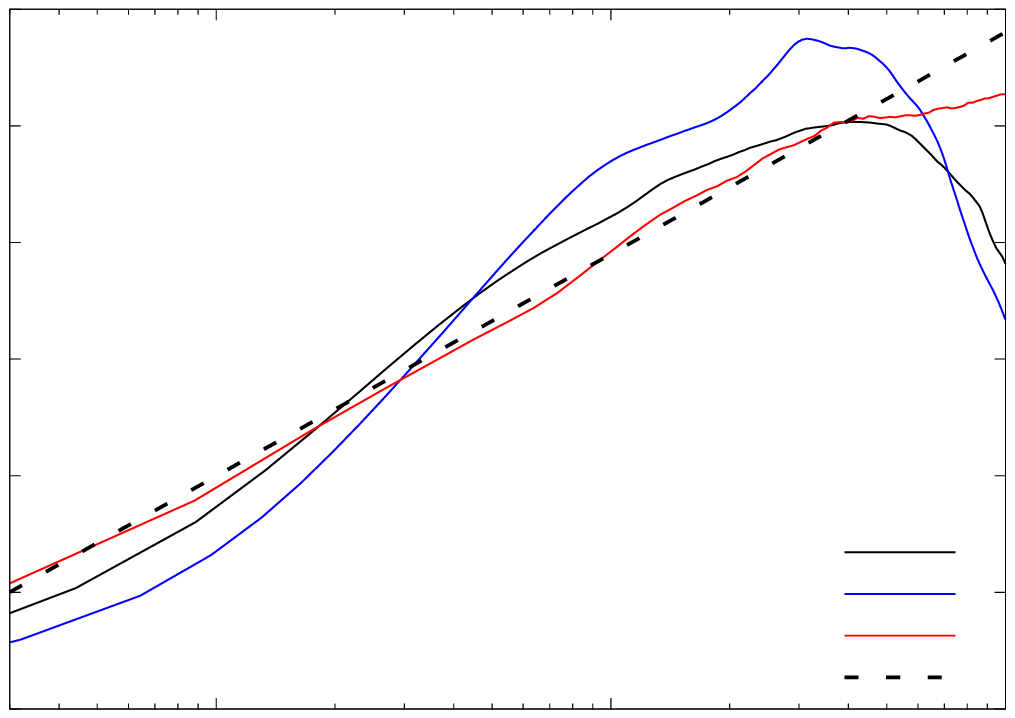}}%
    \gplfronttext
  \end{picture}%
\endgroup

%% file: logBLRe_Ra.tex
% GNUPLOT: LaTeX picture with Postscript
\begingroup
  \fontfamily{Times-Roman}%
  \selectfont
  \makeatletter
  \providecommand\color[2][]{%
    \GenericError{(gnuplot) \space\space\space\@spaces}{%
      Package color not loaded in conjunction with
      terminal option `colourtext'%
    }{See the gnuplot documentation for explanation.%
    }{Either use 'blacktext' in gnuplot or load the package
      color.sty in LaTeX.}%
    \renewcommand\color[2][]{}%
  }%
  \providecommand\includegraphics[2][]{%
    \GenericError{(gnuplot) \space\space\space\@spaces}{%
      Package graphicx or graphics not loaded%
    }{See the gnuplot documentation for explanation.%
    }{The gnuplot epslatex terminal needs graphicx.sty or graphics.sty.}%
    \renewcommand\includegraphics[2][]{}%
  }%
  \providecommand\rotatebox[2]{#2}%
  \@ifundefined{ifGPcolor}{%
    \newif\ifGPcolor
    \GPcolortrue
  }{}%
  \@ifundefined{ifGPblacktext}{%
    \newif\ifGPblacktext
    \GPblacktexttrue
  }{}%
  % define a \g@addto@macro without @ in the name:
  \let\gplgaddtomacro\g@addto@macro
  % define empty templates for all commands taking text:
  \gdef\gplbacktext{}%
  \gdef\gplfronttext{}%
  \makeatother
  \ifGPblacktext
    % no textcolor at all
    \def\colorrgb#1{}%
    \def\colorgray#1{}%
  \else
    % gray or color?
    \ifGPcolor
      \def\colorrgb#1{\color[rgb]{#1}}%
      \def\colorgray#1{\color[gray]{#1}}%
      \expandafter\def\csname LTw\endcsname{\color{white}}%
      \expandafter\def\csname LTb\endcsname{\color{black}}%
      \expandafter\def\csname LTa\endcsname{\color{black}}%
      \expandafter\def\csname LT0\endcsname{\color[rgb]{1,0,0}}%
      \expandafter\def\csname LT1\endcsname{\color[rgb]{0,1,0}}%
      \expandafter\def\csname LT2\endcsname{\color[rgb]{0,0,1}}%
      \expandafter\def\csname LT3\endcsname{\color[rgb]{1,0,1}}%
      \expandafter\def\csname LT4\endcsname{\color[rgb]{0,1,1}}%
      \expandafter\def\csname LT5\endcsname{\color[rgb]{1,1,0}}%
      \expandafter\def\csname LT6\endcsname{\color[rgb]{0,0,0}}%
      \expandafter\def\csname LT7\endcsname{\color[rgb]{1,0.3,0}}%
      \expandafter\def\csname LT8\endcsname{\color[rgb]{0.5,0.5,0.5}}%
    \else
      % gray
      \def\colorrgb#1{\color{black}}%
      \def\colorgray#1{\color[gray]{#1}}%
      \expandafter\def\csname LTw\endcsname{\color{white}}%
      \expandafter\def\csname LTb\endcsname{\color{black}}%
      \expandafter\def\csname LTa\endcsname{\color{black}}%
      \expandafter\def\csname LT0\endcsname{\color{black}}%
      \expandafter\def\csname LT1\endcsname{\color{black}}%
      \expandafter\def\csname LT2\endcsname{\color{black}}%
      \expandafter\def\csname LT3\endcsname{\color{black}}%
      \expandafter\def\csname LT4\endcsname{\color{black}}%
      \expandafter\def\csname LT5\endcsname{\color{black}}%
      \expandafter\def\csname LT6\endcsname{\color{black}}%
      \expandafter\def\csname LT7\endcsname{\color{black}}%
      \expandafter\def\csname LT8\endcsname{\color{black}}%
    \fi
  \fi
    \setlength{\unitlength}{0.0500bp}%
    \ifx\gptboxheight\undefined%
      \newlength{\gptboxheight}%
      \newlength{\gptboxwidth}%
      \newsavebox{\gptboxtext}%
    \fi%
    \setlength{\fboxrule}{0.5pt}%
    \setlength{\fboxsep}{1pt}%
\begin{picture}(7200.00,5040.00)%
    \gplgaddtomacro\gplbacktext{%
      \csname LTb\endcsname%%
      \put(1568,2410){\makebox(0,0)[r]{\strut{}$0.225$}}%
      \csname LTb\endcsname%%
      \put(1568,4719){\makebox(0,0)[r]{\strut{}$0.25$}}%
      \csname LTb\endcsname%%
      \put(1760,704){\makebox(0,0){\strut{}$10^{11}$}}%
      \put(2604,704){\makebox(0,0){\strut{}$10^{12}$}}%
      \put(3448,704){\makebox(0,0){\strut{}$10^{13}$}}%
      \put(4293,704){\makebox(0,0){\strut{}$10^{14}$}}%
      \put(5151,1024){\makebox(0,0)[l]{\strut{}$0.325$}}%
      \put(5151,2872){\makebox(0,0)[l]{\strut{}$0.35$}}%
      \put(5151,4719){\makebox(0,0)[l]{\strut{}$0.375$}}%
    }%
    \gplgaddtomacro\gplfronttext{%
      \csname LTb\endcsname%%
      \put(488,2871){\rotatebox{-270}{\makebox(0,0){\strut{}$\rm d log_{10}(Nu)/d log_{10}(Ra)$}}}%
      \csname LTb\endcsname%%
      \put(6087,2871){\rotatebox{-270}{\makebox(0,0){\strut{}$\rm d log_{10}(Re)/d log_{10}(Ra)$}}}%
      \put(3359,224){\makebox(0,0){\strut{}$\rm{Ra}$}}%
    }%
    \gplbacktext
    \put(0,0){\includegraphics{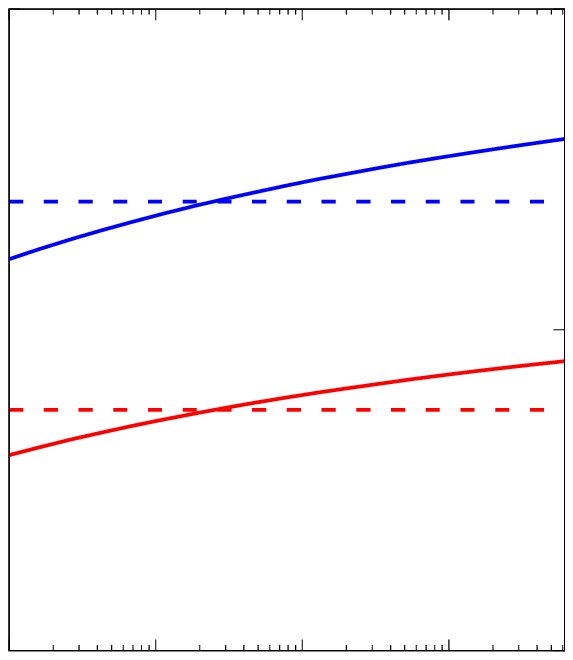}}%
    \gplfronttext
  \end{picture}%
\endgroup

%% file: logBLRe_Pr.tex
% GNUPLOT: LaTeX picture with Postscript
\begingroup
  \fontfamily{Times-Roman}%
  \selectfont
  \makeatletter
  \providecommand\color[2][]{%
    \GenericError{(gnuplot) \space\space\space\@spaces}{%
      Package color not loaded in conjunction with
      terminal option `colourtext'%
    }{See the gnuplot documentation for explanation.%
    }{Either use 'blacktext' in gnuplot or load the package
      color.sty in LaTeX.}%
    \renewcommand\color[2][]{}%
  }%
  \providecommand\includegraphics[2][]{%
    \GenericError{(gnuplot) \space\space\space\@spaces}{%
      Package graphicx or graphics not loaded%
    }{See the gnuplot documentation for explanation.%
    }{The gnuplot epslatex terminal needs graphicx.sty or graphics.sty.}%
    \renewcommand\includegraphics[2][]{}%
  }%
  \providecommand\rotatebox[2]{#2}%
  \@ifundefined{ifGPcolor}{%
    \newif\ifGPcolor
    \GPcolortrue
  }{}%
  \@ifundefined{ifGPblacktext}{%
    \newif\ifGPblacktext
    \GPblacktexttrue
  }{}%
  % define a \g@addto@macro without @ in the name:
  \let\gplgaddtomacro\g@addto@macro
  % define empty templates for all commands taking text:
  \gdef\gplbacktext{}%
  \gdef\gplfronttext{}%
  \makeatother
  \ifGPblacktext
    % no textcolor at all
    \def\colorrgb#1{}%
    \def\colorgray#1{}%
  \else
    % gray or color?
    \ifGPcolor
      \def\colorrgb#1{\color[rgb]{#1}}%
      \def\colorgray#1{\color[gray]{#1}}%
      \expandafter\def\csname LTw\endcsname{\color{white}}%
      \expandafter\def\csname LTb\endcsname{\color{black}}%
      \expandafter\def\csname LTa\endcsname{\color{black}}%
      \expandafter\def\csname LT0\endcsname{\color[rgb]{1,0,0}}%
      \expandafter\def\csname LT1\endcsname{\color[rgb]{0,1,0}}%
      \expandafter\def\csname LT2\endcsname{\color[rgb]{0,0,1}}%
      \expandafter\def\csname LT3\endcsname{\color[rgb]{1,0,1}}%
      \expandafter\def\csname LT4\endcsname{\color[rgb]{0,1,1}}%
      \expandafter\def\csname LT5\endcsname{\color[rgb]{1,1,0}}%
      \expandafter\def\csname LT6\endcsname{\color[rgb]{0,0,0}}%
      \expandafter\def\csname LT7\endcsname{\color[rgb]{1,0.3,0}}%
      \expandafter\def\csname LT8\endcsname{\color[rgb]{0.5,0.5,0.5}}%
    \else
      % gray
      \def\colorrgb#1{\color{black}}%
      \def\colorgray#1{\color[gray]{#1}}%
      \expandafter\def\csname LTw\endcsname{\color{white}}%
      \expandafter\def\csname LTb\endcsname{\color{black}}%
      \expandafter\def\csname LTa\endcsname{\color{black}}%
      \expandafter\def\csname LT0\endcsname{\color{black}}%
      \expandafter\def\csname LT1\endcsname{\color{black}}%
      \expandafter\def\csname LT2\endcsname{\color{black}}%
      \expandafter\def\csname LT3\endcsname{\color{black}}%
      \expandafter\def\csname LT4\endcsname{\color{black}}%
      \expandafter\def\csname LT5\endcsname{\color{black}}%
      \expandafter\def\csname LT6\endcsname{\color{black}}%
      \expandafter\def\csname LT7\endcsname{\color{black}}%
      \expandafter\def\csname LT8\endcsname{\color{black}}%
    \fi
  \fi
    \setlength{\unitlength}{0.0500bp}%
    \ifx\gptboxheight\undefined%
      \newlength{\gptboxheight}%
      \newlength{\gptboxwidth}%
      \newsavebox{\gptboxtext}%
    \fi%
    \setlength{\fboxrule}{0.5pt}%
    \setlength{\fboxsep}{1pt}%
\begin{picture}(7200.00,5040.00)%
    \gplgaddtomacro\gplbacktext{%
      \csname LTb\endcsname%%
      \put(1184,1024){\makebox(0,0)[r]{\strut{}$0$}}%
      \csname LTb\endcsname%%
      \put(1184,1845){\makebox(0,0)[r]{\strut{}$0.1$}}%
      \csname LTb\endcsname%%
      \put(1184,2666){\makebox(0,0)[r]{\strut{}$0.2$}}%
      \csname LTb\endcsname%%
      \put(1184,3487){\makebox(0,0)[r]{\strut{}$0.3$}}%
      \csname LTb\endcsname%%
      \put(1184,4308){\makebox(0,0)[r]{\strut{}$0.4$}}%
      \csname LTb\endcsname%%
      \put(2388,704){\makebox(0,0){\strut{}$10^{1}$}}%
      \put(4512,704){\makebox(0,0){\strut{}$10^{2}$}}%
      \put(5343,1024){\makebox(0,0)[l]{\strut{}$-1.2$}}%
      \put(5343,1948){\makebox(0,0)[l]{\strut{}$-1.1$}}%
      \put(5343,2872){\makebox(0,0)[l]{\strut{}$-1$}}%
      \put(5343,3795){\makebox(0,0)[l]{\strut{}$-0.9$}}%
      \put(5343,4719){\makebox(0,0)[l]{\strut{}$-0.8$}}%
    }%
    \gplgaddtomacro\gplfronttext{%
      \csname LTb\endcsname%%
      \put(288,2871){\rotatebox{-270}{\makebox(0,0){\strut{}$\rm d log_{10}(Nu)/d log_{10}(Pr)$}}}%
      \csname LTb\endcsname%%
      \put(6387,2871){\rotatebox{-270}{\makebox(0,0){\strut{}$\rm d log_{10}(Re)/d log_{10}(Pr)$}}}%
      \csname LTb\endcsname%%
      \put(3263,224){\makebox(0,0){\strut{}$\rm{Pr}$}}%
    }%
    \gplbacktext
    \put(0,0){\includegraphics{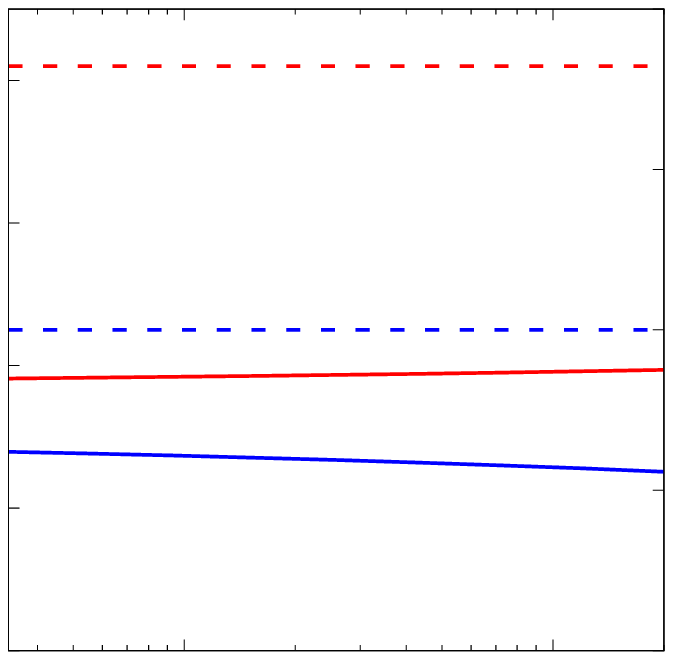}}%
    \gplfronttext
  \end{picture}%
\endgroup

%% file: Ko_Re.tex
% GNUPLOT: LaTeX picture with Postscript
\begingroup
  \fontfamily{Times-Roman}%
  \selectfont
  \makeatletter
  \providecommand\color[2][]{%
    \GenericError{(gnuplot) \space\space\space\@spaces}{%
      Package color not loaded in conjunction with
      terminal option `colourtext'%
    }{See the gnuplot documentation for explanation.%
    }{Either use 'blacktext' in gnuplot or load the package
      color.sty in LaTeX.}%
    \renewcommand\color[2][]{}%
  }%
  \providecommand\includegraphics[2][]{%
    \GenericError{(gnuplot) \space\space\space\@spaces}{%
      Package graphicx or graphics not loaded%
    }{See the gnuplot documentation for explanation.%
    }{The gnuplot epslatex terminal needs graphicx.sty or graphics.sty.}%
    \renewcommand\includegraphics[2][]{}%
  }%
  \providecommand\rotatebox[2]{#2}%
  \@ifundefined{ifGPcolor}{%
    \newif\ifGPcolor
    \GPcolortrue
  }{}%
  \@ifundefined{ifGPblacktext}{%
    \newif\ifGPblacktext
    \GPblacktexttrue
  }{}%
  % define a \g@addto@macro without @ in the name:
  \let\gplgaddtomacro\g@addto@macro
  % define empty templates for all commands taking text:
  \gdef\gplbacktext{}%
  \gdef\gplfronttext{}%
  \makeatother
  \ifGPblacktext
    % no textcolor at all
    \def\colorrgb#1{}%
    \def\colorgray#1{}%
  \else
    % gray or color?
    \ifGPcolor
      \def\colorrgb#1{\color[rgb]{#1}}%
      \def\colorgray#1{\color[gray]{#1}}%
      \expandafter\def\csname LTw\endcsname{\color{white}}%
      \expandafter\def\csname LTb\endcsname{\color{black}}%
      \expandafter\def\csname LTa\endcsname{\color{black}}%
      \expandafter\def\csname LT0\endcsname{\color[rgb]{1,0,0}}%
      \expandafter\def\csname LT1\endcsname{\color[rgb]{0,1,0}}%
      \expandafter\def\csname LT2\endcsname{\color[rgb]{0,0,1}}%
      \expandafter\def\csname LT3\endcsname{\color[rgb]{1,0,1}}%
      \expandafter\def\csname LT4\endcsname{\color[rgb]{0,1,1}}%
      \expandafter\def\csname LT5\endcsname{\color[rgb]{1,1,0}}%
      \expandafter\def\csname LT6\endcsname{\color[rgb]{0,0,0}}%
      \expandafter\def\csname LT7\endcsname{\color[rgb]{1,0.3,0}}%
      \expandafter\def\csname LT8\endcsname{\color[rgb]{0.5,0.5,0.5}}%
    \else
      % gray
      \def\colorrgb#1{\color{black}}%
      \def\colorgray#1{\color[gray]{#1}}%
      \expandafter\def\csname LTw\endcsname{\color{white}}%
      \expandafter\def\csname LTb\endcsname{\color{black}}%
      \expandafter\def\csname LTa\endcsname{\color{black}}%
      \expandafter\def\csname LT0\endcsname{\color{black}}%
      \expandafter\def\csname LT1\endcsname{\color{black}}%
      \expandafter\def\csname LT2\endcsname{\color{black}}%
      \expandafter\def\csname LT3\endcsname{\color{black}}%
      \expandafter\def\csname LT4\endcsname{\color{black}}%
      \expandafter\def\csname LT5\endcsname{\color{black}}%
      \expandafter\def\csname LT6\endcsname{\color{black}}%
      \expandafter\def\csname LT7\endcsname{\color{black}}%
      \expandafter\def\csname LT8\endcsname{\color{black}}%
    \fi
  \fi
    \setlength{\unitlength}{0.0500bp}%
    \ifx\gptboxheight\undefined%
      \newlength{\gptboxheight}%
      \newlength{\gptboxwidth}%
      \newsavebox{\gptboxtext}%
    \fi%
    \setlength{\fboxrule}{0.5pt}%
    \setlength{\fboxsep}{1pt}%
\begin{picture}(7200.00,5040.00)%
    \gplgaddtomacro\gplbacktext{%
      \csname LTb\endcsname%%
      \put(1376,1024){\makebox(0,0)[r]{\strut{}$10^{-3}$}}%
      \put(1376,1763){\makebox(0,0)[r]{\strut{}$10^{-2}$}}%
      \put(1376,2502){\makebox(0,0)[r]{\strut{}$10^{-1}$}}%
      \put(1376,3241){\makebox(0,0)[r]{\strut{}$10^{0}$}}%
      \put(1376,3980){\makebox(0,0)[r]{\strut{}$10^{1}$}}%
      \put(1376,4719){\makebox(0,0)[r]{\strut{}$10^{2}$}}%
      \put(1568,704){\makebox(0,0){\strut{}$10^{0}$}}%
      \put(2290,704){\makebox(0,0){\strut{}$10^{1}$}}%
      \put(3012,704){\makebox(0,0){\strut{}$10^{2}$}}%
      \put(3734,704){\makebox(0,0){\strut{}$10^{3}$}}%
      \put(4457,704){\makebox(0,0){\strut{}$10^{4}$}}%
      \put(5179,704){\makebox(0,0){\strut{}$10^{5}$}}%
      \put(5901,704){\makebox(0,0){\strut{}$10^{6}$}}%
      \put(6623,704){\makebox(0,0){\strut{}$10^{7}$}}%
      \put(2408,1800){\makebox(0,0)[l]{\strut{}$\rm{Ko}\sim \rm{Re}^{-1}$}}%
      \put(4457,4088){\makebox(0,0)[l]{\strut{}$\rm{Ko}\sim \rm{Re}^{-1/2}$}}%
    }%
    \gplgaddtomacro\gplfronttext{%
      \csname LTb\endcsname%%
      \put(288,2871){\rotatebox{-270}{\makebox(0,0){\strut{}$\rm{Re}^{-3} \rm{Ra} \rm{Pr}^{-2}$}}}%
      \put(4095,224){\makebox(0,0){\strut{}$\rm{Re}$}}%
    }%
    \gplbacktext
    \put(0,0){\includegraphics{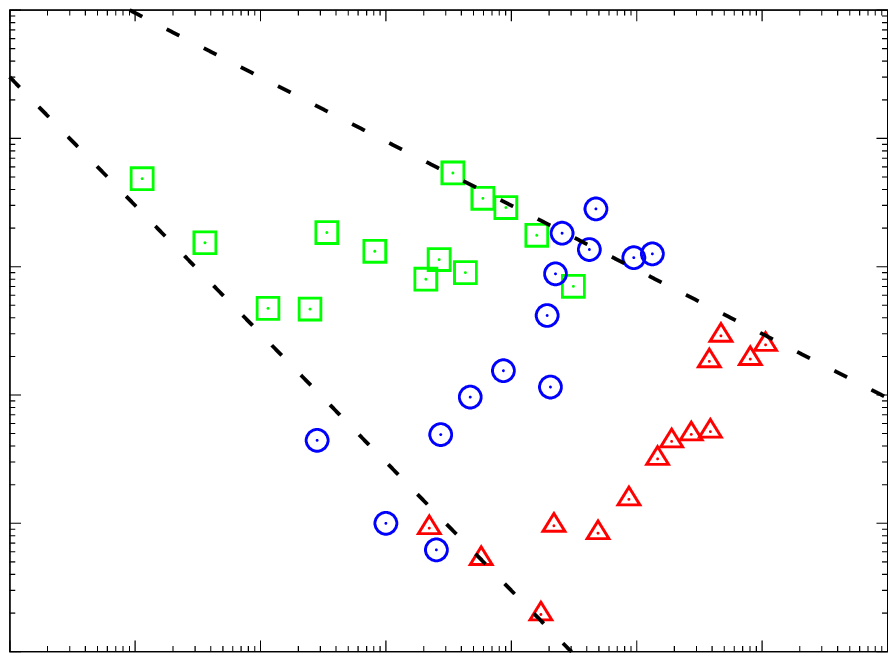}}%
    \gplfronttext
  \end{picture}%
\endgroup

%% file: Ko_Pe.tex
% GNUPLOT: LaTeX picture with Postscript
\begingroup
  \fontfamily{Times-Roman}%
  \selectfont
  \makeatletter
  \providecommand\color[2][]{%
    \GenericError{(gnuplot) \space\space\space\@spaces}{%
      Package color not loaded in conjunction with
      terminal option `colourtext'%
    }{See the gnuplot documentation for explanation.%
    }{Either use 'blacktext' in gnuplot or load the package
      color.sty in LaTeX.}%
    \renewcommand\color[2][]{}%
  }%
  \providecommand\includegraphics[2][]{%
    \GenericError{(gnuplot) \space\space\space\@spaces}{%
      Package graphicx or graphics not loaded%
    }{See the gnuplot documentation for explanation.%
    }{The gnuplot epslatex terminal needs graphicx.sty or graphics.sty.}%
    \renewcommand\includegraphics[2][]{}%
  }%
  \providecommand\rotatebox[2]{#2}%
  \@ifundefined{ifGPcolor}{%
    \newif\ifGPcolor
    \GPcolortrue
  }{}%
  \@ifundefined{ifGPblacktext}{%
    \newif\ifGPblacktext
    \GPblacktexttrue
  }{}%
  % define a \g@addto@macro without @ in the name:
  \let\gplgaddtomacro\g@addto@macro
  % define empty templates for all commands taking text:
  \gdef\gplbacktext{}%
  \gdef\gplfronttext{}%
  \makeatother
  \ifGPblacktext
    % no textcolor at all
    \def\colorrgb#1{}%
    \def\colorgray#1{}%
  \else
    % gray or color?
    \ifGPcolor
      \def\colorrgb#1{\color[rgb]{#1}}%
      \def\colorgray#1{\color[gray]{#1}}%
      \expandafter\def\csname LTw\endcsname{\color{white}}%
      \expandafter\def\csname LTb\endcsname{\color{black}}%
      \expandafter\def\csname LTa\endcsname{\color{black}}%
      \expandafter\def\csname LT0\endcsname{\color[rgb]{1,0,0}}%
      \expandafter\def\csname LT1\endcsname{\color[rgb]{0,1,0}}%
      \expandafter\def\csname LT2\endcsname{\color[rgb]{0,0,1}}%
      \expandafter\def\csname LT3\endcsname{\color[rgb]{1,0,1}}%
      \expandafter\def\csname LT4\endcsname{\color[rgb]{0,1,1}}%
      \expandafter\def\csname LT5\endcsname{\color[rgb]{1,1,0}}%
      \expandafter\def\csname LT6\endcsname{\color[rgb]{0,0,0}}%
      \expandafter\def\csname LT7\endcsname{\color[rgb]{1,0.3,0}}%
      \expandafter\def\csname LT8\endcsname{\color[rgb]{0.5,0.5,0.5}}%
    \else
      % gray
      \def\colorrgb#1{\color{black}}%
      \def\colorgray#1{\color[gray]{#1}}%
      \expandafter\def\csname LTw\endcsname{\color{white}}%
      \expandafter\def\csname LTb\endcsname{\color{black}}%
      \expandafter\def\csname LTa\endcsname{\color{black}}%
      \expandafter\def\csname LT0\endcsname{\color{black}}%
      \expandafter\def\csname LT1\endcsname{\color{black}}%
      \expandafter\def\csname LT2\endcsname{\color{black}}%
      \expandafter\def\csname LT3\endcsname{\color{black}}%
      \expandafter\def\csname LT4\endcsname{\color{black}}%
      \expandafter\def\csname LT5\endcsname{\color{black}}%
      \expandafter\def\csname LT6\endcsname{\color{black}}%
      \expandafter\def\csname LT7\endcsname{\color{black}}%
      \expandafter\def\csname LT8\endcsname{\color{black}}%
    \fi
  \fi
    \setlength{\unitlength}{0.0500bp}%
    \ifx\gptboxheight\undefined%
      \newlength{\gptboxheight}%
      \newlength{\gptboxwidth}%
      \newsavebox{\gptboxtext}%
    \fi%
    \setlength{\fboxrule}{0.5pt}%
    \setlength{\fboxsep}{1pt}%
\begin{picture}(7200.00,5040.00)%
    \gplgaddtomacro\gplbacktext{%
      \csname LTb\endcsname%%
      \put(1376,1024){\makebox(0,0)[r]{\strut{}$10^{-3}$}}%
      \put(1376,1763){\makebox(0,0)[r]{\strut{}$10^{-2}$}}%
      \put(1376,2502){\makebox(0,0)[r]{\strut{}$10^{-1}$}}%
      \put(1376,3241){\makebox(0,0)[r]{\strut{}$10^{0}$}}%
      \put(1376,3980){\makebox(0,0)[r]{\strut{}$10^{1}$}}%
      \put(1376,4719){\makebox(0,0)[r]{\strut{}$10^{2}$}}%
      \put(2956,704){\makebox(0,0){\strut{}$10^{2}$}}%
      \put(4426,704){\makebox(0,0){\strut{}$10^{3}$}}%
      \put(5896,704){\makebox(0,0){\strut{}$10^{4}$}}%
      \put(1928,2502){\makebox(0,0)[l]{\strut{}$\rm{Ko}\sim \rm{Pe}^{-1}$}}%
      \put(4426,4088){\makebox(0,0)[l]{\strut{}$\rm{Ko}\sim \rm{Pe}^{-1/2}$}}%
    }%
    \gplgaddtomacro\gplfronttext{%
      \csname LTb\endcsname%%
      \put(288,2871){\rotatebox{-270}{\makebox(0,0){\strut{}$\rm{Re}^{-3} \rm{Ra} \rm{Pr}^{-2}$}}}%
      \put(4095,224){\makebox(0,0){\strut{}$\rm{Pe}$}}%
    }%
    \gplbacktext
    \put(0,0){\includegraphics{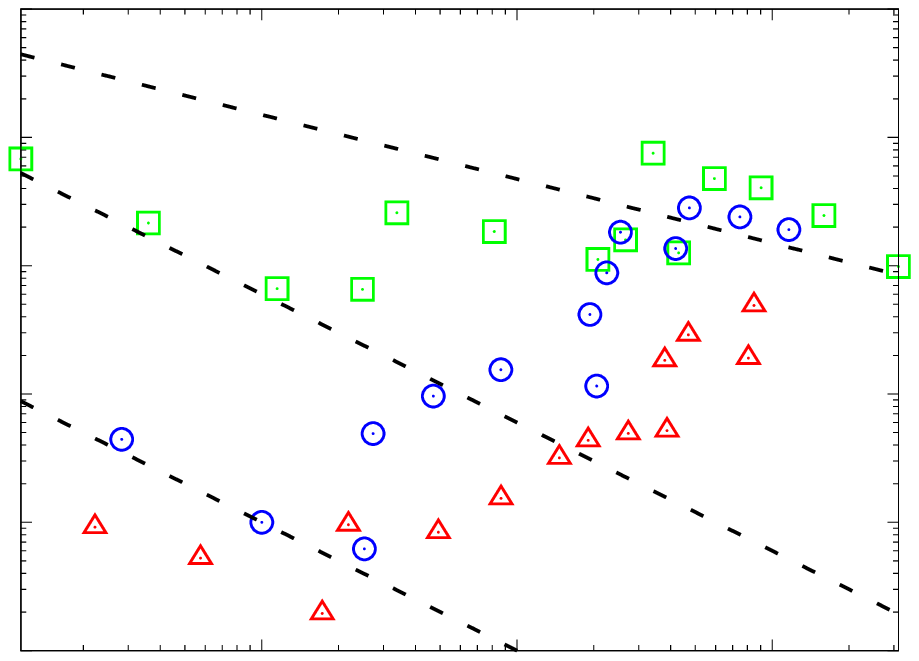}}%
    \gplfronttext
  \end{picture}%
\endgroup

%% file: Ko_Pr_Re.tex
% GNUPLOT: LaTeX picture with Postscript
\begingroup
  \fontfamily{Times-Roman}%
  \selectfont
  \makeatletter
  \providecommand\color[2][]{%
    \GenericError{(gnuplot) \space\space\space\@spaces}{%
      Package color not loaded in conjunction with
      terminal option `colourtext'%
    }{See the gnuplot documentation for explanation.%
    }{Either use 'blacktext' in gnuplot or load the package
      color.sty in LaTeX.}%
    \renewcommand\color[2][]{}%
  }%
  \providecommand\includegraphics[2][]{%
    \GenericError{(gnuplot) \space\space\space\@spaces}{%
      Package graphicx or graphics not loaded%
    }{See the gnuplot documentation for explanation.%
    }{The gnuplot epslatex terminal needs graphicx.sty or graphics.sty.}%
    \renewcommand\includegraphics[2][]{}%
  }%
  \providecommand\rotatebox[2]{#2}%
  \@ifundefined{ifGPcolor}{%
    \newif\ifGPcolor
    \GPcolortrue
  }{}%
  \@ifundefined{ifGPblacktext}{%
    \newif\ifGPblacktext
    \GPblacktexttrue
  }{}%
  % define a \g@addto@macro without @ in the name:
  \let\gplgaddtomacro\g@addto@macro
  % define empty templates for all commands taking text:
  \gdef\gplbacktext{}%
  \gdef\gplfronttext{}%
  \makeatother
  \ifGPblacktext
    % no textcolor at all
    \def\colorrgb#1{}%
    \def\colorgray#1{}%
  \else
    % gray or color?
    \ifGPcolor
      \def\colorrgb#1{\color[rgb]{#1}}%
      \def\colorgray#1{\color[gray]{#1}}%
      \expandafter\def\csname LTw\endcsname{\color{white}}%
      \expandafter\def\csname LTb\endcsname{\color{black}}%
      \expandafter\def\csname LTa\endcsname{\color{black}}%
      \expandafter\def\csname LT0\endcsname{\color[rgb]{1,0,0}}%
      \expandafter\def\csname LT1\endcsname{\color[rgb]{0,1,0}}%
      \expandafter\def\csname LT2\endcsname{\color[rgb]{0,0,1}}%
      \expandafter\def\csname LT3\endcsname{\color[rgb]{1,0,1}}%
      \expandafter\def\csname LT4\endcsname{\color[rgb]{0,1,1}}%
      \expandafter\def\csname LT5\endcsname{\color[rgb]{1,1,0}}%
      \expandafter\def\csname LT6\endcsname{\color[rgb]{0,0,0}}%
      \expandafter\def\csname LT7\endcsname{\color[rgb]{1,0.3,0}}%
      \expandafter\def\csname LT8\endcsname{\color[rgb]{0.5,0.5,0.5}}%
    \else
      % gray
      \def\colorrgb#1{\color{black}}%
      \def\colorgray#1{\color[gray]{#1}}%
      \expandafter\def\csname LTw\endcsname{\color{white}}%
      \expandafter\def\csname LTb\endcsname{\color{black}}%
      \expandafter\def\csname LTa\endcsname{\color{black}}%
      \expandafter\def\csname LT0\endcsname{\color{black}}%
      \expandafter\def\csname LT1\endcsname{\color{black}}%
      \expandafter\def\csname LT2\endcsname{\color{black}}%
      \expandafter\def\csname LT3\endcsname{\color{black}}%
      \expandafter\def\csname LT4\endcsname{\color{black}}%
      \expandafter\def\csname LT5\endcsname{\color{black}}%
      \expandafter\def\csname LT6\endcsname{\color{black}}%
      \expandafter\def\csname LT7\endcsname{\color{black}}%
      \expandafter\def\csname LT8\endcsname{\color{black}}%
    \fi
  \fi
    \setlength{\unitlength}{0.0500bp}%
    \ifx\gptboxheight\undefined%
      \newlength{\gptboxheight}%
      \newlength{\gptboxwidth}%
      \newsavebox{\gptboxtext}%
    \fi%
    \setlength{\fboxrule}{0.5pt}%
    \setlength{\fboxsep}{1pt}%
\begin{picture}(7200.00,5040.00)%
    \gplgaddtomacro\gplbacktext{%
      \csname LTb\endcsname%%
      \put(1376,1024){\makebox(0,0)[r]{\strut{}$10^{-3}$}}%
      \put(1376,1948){\makebox(0,0)[r]{\strut{}$10^{-2}$}}%
      \put(1376,2872){\makebox(0,0)[r]{\strut{}$10^{-1}$}}%
      \put(1376,3795){\makebox(0,0)[r]{\strut{}$10^{0}$}}%
      \put(1376,4719){\makebox(0,0)[r]{\strut{}$10^{1}$}}%
      \put(1942,704){\makebox(0,0){\strut{}$10^{3}$}}%
      \put(3428,704){\makebox(0,0){\strut{}$10^{4}$}}%
      \put(4914,704){\makebox(0,0){\strut{}$10^{5}$}}%
      \put(6400,704){\makebox(0,0){\strut{}$10^{6}$}}%
      \put(2651,2388){\makebox(0,0)[l]{\strut{}$\rm{Ko}\sim \rm{Re}^{-1}$}}%
    }%
    \gplgaddtomacro\gplfronttext{%
      \csname LTb\endcsname%%
      \put(288,2871){\rotatebox{-270}{\makebox(0,0){\strut{}$\rm{Re}^{-3} \rm{Ra} \rm{Pr}^{-2}$}}}%
      \put(4095,224){\makebox(0,0){\strut{}$\rm{Re}$}}%
    }%
    \gplbacktext
    \put(0,0){\includegraphics{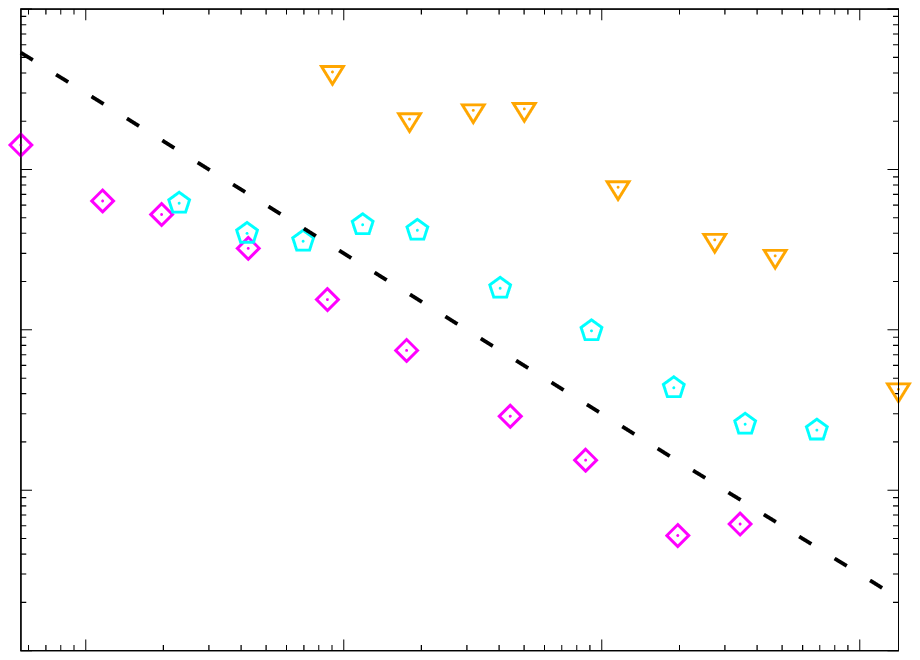}}%
    \gplfronttext
  \end{picture}%
\endgroup

%% file: Ko_Pr_Pe.tex
% GNUPLOT: LaTeX picture with Postscript
\begingroup
  \fontfamily{Times-Roman}%
  \selectfont
  \makeatletter
  \providecommand\color[2][]{%
    \GenericError{(gnuplot) \space\space\space\@spaces}{%
      Package color not loaded in conjunction with
      terminal option `colourtext'%
    }{See the gnuplot documentation for explanation.%
    }{Either use 'blacktext' in gnuplot or load the package
      color.sty in LaTeX.}%
    \renewcommand\color[2][]{}%
  }%
  \providecommand\includegraphics[2][]{%
    \GenericError{(gnuplot) \space\space\space\@spaces}{%
      Package graphicx or graphics not loaded%
    }{See the gnuplot documentation for explanation.%
    }{The gnuplot epslatex terminal needs graphicx.sty or graphics.sty.}%
    \renewcommand\includegraphics[2][]{}%
  }%
  \providecommand\rotatebox[2]{#2}%
  \@ifundefined{ifGPcolor}{%
    \newif\ifGPcolor
    \GPcolortrue
  }{}%
  \@ifundefined{ifGPblacktext}{%
    \newif\ifGPblacktext
    \GPblacktexttrue
  }{}%
  % define a \g@addto@macro without @ in the name:
  \let\gplgaddtomacro\g@addto@macro
  % define empty templates for all commands taking text:
  \gdef\gplbacktext{}%
  \gdef\gplfronttext{}%
  \makeatother
  \ifGPblacktext
    % no textcolor at all
    \def\colorrgb#1{}%
    \def\colorgray#1{}%
  \else
    % gray or color?
    \ifGPcolor
      \def\colorrgb#1{\color[rgb]{#1}}%
      \def\colorgray#1{\color[gray]{#1}}%
      \expandafter\def\csname LTw\endcsname{\color{white}}%
      \expandafter\def\csname LTb\endcsname{\color{black}}%
      \expandafter\def\csname LTa\endcsname{\color{black}}%
      \expandafter\def\csname LT0\endcsname{\color[rgb]{1,0,0}}%
      \expandafter\def\csname LT1\endcsname{\color[rgb]{0,1,0}}%
      \expandafter\def\csname LT2\endcsname{\color[rgb]{0,0,1}}%
      \expandafter\def\csname LT3\endcsname{\color[rgb]{1,0,1}}%
      \expandafter\def\csname LT4\endcsname{\color[rgb]{0,1,1}}%
      \expandafter\def\csname LT5\endcsname{\color[rgb]{1,1,0}}%
      \expandafter\def\csname LT6\endcsname{\color[rgb]{0,0,0}}%
      \expandafter\def\csname LT7\endcsname{\color[rgb]{1,0.3,0}}%
      \expandafter\def\csname LT8\endcsname{\color[rgb]{0.5,0.5,0.5}}%
    \else
      % gray
      \def\colorrgb#1{\color{black}}%
      \def\colorgray#1{\color[gray]{#1}}%
      \expandafter\def\csname LTw\endcsname{\color{white}}%
      \expandafter\def\csname LTb\endcsname{\color{black}}%
      \expandafter\def\csname LTa\endcsname{\color{black}}%
      \expandafter\def\csname LT0\endcsname{\color{black}}%
      \expandafter\def\csname LT1\endcsname{\color{black}}%
      \expandafter\def\csname LT2\endcsname{\color{black}}%
      \expandafter\def\csname LT3\endcsname{\color{black}}%
      \expandafter\def\csname LT4\endcsname{\color{black}}%
      \expandafter\def\csname LT5\endcsname{\color{black}}%
      \expandafter\def\csname LT6\endcsname{\color{black}}%
      \expandafter\def\csname LT7\endcsname{\color{black}}%
      \expandafter\def\csname LT8\endcsname{\color{black}}%
    \fi
  \fi
    \setlength{\unitlength}{0.0500bp}%
    \ifx\gptboxheight\undefined%
      \newlength{\gptboxheight}%
      \newlength{\gptboxwidth}%
      \newsavebox{\gptboxtext}%
    \fi%
    \setlength{\fboxrule}{0.5pt}%
    \setlength{\fboxsep}{1pt}%
\begin{picture}(7200.00,5040.00)%
    \gplgaddtomacro\gplbacktext{%
      \csname LTb\endcsname%%
      \put(1376,1024){\makebox(0,0)[r]{\strut{}$10^{-3}$}}%
      \put(1376,1948){\makebox(0,0)[r]{\strut{}$10^{-2}$}}%
      \put(1376,2872){\makebox(0,0)[r]{\strut{}$10^{-1}$}}%
      \put(1376,3795){\makebox(0,0)[r]{\strut{}$10^{0}$}}%
      \put(1376,4719){\makebox(0,0)[r]{\strut{}$10^{1}$}}%
      \put(2518,704){\makebox(0,0){\strut{}$10^{3}$}}%
      \put(5673,704){\makebox(0,0){\strut{}$10^{4}$}}%
    }%
    \gplgaddtomacro\gplfronttext{%
      \csname LTb\endcsname%%
      \put(288,2871){\rotatebox{-270}{\makebox(0,0){\strut{}$\rm{Re}^{-3} \rm{Ra} \rm{Pr}^{-2}$}}}%
      \put(4095,224){\makebox(0,0){\strut{}$\rm{Pe}$}}%
    }%
    \gplbacktext
    \put(0,0){\includegraphics{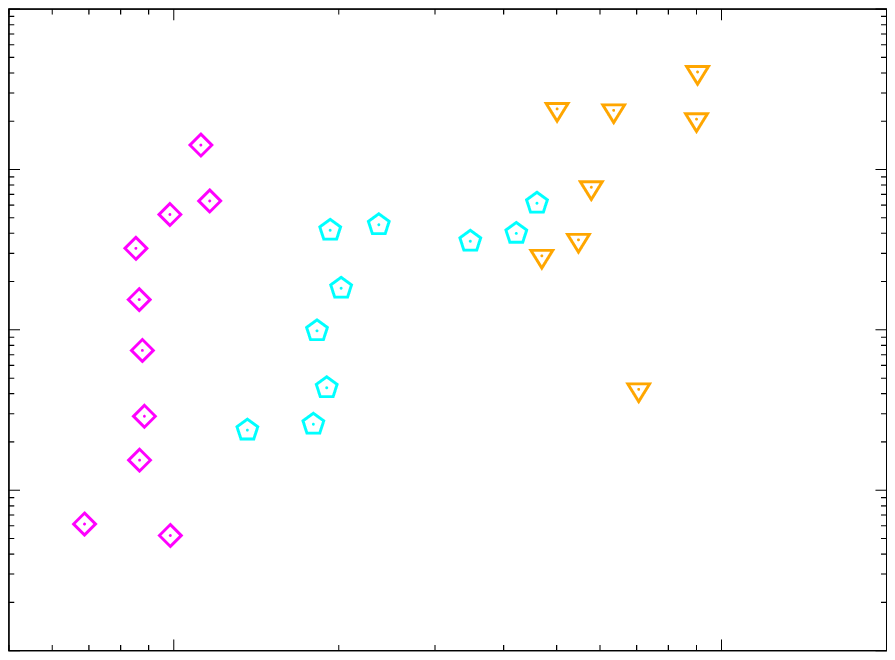}}%
    \gplfronttext
  \end{picture}%
\endgroup